\documentclass[
    openright,titlepage,numbers=noenddot,headinclude,
    cleardoublepage=empty,abstractoff, % <--- obsolete, remove (todo)
    BCOR=5mm,paper=a4,fontsize=11pt,%11pt,a4paper,%
    ngerman,american,%
    ]{scrreprt} % Don't forget to add the twoside option at the beginning here

\PassOptionsToPackage{eulerchapternumbers,listings,%
				 pdfspacing,%floatperchapter,%linedheaders,%
				 beramono,%eulermath,
				parts}{classicthesis}										
\usepackage{ifthen}
\newboolean{enable-backrefs} % enable backrefs in the bibliography
\setboolean{enable-backrefs}{false} % true false
\newcommand{\myTitle}{Ordering Information on Distributions\xspace}

\newcommand{\myName}{John van de Wetering\xspace}
\newcommand{\myProf}{Prof. Bob Coecke\xspace}

\newcommand{\myFaculty}{Wolfson College\xspace}
\newcommand{\myDepartment}{Department of Computer Science\xspace}
\newcommand{\myUni}{University of Oxford\xspace}

\newcommand{\myTime}{July 2016\xspace}
\newcommand{\myVersion}{1st edition\xspace}

\usepackage{amsthm}
\usepackage{chngcntr}
\counterwithin{figure}{chapter}

\newcounter{counter}[section]
\counterwithin{counter}{section}

\newcommand{\benviron}[1]{\ifthenelse{\value{counter}=0}{\refstepcounter{counter}}{}\begin{description}\item [#1 \arabic{chapter}.\arabic{section}.\arabic{counter}:]}
\newcommand{\eenviron}{\end{description}\refstepcounter{counter}}

\newcommand{\bdefin}{\benviron{Definition}}
\newcommand{\blemma}{\benviron{Lemma}}
\newcommand{\bexample}{\benviron{Example}}
\newcommand{\btheor}{\benviron{Theorem}}

\newcommand{\edefin}{\eenviron}
\newcommand{\elemma}{\eenviron}
\newcommand{\eexample}{\eenviron}
\newcommand{\etheor}{\eenviron}

\newcommand{\delim}{~;~}

\newcommand{\bprop}{\begin{prop}}
\newcommand{\bcoro}{\begin{coro}}
\newcommand{\bdefprop}{\begin{defprop}}
\newcommand{\edefprop}{\end{defprop}}
\newcommand{\eprop}{\end{prop}}
\newcommand{\ecoro}{\end{coro}}
\newcommand{\bconj}{\begin{conj}}
\newcommand{\econj}{\end{conj}}
\newcommand{\brem}{\begin{remark}}
\newcommand{\erem}{\endexem\end{remark}}

\newcommand{\sleq}{\sqsubseteq}

\newcounter{dummy} % necessary for correct hyperlinks (to index, bib, etc.)
\providecommand{\mLyX}{L\kern-.1667em\lower.25em\hbox{Y}\kern-.125emX\@}

\usepackage[utf8]{inputenc}
 \usepackage{inputenc}				

 \usepackage{babel}					

\PassOptionsToPackage{square,numbers}{natbib}
 \usepackage{natbib}				

\PassOptionsToPackage{fleqn}{amsmath}		% math environments and more by the AMS 
 \usepackage{amsmath}
 
\usepackage[usenames,dvipsnames]{xcolor}

\PassOptionsToPackage{T1}{fontenc} % T2A for cyrillics
	\usepackage{fontenc}     
\usepackage{textcomp} % fix warning with missing font shapes
\usepackage{scrhack} % fix warnings when using KOMA with listings package          
\usepackage{xspace} % to get the spacing after macros right  
\usepackage{mparhack} % get marginpar right
\PassOptionsToPackage{printonlyused,smaller}{acronym}
	\usepackage{acronym} % nice macros for handling all acronyms in the thesis
\usepackage{tabularx} % better tables
\usepackage{caption}
\usepackage{subcaption}
\usepackage{mathrsfs}
\usepackage{mathptmx}
\usepackage{amsfonts}
\usepackage{amssymb}
\usepackage{xfrac}
\usepackage{sidecap}
\usepackage{listings} 
\PassOptionsToPackage{pdftex,hyperfootnotes=false,pdfpagelabels}{hyperref}
	\usepackage{hyperref}  % backref linktocpage pagebackref
\PassOptionsToPackage{pdftex}{graphicx}
	\usepackage{graphicx} 

\newcommand{\backrefnotcitedstring}{\relax}%(Not cited.)
\newcommand{\backrefcitedsinglestring}[1]{(Cited on page~#1.)}
\newcommand{\backrefcitedmultistring}[1]{(Cited on pages~#1.)}
\ifthenelse{\boolean{enable-backrefs}}%
{%
		\PassOptionsToPackage{hyperpageref}{backref}
		\usepackage{backref} % to be loaded after hyperref package 
		   \renewcommand*{\backref}[1]{}  % disable standard
		   \renewcommand*{\backrefalt}[4]{% detailed backref
		      \ifcase #1 %
		         \backrefnotcitedstring%
		      \or%
		         \backrefcitedsinglestring{#2}%
		      \else%
		         \backrefcitedmultistring{#2}%
		      \fi}%
}{\relax}    

\hypersetup{%
    colorlinks=true, linktocpage=true, pdfstartpage=3, pdfstartview=FitV,%
    breaklinks=true, pdfpagemode=UseNone, pageanchor=true, pdfpagemode=UseOutlines,%
    plainpages=false, bookmarksnumbered, bookmarksopen=true, bookmarksopenlevel=1,%
    hypertexnames=true, pdfhighlight=/O,%nesting=true,%frenchlinks,%
    urlcolor=webbrown, linkcolor=RoyalBlue, citecolor=webgreen, %pagecolor=RoyalBlue,%
    pdftitle={\myTitle},%
    pdfauthor={\textcopyright\ \myName, \myUni, \myFaculty},%
    pdfsubject={},%
    pdfkeywords={},%
    pdfcreator={pdfLaTeX},%
    pdfproducer={LaTeX with hyperref and classicthesis}%
}   

\makeatletter
\@ifpackageloaded{babel}%
    {%
       \addto\extrasamerican{%
				}%
       \addto\extrasngerman{% 
				}%	
    }{\relax}
\makeatother

\usepackage{classicthesis} 
\begin{document}
\frenchspacing
\raggedbottom
\selectlanguage{american} % american ngerman
%\renewcommand*{\bibname}{new name}
%\setbibpreamble{}
\pagenumbering{roman}
\pagestyle{plain}
%********************************************************************
% Frontmatter
%*******************************************************
%\include{FrontBackmatter/DirtyTitlepage}
%*******************************************************
% Titlepage
%*******************************************************
\begin{titlepage}
	% if you want the titlepage to be centered, uncomment and fine-tune the line below (KOMA classes environment)
	\begin{addmargin}[-1cm]{-3cm}
    \begin{center}
        \large  

        \hfill

        \vfill

        \begingroup
            \color{Maroon}\spacedallcaps{\myTitle} \\ \bigskip
        \endgroup

        \spacedlowsmallcaps{\myName}
        
        \spacedlowsmallcaps{Master thesis}
        
         \medskip
        \vfill
        \medskip   
        %\myDegree \\
       \myDepartment \\                            
       \myFaculty \\
        \myUni \\ \bigskip

        \myTime\\

        \vfill

    \end{center}  
  \end{addmargin}       
\end{titlepage}   
\thispagestyle{empty}

\hfill

\vfill

\noindent\myName,
\textit{\myTitle} %\myDegree, 
\textcopyright\ \myTime

\bigskip

\noindent\spacedlowsmallcaps{Supervisor}: \\
\myProf
%\myOtherProf \\ 
%\mySupervisor
%
%\medskip
%
%\noindent\spacedlowsmallcaps{Location}: \\
%\myLocation
%
%\medskip
%
%\noindent\spacedlowsmallcaps{Time Frame}: \\
%\myTime

%\cleardoublepage\include{FrontBackmatter/Dedication}
%\cleardoublepage\include{FrontBackmatter/Foreword}
\cleardoublepage%*******************************************************
% Abstract
%*******************************************************
%\renewcommand{\abstractname}{Abstract}
\pdfbookmark[1]{Abstract}{Abstract}
\begingroup
\let\clearpage\relax
\let\cleardoublepage\relax
\let\cleardoublepage\relax

\chapter*{Abstract}
This thesis details a class of partial orders  on the space of probability distributions and the space of density operators which capture the idea of information content. Some links to domain theory and computational linguistics are also discussed. Chapter 1 details some useful theorems from order theory. In Chapter 2 we define a notion of an information ordering on the space of probability distributions and see that this gives rise to a large class of orderings. In Chapter 3 we extend the idea of an information ordering to the space of density operators and in particular look at the maximum eigenvalue order. We will discuss whether this order might be unique given certain restrictions. In Chapter 4 we discuss a possible application in distributional language models, namely in the study of entailment and disambiguation.

\vspace{2cm}
\chapter*{Acknowledgments}
I'd like to thank Bob Coecke for supervising me for the duration of my Master thesis work and for supplying me with a very interesting problem, Klaas Landsman for putting me in contact with the right people and for agreeing to be the second reader, and the people working at the Quantum Group for helping me with all the questions I had and for supplying a stimulating environment. 

The text, theorems and proofs contained in this thesis are, unless otherwise noted, my own work. Some of the results in this thesis related to distributional semantics have been presented at the 2016 Workshop on the Intersection between NLP, Physics and Cognitive Science and will be published in the proceedings as part of a EPTCS volume \cite{wetering2016}.
\begin{flushright}
-- John van de Wetering

3th of July 2016
\end{flushright}
\vfill

\endgroup			

\vfill
%\cleardoublepage\include{FrontBackmatter/Publication}
%\cleardoublepage\include{FrontBackMatter/Acknowledgments}
\pagestyle{scrheadings}
\cleardoublepage%*******************************************************
% Table of Contents
%*******************************************************
%\phantomsection
\refstepcounter{dummy}
\pdfbookmark[1]{\contentsname}{tableofcontents}
\setcounter{tocdepth}{2} % <-- 2 includes up to subsections in the ToC
\setcounter{secnumdepth}{3} % <-- 3 numbers up to subsubsections
\manualmark
\markboth{\spacedlowsmallcaps{\contentsname}}{\spacedlowsmallcaps{\contentsname}}

\tableofcontents 

\automark[section]{chapter}
\renewcommand{\chaptermark}[1]{\markboth{\spacedlowsmallcaps{#1}}{\spacedlowsmallcaps{#1}}}
\renewcommand{\sectionmark}[1]{\markright{\thesection\enspace\spacedlowsmallcaps{#1}}}
%*******************************************************
% List of Figures and of the Tables
%*******************************************************
\clearpage

\begingroup 
    \let\clearpage\relax
    \let\cleardoublepage\relax
    \let\cleardoublepage\relax
    %*******************************************************
    % List of Figures
    %*******************************************************    
    \phantomsection 
    \refstepcounter{dummy}
    \addcontentsline{toc}{chapter}{\listfigurename}
    \pdfbookmark[1]{\listfigurename}{lof}
    \listoffigures

    \vspace*{8ex}

    %*******************************************************
    % List of Tables
    %*******************************************************
    %\phantomsection 
    %\refstepcounter{dummy}
    %\addcontentsline{toc}{chapter}{\listtablename}
    %\pdfbookmark[1]{\listtablename}{lot}
    %\listoftables
        
    %\vspace*{8ex}
%   \newpage
    
    %*******************************************************
    % List of Listings
    %*******************************************************      
	  %\phantomsection 
    %\refstepcounter{dummy}
    %\addcontentsline{toc}{chapter}{\lstlistlistingname}
    %\pdfbookmark[1]{\lstlistlistingname}{lol}
    %\lstlistoflistings 

    %\vspace*{8ex}
       
    %*******************************************************
    % Acronyms
    %*******************************************************
    %\phantomsection 
    %\refstepcounter{dummy}
    %\pdfbookmark[1]{Acronyms}{acronyms}
    %\markboth{\spacedlowsmallcaps{Acronyms}}{\spacedlowsmallcaps{Acronyms}}
    %\chapter*{Acronyms}
    %\begin{acronym}[UML]
      %  \acro{UML}{Unified Modeling Language}
        %\acro{LQG}{Loop Quantum Gravity}
       % \acro{GR}{General Relativity}
       
    %\end{acronym}                     
\endgroup

\cleardoublepage

\pagenumbering{arabic}
%\setcounter{page}{90}
% use \cleardoublepage here to avoid problems with pdfbookmark
\cleardoublepage
\chapter*{Introduction}
\addtocontents{toc}{\protect\enlargethispage{3cm}}
\addcontentsline{toc}{chapter}{\tocEntry{Introduction}}
\pdfbookmark[1]{Introduction}{Introduction}
\markboth{Introduction}{Introduction}

One of the most fundamental ideas in mathematics (if not \emph{the} most) is that of ordering objects. Even before you can count numbers, you have to be able to say which number is bigger than another (idea stolen from \cite{coecke2013}).

One of the most fundamental ideas in science (if not \emph{the} most) is that of information. Science is ultimately about the pursuit of more accurate knowledge about the world and this knowledge is gained somehow via information transfer.

Combining these two fundamental idea's then gives rise to a natural question: what kind of order structure exists with relation to information content?

To start answering this question we have to precisely define what we mean by 'order structure' and 'information content'. The order structure we will take to be a \emph{partial order}, some properties of which we will look at in Chapter 1. Information content we will take to be a collection of certain properties of a \emph{state} an agent can be in. That is: certain states are more informative than others. The agent has a preference of being in the more informative state. The precise question we wish to answer in this thesis will then be: is there a natural choice of partial order on the space of states (either classical or quantum) that orders the states according to their information content? 

Note that we haven't actually defined yet what we mean by information content. We will actually not give a complete definition of that in this thesis. We will skirt the issue by specifying some minimal set of properties that a notion of information content should satisfy in Chapter 2, and look at what kind of partial orders are compatible with these properties. An exact definition of information content is left as an exercise to the reader.

In classical physics, a state can be represented by a probability distribution over the different definite (\emph{pure}) states a system can be in. A partial order over classical states will thus be a partial order on the space of probability distributions. In a similar vein in quantum physics a state can be represented by a density matrix. We will be looking at classical states (probability distributions) in Chapter 2, and at quantum states (density matrices) in Chapter 3.

A possible application of this theory is in computation. When we want to know if a certain computation is producing valuable output we might want to check whether the information content of the state the process is in is actually increasing or not. A powerful way to study the behaviour of processes is by using a special kind of partial order called a \emph{domain}. For this reason we will also show properties related to domain theory. Another application is in computational linguistics. Some concepts in language are related to the information content present in words. This is studied in detail in Chapter 4.
\chapter{Orderings}

\section{Definitions}

We'll start with the basic definitions related to orders.

\bdefin
    A \emph{preorder} $\sleq$ on a set $P$ is a binary relation which is 
    \begin{itemize}
    \item Reflexive: $\forall x \in P: x \sleq x$.
    \item Transitive: $\forall x,y,z \in P: x \sleq y \text{ and }y \sleq z \implies x\sleq z$.
    \end{itemize}
    A \emph{partial order} is a preorder that is also antisymmetric: 
    \begin{itemize}
    \item if $x\sleq y$ and $y\sleq x$ then $x=y$.
    \end{itemize}
    A set which has a partial order defined on it is called a \emph{poset} and is denoted as $(P,\sleq)$ or just as $P$ when it is clear which partial order we are referring to.
\edefin

\bdefin
    For an element $x$ in a poset $(P,\sleq)$ we define the \emph{upperset} of $x$ as $\uparrow x = \{y \in P\delim x\sleq y\}$ and conversely the \emph{downset} of $x$ as $\downarrow x = \{z \in P \delim z\sleq x\}$.
\edefin

\bdefin 
    Let $S\subseteq P$ be a subset of a poset. The \emph{join} (or \emph{supremum}) of $S$ if it exists is the smallest upper bound of $S$ and is denoted as $\vee S$. Conversely the \emph{meet} (or \emph{infinum}) of $S$ if it exists is the largest lower bound of $S$ and is denoted as $\wedge S$.
\edefin
So if the meet and join of $S$ exist we have for all $s\in S$ $s\sleq \vee S$ ($\vee S$ is an upperbound) and for all $p\in P$ that are upperbounds of $S$ $\vee S \sleq p$ ($\vee S$ is minimal), and the same with the directions reversed for $\wedge S$.

A specific kind of particularly nice type of poset is a \emph{domain}. In order to define what a domain is we need some further definitions.

\bdefin
    A subset $S\subseteq P$ of a poset $P$ is called (upwards) \emph{directed} iff for all $x,y \in S$ there is a $z\in S$ such that $x\sleq z$ and $y\sleq z$. A particular kind of directed subset is an \emph{increasing sequence}. This is a set of elements $(a_i)_{i\in \mathbb{N}}$ such that $a_i\sleq a_j$ for $i\leq j$.
\edefin

\bdefin
    For $x,y \in P$ we define $x\ll y$ iff for all directed subsets $S\subseteq P$ with existing supremum we have that when $y \sleq \vee S$ then there exists an $s\in S$ such that $x\sleq s$. We call $\ll$ the \emph{approximation} relation and say that $x$ approximates $y$. We denote Approx$(y) = \{x\in P \delim x\ll y\}$. We call the poset $P$ \emph{continuous} iff Approx$(y)$ is directed with supremum $y$, for all $y\in P$.
\edefin

\bdefin
    If all directed subsets of a poset $P$ have a join we call $P$ \emph{directed complete} and say that $P$ is a directed complete poset which we will abbreviate to \emph{dcpo}.
\edefin

\bdefin
    A poset $P$ is a \emph{domain} if it is a dcpo and continuous.
\edefin

Since we will often be talking about different partial orders on the same space, we will often say that a partial order itself is a dcpo/domain when it turns the underlying set into a dcpo/domain.

Domains are spaces that allow a natural way to talk about continuous approximation of elements\cite{abramsky1994}. This is why they are used when talking about for instance formal semantics of programming languages such as in \cite{winskel1993}. We will not specifically use the theory of domains, but we will note it when certain partial orders have a dcpo or domain structure.

When talking about mathematical structures we are of course interested in the structure preserving maps.
\bdefin
    A map $f: (S,\sleq_S) \rightarrow (P,\sleq_P)$ between posets (or preorders) is called \emph{monotone} if for all $a,b\in S$ with $a\sleq_S b$ we have $f(a)\sleq_P f(b)$. The map is called \emph{Scott-continuous} iff it preserves all directed joins. That is, if we have a directed subset of $S$ called $D$ whose join exists we have $\vee f(D) = f(\vee D)$.
\edefin
The relevant morphisms for posets are monotone maps, and for dcpo's they are Scott-continuous maps. Note that Scott-continuous maps are always monotone. If a monotone map $f$ is bijective and its inverse is also monotone then $f$ is called an \emph{order isomorphism} and $S$ and $P$ are called order isomorphic.

\bdefin
    Let $f: (S,\sleq_S) \rightarrow (P,\sleq_P)$ be a monotone map from a preordered set $S$ to a poset $P$. We call $f$ \emph{strict monotone} iff for all $a,b\in S$ with $a\sleq b$ and $f(a)=f(b)$ then $a=b$. If $f$ is furthermore Scott-continuous and $P$ a dcpo then we call $f$ a \emph{measurement}.
\edefin

Note that if $S$ is a preorder and it allows a strict monotone map to a poset, then $S$ is a partial order. Because $a\sleq_S b$ implies $f(a)\sleq_P f(b)$ and $b\sleq_S a$ implies $f(b)\sleq_P f(a)$, and since $\sleq_P$ is antisymmetric we have $f(a)=f(b)$ which by strictness implies $a=b$ so that $\sleq_S$ is also antisymmetric. Any injective monotone map is also strict monotone, so strict monotonicity can be seen as a generalisation of injectivity.

\section{Some examples}
Partial orders occur everywhere in mathematics, so we could list hundreds of examples, but a few will hopefully suffice.

\bexample
For any set $X$ the powerset $P(X)$ is a poset with the partial order given by inclusion. The maximal element is $X$ and the minimal element is the empty set. $P(X)$ is in fact a \emph{complete lattice}: all joins and meets exist and are given respectively by the union of the sets, and the intersection of the sets.
\eexample

\bexample
The real line $\mathbb{R}$ is a poset with $x\sleq y$ iff $x\leq y$. In fact, it is \emph{totally ordered}: for all $x\neq y$ in $\mathbb{R}$ we have either $x\sleq y$ or $y\sleq x$. $\mathbb{R}$ is also a lattice with the join and meet of finite sets given by the maximal and minimal elements of the set. For any poset $(P,\sleq)$ we denote the \emph{dual} order as $(P^*,\sleq^*)$, which is given by $x\sleq^* y$ iff $y\sleq x$. Let $[0,\infty)^*$ be the restriction of $\mathbb{R}$ to the positive reals with the reversed order. The maximal element is 0 and any directed set is an decreasing sequence in $\mathbb{R}$ bounded by 0, so the supremum is well defined. So $[0,\infty)^*$ contains all joins of directed sets, so it is a dcpo. We furthermore have $x\ll y$ iff $y<x$ which means that it is continuous, so $[0,\infty)^*$ is a domain.
\eexample

\bexample
For a locally compact space $X$ its \emph{upper space} is given by
\begin{equation*}
    UX = \{K\subseteq X\delim \emptyset \neq K \text{ compact}\}.
\end{equation*}
When equipped with the reversed inclusion order: $A\sleq B$ iff $B\subseteq A$ it is a continuous dcpo with the join of a directed subset given by the intersection (which is again a compact set, and garantueed non-empty because of directedness) and $A \ll B$ iff $B\subseteq $int$(A)$, where int$(A)$ denotes the interior of a set. The maximal elements are the singletons, and $UX$ has a minimal element if and only if $X$ is compact, in which case $X$ is the minimal element.
\eexample

If $X$ is compact then $UX$ is compact, and if $X$ is a compact metric space then $UX$ is also a compact metric space with the metric given by
\begin{equation*}
    d_{UX}(A,B) = \max\left\{\sup_{a\in A}\{d_X(a,B)\}, \sup_{b\in B}\{d_X(A,b)\}\right\}.
\end{equation*}

\section{Proving directed completeness}

There are some general methods to show that a partial order is directed complete. A useful one was given in \cite[Theorem 2.2.1]{martinphd}:

\btheor Given a poset $(P,\sleq)$ and a map $\mu: P \rightarrow [0,\infty)*$ that is strict monotone and preserves the joins of directed sequences we have the following:
\begin{itemize}
    \item P is a dcpo.
    \item $\mu$ is Scott continuous
    \item Every directed subset $S\subseteq P$ contains an increasing sequence whose supremum is $\vee S$.
    \item For all $x,y \in P$, $x\ll y$ iff for all increasing sequences $(a_i)$ with $y\sleq \vee a_i$ then there is an $n$ such that $x\sleq a_n$.
    \item For all $x\in P$ Approx$(x)$ is directed with supremum $x$ if and only if it contains an increasing sequence with supremum $x$.
\end{itemize}
\label{theor:martinphd}
\etheor

In short, such a map $\mu$ makes sure $P$ is a dcpo and that wherever you normally have to work with a directed set you can instead simplify to working with an increasing sequence.

We will now show that a certain class of topological posets has the same sort of properties. Results very similar to these can be found in \cite[Chapter VI]{scottbook2003} although the results proven here are sometimes slightly more general. We also use different terminology\footnote{The proofs given here are original because the author wasn't aware that these statements were already proven. Since the source mentioned is behind a paywall, the proofs here can be seen as a public service.}.

\bdefin
Let $X$ be a Haussdorff topological space. It is called \emph{first countable} if it admits a countable neighbourhood basis. It is called \emph{separable} if it contains a countable dense subset, and it is called \emph{sequentially compact} iff any sequence contains a convergent subsequence.
\edefin
$X$ being Haussdorf means that limits of nets and sequences are unique when they exist. First countable means that the topology can be understood in terms of sequences, instead of the more general nets. Separable ensures that the space isn't too large. Sequentially compact is a different notion of compactness. For metric spaces it is equivalent to the requirement of compactness.

\bdefin
Let $(X,\sleq)$ be a poset with $X$ first countable Haussdorff. We call $\sleq$ \emph{upwards small} iff for every $x\in X$, $\uparrow x$ is sequentially compact and $\downarrow x$ is closed. Dually, it is called downwards small if $\downarrow x$ is sequentially compact and $\uparrow x$ is closed. It is called \emph{small} iff it is downwards small and upwards small.
\edefin

An upwards small poset has uppersets that are bounded in a certain sense. Indeed for $X$ a subset of Euclidean space, sequentially compact is equivalent to bounded and closed. Note that a sequentially compact subspace is always closed. Closedness of uppersets (or downsets) means that if we have a convergent sequence $a_n \rightarrow a$ and $x\sleq a_n$ (or $a_n\sleq x$) for all $n$, then $x\sleq a$ (or $a\sleq x$). Note that if $X$ is sequentially compact, then a poset is upwards small if and only if it is downwards small. This definition als works for preorders, since the required properties have nothing to do with antisymmetry, but when not otherwise specified we will assume $\sleq$ to be a partial order. Upwards small partial orders turn out to interact really nicely with the topology. They are a special case of what in the literature is known as a \emph{pospace}: a topological poset $(X,\sleq)$ where the graph of $\sleq$ is a closed subset of $X^2$.

\blemma
Let $(X, \sleq)$ be a first countable Haussdorff space with $\sleq$ upwards small, then
\begin{itemize}
    \item All increasing sequences have a join.
    \item All increasing sequences converge.
    \item the join of an increasing sequence is equal to its limit.
\end{itemize}
\elemma
\begin{proof}
Let $(x_i)$ be an increasing sequence in $X$. We have $x_1\sleq x_i$ for all $i$, so $x_i\in \uparrow x_1$. The sequence lies in $\uparrow x_1$ which is sequentially compact, so there exists a convergent subsequence $x_{m_j} \rightarrow x$. Since we have $x_{m_j}\sleq x_{m_k}$ for $j\leq k$, we can take the limit on the right side and use the closedness of the uppersets to get $x_{m_j}\sleq x$. So $x$ is an upper bound of $(x_{m_j})$. Suppose $y$ is also an upper bound. Then we get $x_{m_j}\sleq y$ for all $j$. By the closedness of lowersets we get $x\sleq y$, so that $x$ is the least upper bound of $(x_{m_j})$. Now, since for any $x_i$ we can find a $j$ such that $x_i\sleq x_{m_j}$ we see that $x$ is also an upper bound of $x_i$. Any upper bound of $(x_i)$ will also be an upper bound of $(x_{m_j})$, so that $x$ is the least upper bound of $(x_i)$. By the same argument as above, any convergent subsequence of $(x_i)$ will converge to a least upper bound of $(x_i)$. A least upper bound is unique, so they all converge to $x$. Therefore $(x_i)$ is convergent and we have
\begin{equation*}
    \vee x_i = \lim_{i\rightarrow \infty} x_i
\end{equation*}
\end{proof}

\blemma
Let $(X,\sleq)$ be a first countable Haussdorff space with $\sleq$ upwards small and let $S$ be a directed set. $\overline{S}$ denotes the closure of $S$. Then
\begin{itemize}
    \item $\overline{S}$ is directed.
    \item $S$ has a least upper bound if and only if $\overline{S}$ has a least upper bound.
    \item If either $S$ or $\overline{S}$ has a least upper bound, then $\vee S = \vee \overline{S}$.
\end{itemize}
\label{lemma:directed}
\elemma
\begin{proof}
An element in $\overline{S}$ is the limit of a sequence in $S$, so take $s_i \rightarrow s \in \overline{S}$ and $t_i \rightarrow t \in \overline{S}$, where $s$ and $t$ are arbitrary elements in $\overline{S}$. Because of the directedness of $S$ we can find $z_i$ such that $s_i,t_i \sleq z_i$. Let $z_1^\prime = z_1$ and let $z_i^\prime$ be chosen so that it is bigger than $z_i$ and $z_{i-1}^\prime$, then $(z_i^\prime)$ is an increasing sequence. By the lemma above it is convergent, so that the limit/join $z_i^\prime \rightarrow z$ lies in $\overline{S}$ and we have $s_i\sleq z$ for all $i$. Because downsets are closed we then get $s\sleq z$, and the same for $t \sleq z$. So for any $s,t\in \overline{S}$ we can find $z\in\overline{S}$ such that $s,t\sleq z$. So if $S$ is directed, then $\overline{S}$ is directed as well.

Suppose $S$ has a join: $\vee S = x$. Then $S \subseteq \downarrow x$. Taking the closure on both sides and using that downsets are closed we get $\overline{S} \subseteq \downarrow x$, so $x$ is an upper bound of $\overline{S}$. Suppose $y$ is another upper bound of $\overline{S}$, then we get $S\subseteq \overline{S} \subseteq \downarrow y$, so that $y$ is an upper bound of $S$ as well, so that $y\sleq x$. The other direction works similarly.
\end{proof}

If we also require that $X$ is separable we can do a bit more than this. We get an analog of Theorem 1.3.1:
\btheor
Let $(X, \sleq)$ be a first countable separable Haussdorff space with $\sleq$ upwards small, then 
\begin{enumerate}
    \item $X$ is a dcpo.
    \item Every directed set $S\subseteq X$ has a join in its closure: $\vee S = \vee \overline{S} \in \overline{S}$.
    \item Every directed set $S\subseteq X$ contains an increasing sequence with the same join.
    \item For all $x,y \in X$, $x\ll y$ iff for all increasing sequences $(a_i)$ with $y\sleq \vee a_i$ there is an $n$ such that $x\sleq a_n$.
    \item For all $x\in X$, Approx$(x)$ is directed with supremum $x$ if and only if it contains an increasing sequence with supremum $x$.
\end{enumerate}
\etheor
\begin{proof}
Let $S\subseteq X$ be a directed subset. If it is finite then it contains its maximal element and we are done, so suppose it is infinite. Because of the previous lemma we can take $S$ to be closed.

Let $(s_i)$ be any countable collection of elements in $S$. Then set $x_1 = s_1$. By directedness, there exists an element $x_2$ in $S$ such that $x_1,s_2\sleq x_2$. Continuing this procedure we construct an increasing sequence $(x_i)$ in $S$ such that $s_i\sleq x_i$. We know that $(x_i)$ has a join equal to its limit. Since $x_i\in S$ and $S$ is closed, the limit is an element $s\in S$. So we know that $s$ is an upperbound of $(s_i)$. We conclude that any countable subset of $S$ contains an upperbound in $S$.

Now because $X$ is separable, there exists a countable dense subset $A\subseteq X$. Dense here means that $\overline{A} = X$. $A\cap S$ is a dense subset of $S$, because $S$ is closed. Since $A\cap S$ is a countable subset of a directed set, it has an upper bound $x\in S$. So we have $A\cap S \subseteq \downarrow x$. Now taking the closure on both sides we get $\overline{A\cap S} = \overline{A} \cap \overline{S} = X\cap S = S \subset \overline{\downarrow x} = \downarrow x$. So $x$ is an upperbound of $S$. Now suppose $y$ is another upper bound of $S$. Since $x\in S$, we must have $x\sleq y$. So $x$ is the least upper bound of $S$.

If $S$ is not necessarily closed, then we can use the above argument for $\overline{S}$, to find a least upper bound $x\in\overline{S}$ that is also a least upper bound of $S$.

Since the least upper bound of any directed set $S$ is contained in its closure there is a convergent sequence $s_i \rightarrow \vee S$. Set $x_1 = s_1$, and let $x_2$ be such that $x_1,s_2\sleq x_2$, then we construct an increasing sequence $(x_i)$ in $S$. This sequence is necessarily convergent: $x_i \rightarrow y$ and we have $s_i\sleq y$ for all $i$, so we also have $\lim s_i = \vee S \sleq y$, but since $\vee S$ is the least upper bound of $\overline{S}$ we get $y\sleq \vee S$, so that $y=\vee S$.

For the last two points we only need to prove the only if part. Suppose we have $x$ and $y$ such that for all directed sequences $(a_i)$ with $y\sleq \vee a_i$ there is an $n$ such that $x\sleq a_n$. Let $S$ be a directed subset such that $y\sleq \vee S$. There is a directed sequence $(s_i)$ in $S$ such that $s_i \rightarrow \vee S$. So $y\sleq \lim s_i = \vee s_i$, but then there is an $n$ such that $x\sleq s_n$. Since $s_n\in S$, we have $x\ll y$.

For the last point, suppose Approx$(x)$ contains an increasing sequence with join $x$. Call this sequence $(x_i)$. Let $z_1,z_2\in$ Approx$(x)$. We have $x\sleq x = \lim x_i = \vee x_i$, so there are $n$ and $m$ such that $z_1\sleq x_n$ and $z_2\sleq x_m$. Let $k=\max\{n,m\}$. Then $z_1,z_2\sleq x_k \in $Approx$(x)$. So Approx$(x)$ is directed. Since it is contained inside of $\downarrow x$, its join is smaller than $x$, and because $\vee x_n = x$, the join is exactly $x$.
\end{proof}

So we see that these type of spaces are really well behaved. The closedness of the upper and lowersets ensures that the joins of directed sets are ``nearby'' to the set (in the closure of the set). 

We can actually weaken the requirements on $X$ a bit further, by moving the separable condition from $X$ to the directed sets in $X$. For instance, we have the following:
\btheor
Let $(X,\sleq)$ be a metric space with an upwards small partial order $\sleq$. Then the above theorem still holds.
\etheor
\label{theor:separable}
\begin{proof}
Since $X$ is a metric space, it is first countable and Haussdorff so we can use the lemma's proved earlier. For a metric space sequential compactness is equivalent to compactness. 

Let $S$ be a closed directed set in $X$. Pick an arbitrary $s\in S$. Then $\uparrow s$ is a compact set and $\uparrow s \cap S$ is the intersection of a closed set with a compact set, so is compact. Furthermore $\uparrow s \cap S$ is a directed set. A compact subset of a metric space inherits the metric, so is a compact metric space. Compact metric spaces are always separable, so we can use the previous theorem to find a least upper bound of $\uparrow s \cap S$. Call this $x$. Let $t\in S$. By directedness there is a $z \in S$ such that $t,s\sleq z$. But then $z\in \uparrow S$, so $t\sleq z\sleq x$. So $x$ is also an upper bound for $S$ and if $y$ is another upper bound of $S$ then it must also be an upper bound of $\uparrow s \cap S$, so that $x\sleq y$. For $S$ not closed, we can use Lemma \ref{lemma:directed}.
\end{proof}

Due to these theorems, directedness can be understood in terms of increasing sequences:

\blemma
Let $X$ and $P$ be first countable separable Haussdorff with partial orders that are upwards small, then for a monotone map $f: X\rightarrow P$ we have that it is Scott-continuous iff it preserves suprema of increasing sequences.
\elemma
\begin{proof}
A Scott-continuous map obviously preserves suprema of increasing sequences, so we only have to check the other direction. Let $f$ be a map that preserves the join of increasing sequences and let $S$ be a closed directed set. Then $f(S)$ is also a directed set by monotonicity of $f$. Then there is an increasing sequence $f(s_i)$ in $f(S)$ with $f(s_i) \rightarrow \vee f(S)$, and $s_i$ is a sequence in $S$, and we can construct an increasing sequence $x_i \rightarrow x$ in $S$ with $s_i\sleq x_i\sleq x$. Then $f(s_i)\sleq f(x)$ and by closedness $\lim f(s_i) = \vee f(S) \sleq f(x)$, but because $S$ is closed we have $x\in S$, so $f(x)\in f(S)$, so $f(x)\sleq \vee f(S)$, so $f(x) = \vee f(S)$. Since $x\sleq \vee S$ we have so $f(x)=\vee f(S) \sleq f(\vee S)$. For the other direction: there exists an increasing convergent sequence $s_i \rightarrow \vee S$. Then $f(\vee S) = f(\vee s_i) = \vee f(s_i) \sleq \vee f(S)$, because we are taking the join over a subset of $f(S)$ so that the join will be smaller. We conclude that $f(\vee S) = \vee f(S)$.

Now suppose $S$ is not necessarily closed. We know that $f(\vee S) = f(\vee \overline{S}) = \vee f(\overline{S})$ and since $S\subseteq \overline{S}$ we have $\vee f(S) \sleq \vee f(\overline{S}) = f(\vee S)$. In the other direction we again have $f(\vee S) = f(\vee s_n) = \vee f(s_n) \sleq \vee f(S)$, so $f(\vee S) = \vee f(S)$.
\end{proof}

From now on, when we are talking about a set with an upwards small partial order, it is to be understood that the space is first countable separable Haussdorff.

\blemma
Let $X$ and $P$ be posets with an upwards small partial order. Any continuous (in the topologies of $X$ and $P$) monotone map $f: X \rightarrow P$ is Scott-continuous.
\label{lemma:scottcont}
\elemma
\begin{proof}
Due to the previous lemma we only have to prove that $f$ preserves suprema of increasing sequences. But if $(x_i)$ is an increasing sequence in $X$, then it is also convergent, and the join is equal to the limit. Note that for any continuous map $f(\lim x_n) = \lim f(x_n)$, so we get
\begin{equation*}
    f(\vee x_n) = f(\lim x_n) = \lim f(x_n) = \vee f(x_n).
\end{equation*}
\end{proof}

This is not too surprising. In fact a map between posets is Scott-continuous when it is continuous with respect to their Scott topology. The closed sets in this topology are given by Scott downsets, a basis of which is given by the downsets of all elements in $x$. Since all these downsets are closed with respect to the original topology, we see that the original topology is finer than the Scott topology. So if we have a continuous map from $X$ in its Scott topology to $P$ in its original topology (a monotone continuous map) we can replace the topology on $P$ with a coarser topology, in this case the Scott topology, and we are left with a Scott continuous map. For this reason such a partial order is also called \emph{compatible} (with the topology) in the literature.

\section{Examples of upwards small posets}
\bexample
Any metric space satisfies the necessary topological conditions for Theorem \ref{theor:separable}, so for $X$ a metric space the requirement for upwards smallness becomes that uppersets are compact and the downsets are closed. If $X$ is compact itself, it is enough to require that uppersets and downsets are closed. We call a partial order whose uppersets and downsets are closed, a \emph{closed} partial order.

For any (sequentially) compact, first countable, separable Haussdorff space $X$ with partial order $\sleq$. The closedness, upward smallness and downward smallness of $\sleq$ are equivalent.
\eexample

\bexample
The real line with its standard ordering is a closed poset. So any closed subset of the real line that is bounded from above is upwards small. When equipped with the reversed ordering, any closed subset bounded from below is upwards closed. For example: $[0,\infty)^*$.
\eexample

\bexample
Any poset $(X,\sleq)$ where we equip $X$ with the discrete topology is closed. It is upwards small iff all the uppersets are finite. Since any discrete space can be seen as a metric space with the discrete metric, Theorem \ref{theor:separable} applies. For instance $\mathbb{N}$ with $a\sleq b$ iff $b\leq a$ is upwards small. Similarly the set of finite subsets of $\mathbb{N}$ with the reversed inclusion order and the topology of the upper space is upwards small.
\eexample

\bexample
Define $PO(n) = \{A\in M^{n\times n}(\mathbb{C})\delim A^\dagger = A, \forall v\in \mathbb{C}^n: v^\dagger A v \geq 0\}$ the space of positive operators on $\mathbb{C}^n$. We write the condition that $v^\dagger A v\geq 0$ for all $v$ as $A\geq 0$. $PO(n)$ allows a natural choice for a partial order called the Löwner order: $A\sleq_L B$ iff $B-A\geq 0$. $PO(n)$ can be seen as a subset of $\mathbb{C}^{n^2}$, so it satisfies the necessary topological conditions. This means that the map $f_v(A) = v^\dagger A v$ is continuous for any $v\in\mathbb{C}^n$. This ensures that $\sleq_L$ is closed. Note that all the elements in $\downarrow A$ have a smaller trace than $A$, so $\downarrow A$ is bounded, which together with its closedness makes it compact, so $\sleq_L$ is downwards small. Any compact subset of $PO(n)$ will therefore be a dcpo, and the dual order of $\sleq_L$ on $PO(n)$ is upwards small so that it is a dcpo. More on this in Chapter 3.
\eexample

We can construct new upwards small partial orders from other ones.

\blemma
For $1\leq i \leq k$ let $(P_i,\sleq_i)$ be a finite collection of closed/upwards small posets/preorder spaces, then $P = \prod_i P_i$ is a a closed/upwards small poset/preorder space, given by the product partial order/preorder.
\elemma
\begin{proof}
We equip $P$ with the regular product topology. Since we are considering finite products, this preserves the Haussdorff, separable, first countable and possibly (sequential) compactness properties. For $(x_i),(y_i) \in P$ where $x_i,y_i\in P_i$ define $(x_i)\sleq (y_i)$ iff for all $1\leq i \leq k$ $x_i\sleq_i y_i$. Transitivity and reflexitivity (and antisymmetry in the case of a partial order) carry over from the $\sleq_i$. And we have $\uparrow (x_i) = \prod_i \uparrow x_i$ and the same for downsets, so closedness and sequential compactness are preserved.

Note also that the projection maps $\pi_i: P \rightarrow P_i$ are monotone and continuous, so they are Scott-continuous (when all the $P_i$ are partial orders).
\end{proof}

\btheor
Let $X$ be a sequentially compact first countable separable Haussdorff space and $P$ some first countable Haussdorff space with a closed partial order (respectively a preorder) $\sleq$ and $f: X \rightarrow P$ continuous and injective, then $X$ inherits a small partial order (respectively a preorder) from $P$ via $f$.
\label{theor:inducedpo}
\etheor
\begin{proof}
Define the partial order $\sleq_f$ for $x,y \in X$ as $x\sleq_f y$ iff $f(x)\sleq f(y)$. Reflexivity and transitivity follow from those properties on $\sleq$ and antisymmetry follows from injectivity of $f$. Suppose $a_i\rightarrow a$ is a convergent sequence in $X$ and $x\sleq_f a_i$. Then $f(x)\sleq f(a_i)$ and because of the continuity of $f$ $f(a_i)\rightarrow f(a)$, so by the closedness of $\sleq$ we get $f(x)\sleq f(a)$ so that $x \sleq a$. We can do the same thing for sequences with $a_i\sleq x$, so the partial order $\sleq_f$ is closed under limits, so it is closed. Since $X$ is sequentially compact, $\sleq_f$ is also small.
\end{proof}

We also have a theorem that classifies closed partial orders on compact metric spaces \cite[Excercise VI-1.18]{scottbook2003}:
\btheor
Urysohn-Carruth Metrization Theorem. Let $(X,\sleq)$ be a compact metrizable pospace (poset with closed uppersets and downsets), then there is a homeormorphic order-isomorphic  map $g: X \rightarrow g(X)\subseteq [0,1]^\mathbb{N}$ which induces a radially convex metric on $X$.
\label{theor:urysohn}
\etheor
A radially convex metric $d$ on a pospace $X$ is a metric such that when $x\sleq z \sleq y$ we have $d(x,y) = d(x,z)+d(z,y)$.
The important part of this theorem is usually considered to be that each compact metric pospace can be equipped with a radially convex metric. The part we are interested in in this paper however is that we can understand any closed partial order on a compact metric space to be induced by a continuous injective map $g:X\rightarrow [0,1]^\mathbb{N}$.

\section{The upperset map}
So far we have shown that there is a class of partial orders called upwards small that are all dcpo's. Can we also say something about their approximation structure? Recall that a poset is called continuous when Approx$(x)$ is a directed set with supremum $x$, for all $x$.

We will work with a slightly stricter type of space. Namely, we require that $X$ is locally compact in addition to it being Haussdorff, first countable and separable. In addition we require that our uppersets are compact instead of sequentially compact. Note that in first countable spaces compactness implies sequential compactness, so the partial order is still upwards small.

Recall that we defined $K(X) = \{A\subseteq X; \emptyset\neq A \text{ compact}\}$, which when equiped with the reversed inclusion order is a domain. Since we required that the uppersets in $X$ are compact, we have for every $x\in X$: $\uparrow x \in K(X)$. In fact suppose we have $x \sleq y$, then for any $z\in \uparrow y$, we have $x\sleq y \sleq z$ so that $z\in \uparrow x$, which gives $\uparrow y \subseteq \uparrow x$. The converse is true as well, so we have $x\sleq y$ if and only if $\uparrow y \subseteq \uparrow x$. 

This means that the \emph{upperset map}
\begin{equation*}
    \uparrow: X \rightarrow K(X)
\end{equation*}
is a strict monotone map. It is also injective, so $\uparrow(X)$ is embedded into $K(X)$, and in fact $\uparrow(X)$ is order isomorphic to $X$. Because it is an order isomorphism it is also Scott-continuous. It is helpful to show this explicitly though. Let $(x_i)\rightarrow x $ be an increasing sequence convergent to its join $x$. Since $x_i \sleq x$ we have $\uparrow x\subseteq \uparrow x_i$ for all $x_i$, so in fact $\uparrow x \subseteq \bigcap_i \uparrow x_i$.
For the other direction, let $y\in \bigcap_i\uparrow x_i$, then $x_i\sleq y$, and by closedness of $\sleq$ we get $x\sleq y$, so that $y\in \uparrow x$, so $\uparrow x = \bigcap_i \uparrow x_i$. So we in fact have $\uparrow(\vee x_i) = \uparrow(\lim x_i) = \uparrow(x) = \vee \uparrow(x_i) = \bigcap_i \uparrow x_i$.

We can view $X$ as a subset of $K(X)$ with the reversed inclusion order using the upperset map. So when we have $\uparrow(x)\ll \uparrow(y)$ in $K(X)$ we also have $x\ll y$. The converse is not true, since in general it will be easier to satisfy the $\ll$ condition in $X$ as it is a smaller space that allows a smaller set of increasing sequences. Recall that $A\ll B$ in $K(X)$ iff $B\subseteq $int$(A)$, so when $\uparrow y \subseteq $int$(\uparrow x)$ we have $x \ll y$.

If $X$ is compact, then we can also look at $K(X)$ with the regular (not reversed) inclusion order. So $A\sleq B$ iff $A\subseteq B$. The join of an increasing sequence is then $\vee S = \overline{\bigcap S}$. Note that $X$ compact ensures that this is again a compact space, since any closed subspace of a compact space is again compact. We also have $A\ll B$ iff $A\subseteq $int$(B)$, but note that $K(X)$ is no longer a domain as any singleton $\{x\}$ has Approx$(\{x\}) = \emptyset$.

In a similar way as before we can order embed $X$ into $K(X)$ with the \emph{downset map} $\downarrow: X \rightarrow K(X)$. So then we see that when $\downarrow(x)\subseteq $int$(\downarrow(y))$ we have $x\ll y$.

Unfortunately it doesn't seem we can say much more about the approximations of an upwards closed poset without restricting ourselves further.

\section{Convex uppersets}
The condition for upwards smallness refers to the uppersets \emph{and} downsets of elements. For a certain well behaved class of partial orders, it is enough to say something about only the uppersets.

Let $X\subseteq \mathbb{R}^n$ be a compact subset of Euclidean space. And denote $KC(X)\subseteq K(X)$ the space of compact and convex subsets of $X$. Because $X$ is a compact metric space, $KC(X)$ is one as well. $KC(X)$ inherits the domain structure from $K(X)$. The reversed inclusion order is closed, so the order on $KC(X)$ is upwards small. 

Let $L^n$ denote the Lebeque measure on $\mathbb{R}^n$. Denote the interior of a set $A$ as $A^o$.
\blemma
Let $A,B\subseteq \mathbb{R}^n$ such that $L^n(A),L^n(B)\neq 0$ and both finite. Then if $A\subseteq B$ we have $L^n(A)\leq L^n(B)$ (monotonicity). If furthermore $\overline{A^o}=A$ and $\overline{B^o}=B$ then if $A\subseteq B$ and $L^n(A)=L^n(B)$ we have $A=B$ (strict monotonicity).
\elemma
\begin{proof}
The first statement on monotonicity follows directly from the definition of the Lebeque measure. For the second statement: Let $A$ and $B$ as described. Since $A\subseteq B$ we can write $B= B\backslash A \cup A$, so that $L^n(B) = L^n(B\backslash A) + L^n(A)$. From $L^n(B)=L^n(A)$ we then get $L^n(B\backslash A)=0$. Then we can write $B^o = (B\backslash A \cup A)^o = (B\backslash A)^o \cup A^o = A^o$ because the interior of any zero measure set is empty. Since $B^o = A^o$ we can take the closure on both sides: $B = \overline{B^o} = \overline{A^o} = A$ and we are done.
\end{proof}

Any closed and bounded (so compact) convex subset $A$ with non-empty interior has this property that $\overline{A^o} = A$.\footnote{It would also hold for slightly more complex spaces, such as a finite union of convex spaces.} If the interior of such a set is empty then there is a hyperplane into which the convex set embeds, so that we can view the set as `lower dimensional': there is an isometric homeomorphic map $f: U \rightarrow A\subseteq \mathbb{R}^n$ where $U\subseteq \mathbb{R}^k$ is a compact convex set in $\mathbb{R}^k$ with nonempty interior. We can then define $L^k(A) = L^k(U)$, its $k$-dimensional measure. We can also do a similar thing when the interior of $A$ is not empty:
\bdefin
    Define the $k$-dimensional measure of a compact convex set $A$ as
    \begin{equation*}
        \mu^k(A) = \sup\{L^k(U)\delim f:U\subseteq \mathbb{R}^k \rightarrow A, f\text{ isometric homeomorfic}\}.
    \end{equation*}
\edefin
It then follows that when $A$ embeds into $\mathbb{R}^k$ we have $L^k(A) = \mu^k(A)$. Furthermore we also have that $A\subseteq B$ implies that $\mu^k(A) \leq \mu^k(B)$. 
\blemma
    Let $\mu: KC(X) \rightarrow [0,\infty)^*$ be the \emph{total measure} defined as $\mu(A) = \sum_{k=1}^n \mu^k(A)$. $\mu$ is a strict monotonic continuous and thus Scott-continuous map.
\elemma
\begin{proof}
Let $A,B\in KC(X)$ be compact convex sets and $B\sleq A$ so $A\subseteq B$. Then for all $k$, $\mu^k(A)\leq \mu^k(B)$, so also $\mu(A) = \sum_k \mu^k(A) \leq \sum_k \mu^k(B) = \mu(B)$. Suppose furthermore that $\mu(A)=\mu(B)$. This is only possible when $\mu^k(A)=\mu^k(B)$ for all $k$. Pick the highest $k$ such that $\mu^k(A)=\mu^k(B)\neq 0$, then $A$ and $B$ embed into $\mathbb{R}^k$ and we can use the above lemma to conclude that $A=B$. Continuity follows because the Lebeque measure is continuous and the fact that the supremum in the definition of $\mu^k$ changes smoothly when the set is slightly changed. Scott-continuity then follows because both $KC(X)$ and $[0,\infty)^*$ are upwards small.
\end{proof}

Note also that the maximal elements of $KC(X)$ are precisely the singletons, so we have $A\in KC(X)$ maximal iff $\mu(A)=0$.

Now let $(X,\sleq)$ be a compact subset of Euclidean space and let $\sleq$ be such that the upperset of any element in $X$ is a closed convex set. Then the uparrow map maps order isomorphically into a subset of $KC(X)$. Specifically, this map preserves joins of directed sets.

\btheor
If $(X,\sleq)$ is a compact subset of Euclidean space and the upperset of any element is a closed convex subspace, then $\sleq$ is a dcpo, and Theorem \ref{theor:martinphd} applies.
\label{theor:convex}
\etheor
\begin{proof}
In this case $\uparrow: X \rightarrow KC(X)$ is a strictly monotone map that preserves joins of increasing sequences. $\mu: KC(X)\rightarrow [0,\infty)^*$ is a Scott-continuous strict monotone map. In particular it preserves joins of increasing sequences. Then the composition $\mu\circ\uparrow:X\rightarrow [0,\infty)^*$ is also strict monotone and preserves joins of increasing sequences so Theorem \ref{theor:martinphd} applies.
\end{proof}

Note that this theorem also holds when the uppersets of $X$ are slightly more complex, for instance when they are finite unions of closed convex sets. Then the upperset map no longer maps to $KC(X)$, but we can still use the map $\mu$ in the same capacity.
\chapter{Ordering Distributions}
In this chapter we will postulate a minimal set of conditions that any order of information content should satisfy. We will study a subset of these dubbed \emph{restricted information orders} in great detail. In this chapter we restrict ourselves to studying classical states.

\section{Information Orders}
A finite classical state is given by a probability distribution on some finite amount of points, which we will label $n\in\mathbb{N}_{>0}$. Such a probability distribution $x$ is a set of $n$ real numbers $x_i$ such that for all $1\leq i \leq n$ $x_i\geq 0$ and $\sum_i x_i = 1$. In this case $x$ refers to the entire probability distribution, while $x_i$ referred to a specific \emph{component} or \emph{coordinate} of $x$.

\bdefin
The space of probability distributions on $n$ points is
\begin{equation*}
    \Delta^n = \{x \in \mathbb{R}^n\delim x_i\geq 0, \sum_i x_i = 1\}.
\end{equation*}
\edefin

Geometrically $\Delta^n$ can be interpreted as the $(n-1)$-simplex. So $\Delta^2$ is a line while $\Delta^3$ is a triangle. The figures in this chapter will often use that depiction of $\Delta^3$. $\Delta^n$ is a compact convex subspace of $\mathbb{R}^n$:

\bdefin
A subspace $A\subseteq \mathbb{R}^n$ is called \emph{convex} iff for all $a,b\in A$ and $0\leq t \leq 1$ we have $(1-t)a+tb \in A$. Informally, $A$ contains the line connecting two arbitrary points in $A$. A point $a\in A$ is called an \emph{extremal convex} point when for any $x,y \in A$, if $a = (1-t)x+ty$ for a $0<t<1$, then $x=y=a$. Intuitively this means that $a$ is in a corner of $A$.
\edefin

The extremal convex points of $\Delta^n$ are precisely the \emph{pure} distributions $\top_i = (0,\ldots,0,1,0,\ldots,0)$. The distributions that are $1$ on their $i$-th component, and $0$ everywhere else.

There is another special distribution in $\Delta^n$: the \emph{uniform} distribution $\bot_n = \frac{1}{n}(1,\ldots,1)$. This is the unique probability distribution with $x_i=x_j$ for all $i$ and $j$. It can be seen as lying in the `middle' of $\Delta^n$.

The pure distributions and the uniform distribution have a special role concerning information content. Namely, let us define the standard notion of information content:
\bdefin
The \emph{Shannon entropy} of a probability distribution $x$ is given by $\mu_S(x) = -\sum_i x_i \log(x_i)$.
\edefin
The Shannon entropy is a positive value corresponding to the uncertainty that a distribution represents. For instance, the only distributions that have $\mu_S(x) = 0$, no uncertainty, are the pure distributions $\top_i$. The unique distribution on $n$ points with the highest entropy (the most uncertainty) is $\bot_n$.

This intuitively makes a lot of sense. Suppose you have $n$ boxes, where there is a bar of gold in precisely one of the boxes. You are also given a probability distribution that tells you the probability that a certain box contains this gold bar. What kind of probability distributions would you like to be given? Of course you would be really happy with a pure distribution, as you would be sure to get the gold (if we assume that the probability distribution actually describes the real world, and you haven't been lied to). You would be the least happy with the uniform distribution as that doesn't give you \emph{any} extra information regarding which box contains the gold. The Shannon entropy of a distribution is a measure of how happy you should be when given that distribution in this scenario (with lower values corresponding to more happiness).

For our information orders it would therefore make sense to require that $\bot_n$ is the least element of the partial order, and that the $\top_i$ are the maximal elements. Since you would be happier with replacing $\bot_n$ with any state $x$ we say that $\bot_n\sleq x$ for all $x$. For the $\top_i$ the situation is slightly more complicated, since we would still be happier when our distribution were a pure distribution, but it would still have to represent compatible information: $\top_1$ and $\top_2$ are obviously incompatible. Since $\top_1$ would tell you that the gold is \emph{definitely} in box 1, while $\top_2$ would tell you that the gold is \emph{definitely} in box 2. We will therefore require that each $x$ is smaller than \emph{at least one} pure state $\top_i$.

Suppose we have $x,y\in \Delta^n$ and some partial order $\sleq$ that captures the idea of information content on $\Delta^n$, so that $x\sleq y$ would mean ``$y$ contains at least all the information $x$ has''. Suppose now that we have two other distributions $x^\prime$ and $y^\prime$ that are the same as $x$ and respectively $y$, but with their first and second components interchanged. We didn't really change any of the information content, we just changed in which order the coordinates were presented, so we would assume that $x^\prime \sleq y^\prime$.

\bdefin
    Let $S^n$ be the permutation group on $n$ points. So $\sigma\in S^n$ is a bijection $\sigma:\{1,\ldots,n\}\rightarrow \{1,\ldots,n\}$. Then we define for $x\in \Delta^n$: $\sigma(x)_i = x_{\sigma(i)}$. We call a partial order on $\Delta^n$ \emph{permutation invariant} when for all $x,y\in \Delta^n$ and $\sigma\in S^n$ we have $x\sleq y \iff \sigma(x)\sleq \sigma(y)$.
\edefin

We need an additional ingredient to arrive at our minimal definition of an information order. Suppose we have a state $x$ that has less information than a state $y$. Now we can imagine a process that transforms the state $x$ into $y$, thereby gaining information. The most simple mix of states would be $(1-t)x+ty$ for any $0\leq t\leq 1$, which is guaranteed to be a distribution by convexity of $\Delta^n$.

\bdefin
    We say that a partial order $\sleq$ on a convex space $C$ allows \emph{mixing} iff for all $x,y\in C$ we have $x\sleq y \implies x\sleq (1-t)x+ty\sleq y$ for all $0\leq t\leq 1$.
\edefin

And finally, we have the logical inclusion $\Delta^k\subseteq \Delta^n$ by sending $(x_1,\ldots,x_k)$ to $(x_1,\ldots,_k,0,\ldots,0)$. We want the restriction of an information order on $\Delta^n$ to $\Delta^k$ to still be an information order. 

We now have all we need to make a basic definition of what we call an information order.
\bdefin
    Let $\sleq$ be a partial order on $\Delta^n$. We call $\sleq$ an \emph{information order} if and only if all the following hold
    \begin{itemize}
        \item $\sleq$ is permutation invariant.
        \item $\sleq$ allows mixing.
        \item $\bot_n\sleq x$ for all $x\in\Delta^n$.
        \item The $\top_i$ are maximal and for all $x\in\Delta^n$: $x\sleq\top_i$ for some $i$.
        \item $\sleq$ restricted to $\Delta^k$ is also an information order.
    \end{itemize}
    \label{def:informationorder}
\edefin

\section{Examples}
We'll start with giving the example that guided this minimal definition for an information order.

\subsection{Bayesian order}

The Bayesian order was defined in \cite{Coecke2010book} as an example of a partial order on $\Delta^n$ that captures the idea of information content. It is defined inductively using the example above about the boxes and the bar of gold. 

Suppose we have Alice and Bob that each have some information about where the gold is, which is captured by probability distributions $x$ and $y$. Suppose that for some definition $y$ has more information than $x$ which we denote as $x\sleq y$. Now, if someone who knew where the gold was revealed that  box $k$ didn't contain it, then $x$ and $y$ can now be updated to reflect that they have certainty that $k$ doesn't contain the gold. So $x_k=y_k=0$, and the rest of the probabilities is rescaled. Then we would still expect the updated probabilities $\overline{y}$ to have more information than $\overline{x}$: $\overline{x}\sleq \overline{y}$. We can then define $\sleq$ inductively in the following way: Let $x,y \in \Delta^n$, and supposed $\sleq^{n-1}$ is defined for $\Delta^{n-1}$. Then set $x\sleq^n y$ if and only if for all $i\leq n$ such that $x_i\neq 1$ and $y_i \neq 1$ we have
\begin{equation*}
 \frac{1}{1-x_i}(x_1,\ldots,x_{i-1},x_{i+1},\ldots,x_n)\sleq^{n-1} \frac{1}{1-y_i}(y_1,\ldots,y_{i-1},y_{i+1},\ldots,y_n)
\end{equation*}
where we define $\sleq^2$ in the only sensible way possible: $x\sleq^2 y$ iff $\mu_S(x)\geq \mu_S(y)$, or equivalently, $(x_1,x_2)\sleq^2 (y_1,y_2)$ iff $x_1\geq x_2$ and $y_1\geq y_2$ and $x_1\leq y_1$, or $x_2 \geq x_1$ and $y_2 \geq y_1$ and $x_2 \leq y_2$.

As is shown in \cite{Coecke2010book}, this inductive procedure gives you a valid partial order. To give a concise description of this order we need an extra definition.

\bdefin
The \emph{monotone sector} of $\Delta^n$ is $\Lambda^n = \{x \in \Delta^n \delim \forall i: x_i\geq x_{i+1}\}$. A \emph{sector} of $\Delta^n$ is a region equal to $\sigma(\Lambda^n)$ for some $\sigma\in S^n$. We have $\Delta^n = \bigcup_{\sigma \in S^n} \sigma(\Lambda^n)$. The boundary of the montone sector consists of the \emph{degenerated} distributions: $\partial \Lambda^n = \{x\in \Lambda^n\delim \exists i: x_i = x_{i+1}\}$
\edefin

\bdefin
The \emph{Bayesian order} on $\Delta^n$ is given by $x\sleq_B y$ if and only if there exists $\sigma \in S^n$ such that $\sigma(x),\sigma(y)\in \Lambda^n$ and 
\begin{equation*}
    \forall 1\leq k \leq n-1: \sigma(x)_{k}\sigma(y)_{k+1} \leq \sigma(y)_{k}\sigma(x)_{k+1}.
\end{equation*}
\edefin

So two distributions $x$ and $y$ are comparable in the Bayesian order when they belong to the same sector of $\Delta^n$. Lets take $x$ and $y$ to belong to $\Lambda^n$ so we can take $\sigma = id$, and suppose $x_i,y_i\neq 0$ for all $i$. Then we can write the comparisons in the Bayesian order as
\begin{equation*}
    \forall 1\leq k \leq n-1: \frac{x_k}{x_{k+1}} \leq \frac{y_k}{y_{k+1}}.
\end{equation*}
We see that $x\sleq_B y$ means that the coordinates of $x$ and $y$ can be simultaneously ordered and that the coordinates of $y$ are more sharply decreasing. It is routine to check that the Bayesian order as given in this form satisfies the properties for an information order outlined in Definition \ref{def:informationorder}. 

\begin{figure}[htb!]
    \centering
    \includegraphics[width=0.6\textwidth]{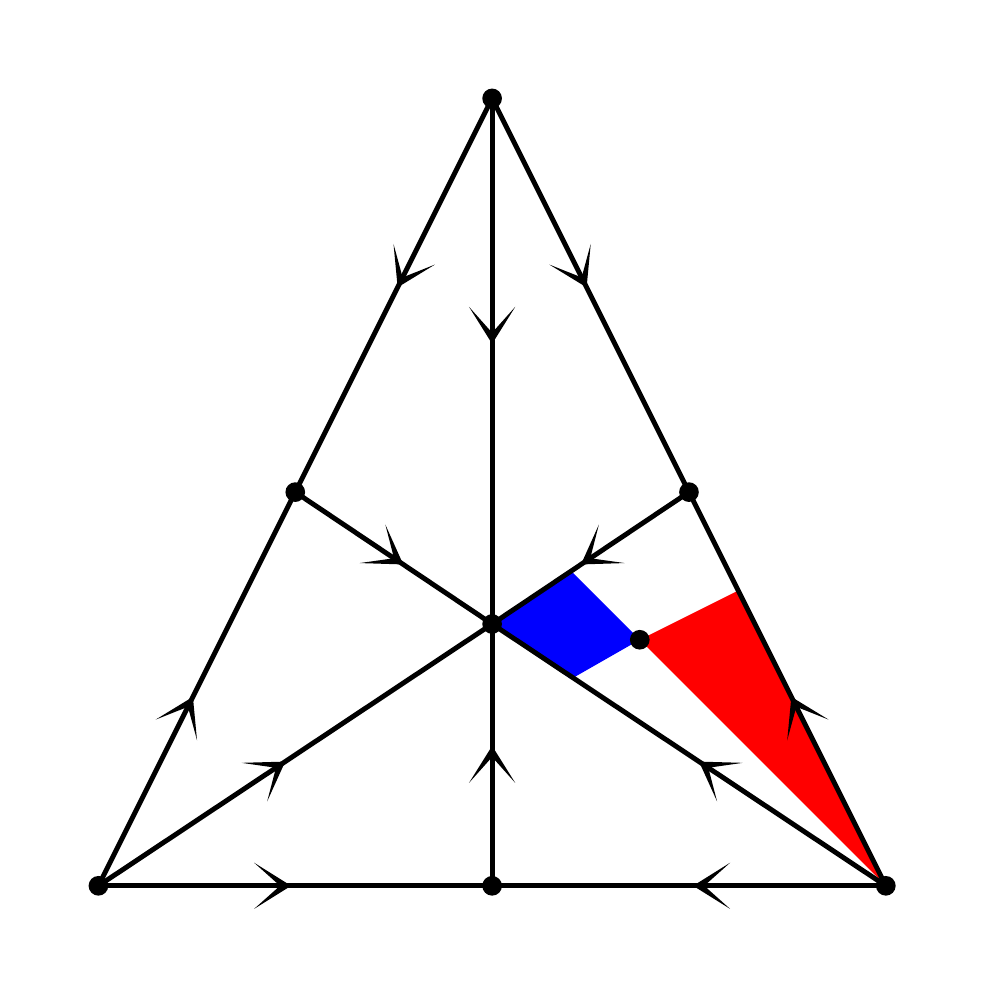}
    \caption[Bayesian order on $\Delta^3$]{Illustration of the upperset (in red) and downset (blue) of a specific point in $\Lambda^3$ in the Bayesian order.}
    \label{fig:bayes}
\end{figure}

In Figure \ref{fig:bayes} you can see an illustration of the upperset and downset of a specific point in $\Lambda^3$. As you can see, the upper- and downset are contained inside of $\Lambda^3$.

\subsection{Renormalised Löwner orders}
In Chapter 3 we will be looking at partial orders on density operators, which come from the Löwner order, which is the `standard' partial order structure that the positive operators carry. Its restriction to diagonal matrices, that is, the space $\mathbb{R}^n$ is just $v\sleq_L w$ iff $v_i\leq w_i$ for all $i\leq n$. If we put this order on $\Delta^n$ and we have $x\sleq_L y$, then $x_i\leq y_i$, but also $\sum_i x_i = \sum_i y_i = 1$, so we must have $x_i = y_i$, which gives $x=y$. So this partial order reduces to $x\sleq_L y$ iff $x=y$.

There are ways to change this partial order into something nontrivial. The two ways we will show here are what we refer to as \emph{eigenvalue renormalisations}. The fact that $\sleq_L$ is trivial on $\Delta^n$ is a result of the normalisation of the elements in $\Delta^n$. By changing the normalisation, we can create a nontrivial partial order.

Define $x^+ = \max\{x_i\}$ the maximum eigenvalue (or coordinate) of $x$. So the normalisation of $x$ such that $x^+ = 1$ is given by $\frac{x}{x^+}$. If we use $\sleq_L$ on the maximum eigenvalue renormalised distributions and switch to the dual of $\sleq_L$ we get the partial order
\begin{equation*}
    x\sleq^+ y \iff \forall k: x^+ y_k \leq y^+ x_k.
\end{equation*}
Transitivity and reflexivity follow easily. Antisymmetry is a result of the normalisation of distributions in $\Delta^n$: if $x\sleq^+ y$ and $y\sleq^+ x$ then $\frac{x^+}{y^+} = \frac{x_k}{y_k}$ for all $k$. So if $x^+> y^+$ then also $x_k>y_k$ for all $k$, which breaks the normalisation. So $x_k=y_k$ for all $k$. That $\bot_n$ is minimal and that the $\top_i$ are maximal is also easily checked. Permutation invariance follows because a permutation would just switch around the order of the inequalities.

Mixing is slightly more tricky. First, suppose $x\sleq^+ y$, and suppose $y_k = y^+$, then $x^+ y_k = x^+y^+\leq y^+ x_k$. So $x^+\leq x_k$, but also $x_k\leq x^+$ by definition, so $x_k = x^+$. So, when $x\sleq^+ y$ there is a $k$ such that $y_k = y^+$ and $x_k = x^+$. Now define $z=(1-t)x+ty$. Then $z^+ \leq (1-t)x^+ +ty^+ = (1-t)x_k + ty_k = z_k$, so $z_k = z^+$ and $z^+ = (1-t)x^+ + ty^+$. That $x\sleq^+ z \sleq^+ y$ then follows easily by just substiting in the definition of $z$ into the inequalities.

$\sleq^+$ satisfies all the conditions of Definition \ref{def:informationorder}. Since all the inequalities involved are continuous it is also not hard to see that the uppersets and downsets are closed. Because $\Delta^n$ is a compact set, $\sleq^+$ is upwards small. So $\sleq^+$ is a dcpo.

We can do a similar sort of construction with a renormalisation to the lowest eigenvalue, although we do run into some extra difficulties here. First, when $x$ and $y$ contain no zeroes, we can simply take $x^- = \min\{x_i\}$, and set $x\sleq^- y$ iff for all $k$: $x_ky^- \leq y_k x^-$. It then follows in the same way as above that this has all the correct properties. Now, if $x$ and $y$ contain some zeroes, but we have $x_k=0$ if and only if $y_k = 0$, then we can simply ignore these zeroes, and set $x^- = \min\{x_i\delim x_i \neq 0\}$. We then still get the definition above and everything works out. But now suppose that $y$ contains more zeroes than $x$. In the case that there is a $k$ such that $y_k = 0$ while $x_k = x^-\neq 0$ the renormalisation to the $k$th coordinate would `blow up' $y$ to infinity while $x$ stays bounded. So we simply define $x\sleq^- y$. If however there is no such $k$ then there is no easy choice of normalisation, so we say that $x$ and $y$ are incomparable.

Now let $Z(x) = \{k\delim x_k = 0\}$ be the set of zeroes of $x$ and $x^-$ the lowest nonzero coordinate and set $x\sleq^- y$ if and only if one of the following (mutually exclusive options) holds:
\begin{itemize}
    \item $Z(x) = Z(y)$ and for all $k: x_ky^- \leq y_k x^-$.
    \item $Z(x)\subset Z(y)$ and there exists a $k$ such that $y_k = 0$ and $x_k = x^-$.
\end{itemize}

\blemma
$\sleq^-$ as defined above is an information order.
\elemma
\begin{proof}
Reflexivity is trivial. Antisymmetry follows because in that case we must have $Z(x) = Z(y)$, so that it reduces to the same problem as for $\sleq^+$. For transitivity we distinguish 4 cases. Let $x\sleq^-y$ and $y \sleq^- z$. If $Z(x) = Z(y) = Z(z)$, then it follows easily. If $Z(x) = Z(y)$ then we note that in the same way as for $\sleq^+$, when $y_k = y^-$ we also have $x_k = x^-$. So if $Z(y)\subset Z(x)$, then there is a $k$ such that $z_k = 0$ and $y_k = y^-$, but then $x_k = x^-$ as well, so $x\sleq^- z$. If instead $Z(x)\subset Z(y)=Z(z)$, then there is a $k$ such that $y_k = 0$ and $x_k = x^-$, in which case $z_k=0$ as well, so that $x\sleq^- z$. Suppose $Z(x)\subset Z(y)\subset Z(z)$, so there is a $k$ such that $y_k=0$ and $x_k=x^-$, but then also $z_k=0$ so again $x\sleq^- z$.

The minimality and maximality of $\bot_n$ and $\top_i$ can be directly checked, and permutation invariance is also clear from the definition. For mixing we split into two different cases. Either $Z(x) = Z(y)$ in which case we can use the same argument as for $\sleq^+$, or $Z(x)\subset Z(y)$, in which case there is a $k$ such that $y_k=0$ and $x_k = x^-$. Let $z = (1-t)x + ty$, then $z_k = (1-t)x_k + ty_k = (1-t)x^- = z^-$. For $t>0$ we have $Z(x) = Z(z) \subset Z(y)$, so we immediately have $x\sleq^- z$, and since $y_k = 0$ and $z_k = z^-$ we also have $z\sleq^- y$.
\end{proof}

$\sleq^-$ is also a dcpo: any element which has a nondegenerated lowest nonzero coordinate has a closed convex upperset. If it is degenerated then the upperset will be a finite union of convex spaces, so we can use Theorem \ref{theor:convex} to prove directed completeness.

\begin{figure}[htb!]
    \centering
    \begin{subfigure}[b]{0.48\textwidth}
    \centering
        \includegraphics[width=0.8\textwidth]{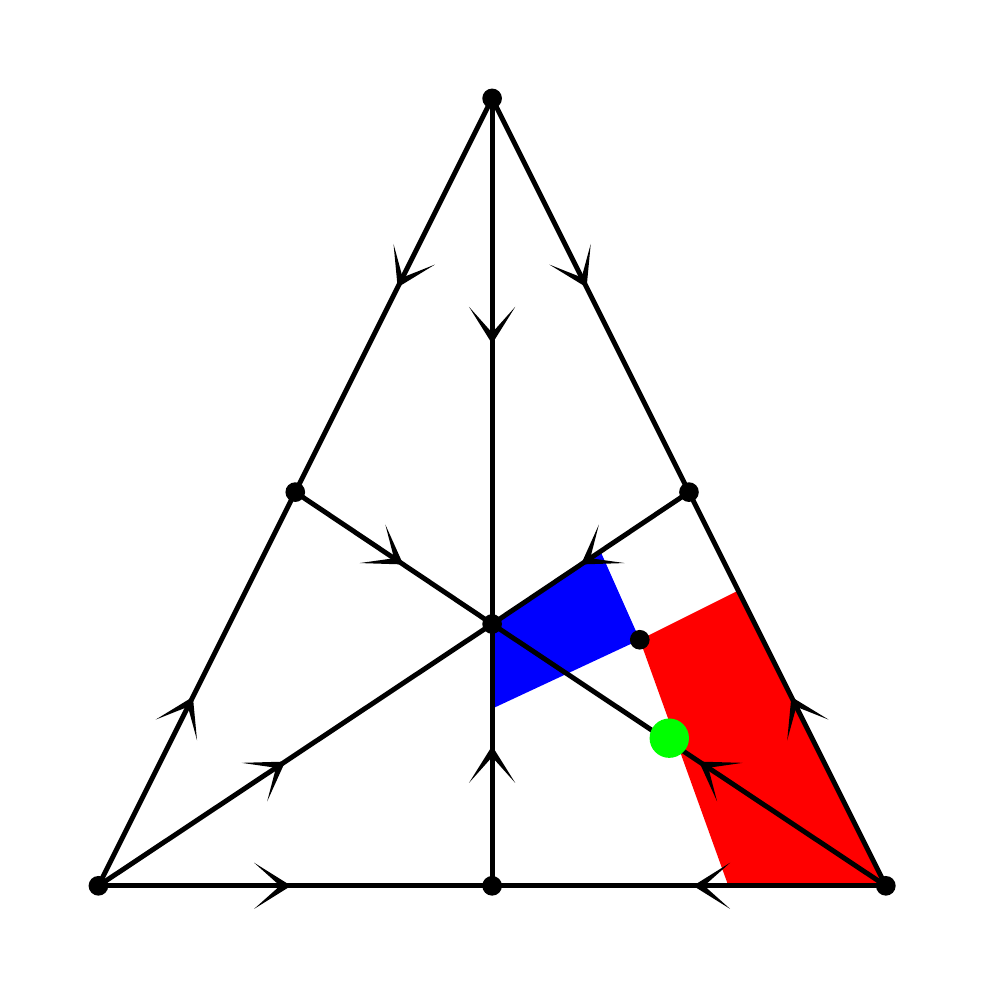}
        \caption{$\sleq_L^+$.}
    \end{subfigure}
    ~
    \begin{subfigure}[b]{0.48\textwidth}
    \centering
        \includegraphics[width=0.8\textwidth]{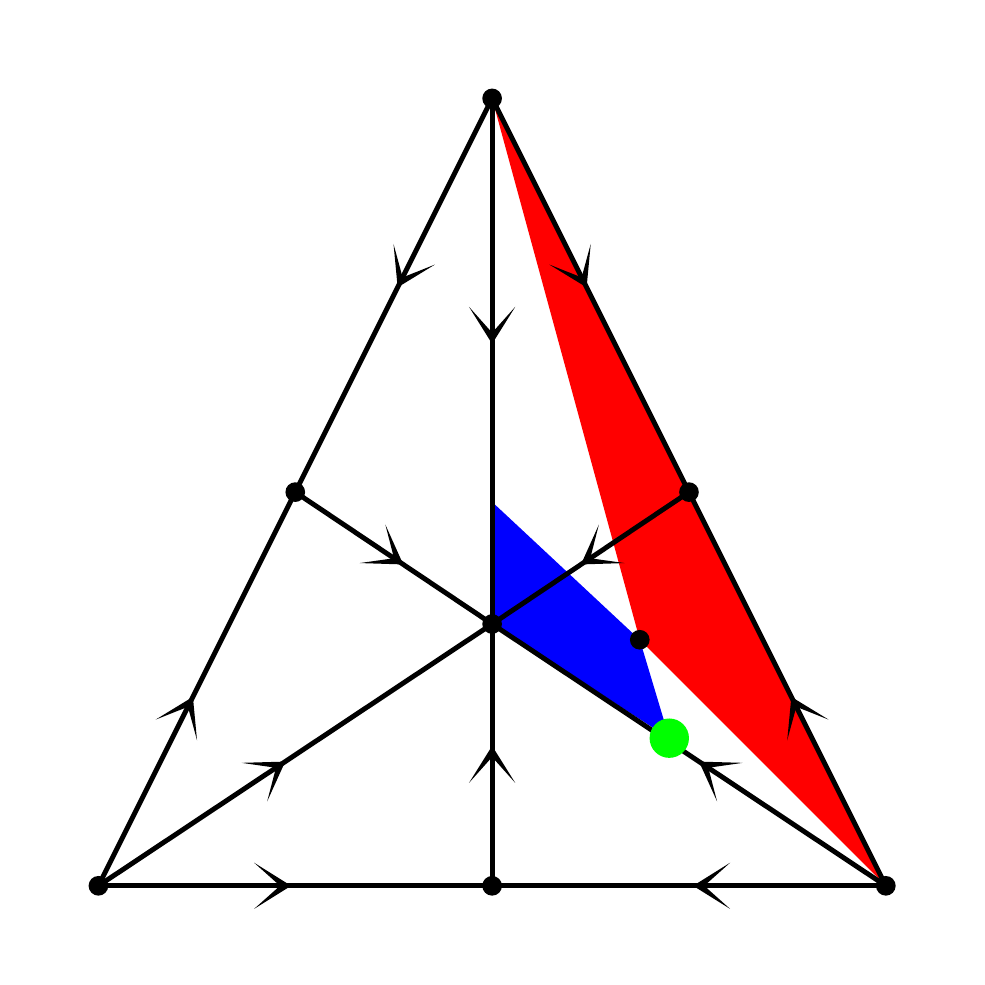}
        \caption{$\sleq_L^-$.}
    \end{subfigure}
    \caption[Renormalised Löwner orders in $\Delta^3$]{Upperset (red) and downset (blue) of the distribution $y=\frac{1}{30}(15,10,5)$ with respect to the renormalised Löwner orders. The point $x=\frac{1}{10}(6,2,2)$ is denoted in green.}
    \label{fig:lowner}
\end{figure}

We now already have 3 very different orders that satisfy the information order conditions. In fact, in Figure \ref{fig:lowner} it is easily seen how different the two renormalised Löwner orders are. In fact, the two orders are \emph{contradicting}: The illustrated points $x=\frac{1}{10}(6,2,2)$ and $y=\frac{1}{30}(15,10,5)$ have $x\sleq_L^+ y$ and $y\sleq_L^-x$. In so far as Definition \ref{def:informationorder} defines a notion of information content, the conditions are not strong enough to define a \emph{unique} direction of information content, as illustrated by this example.

\section{Basic properties}
Information orders (partial orders satisfying Definition \ref{def:informationorder}) have a certain kind of minimal structure, which we will look at in detail in this section. Let $\sleq$ denote an information order for the duration of this section.

\blemma
Let $x\in \Delta^n$ and $\sigma\in S^n$ such that $\sigma(x)\sleq x$ or $x\sleq \sigma(x)$. Then $\sigma(x)=x$.
\label{lemma:perm}
\elemma
\begin{proof}
Suppose $\sigma(x)\sleq x$. We have $\sigma^k = id$ for some $k$, so by permutation invariance we get $x=\sigma^k(x) = \sigma^{k-1}(\sigma(x))\sleq \sigma^{k-1}(x) = \sigma^{k-2}(\sigma(x)) \sleq \ldots \sleq \sigma(x)$. So $x\sleq \sigma(x)$, but also $\sigma(x)\sleq x$, so by antisymmetry of $\sleq$ we have $\sigma(x) = x$. The other case follows analogously.
\end{proof}

\begin{figure}[htb!]
\centering
\includegraphics{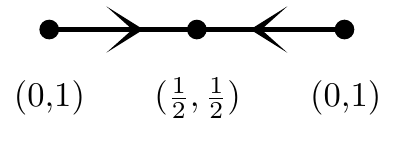}
\caption[The unique information order on $\Delta^2$]{The unique information order on $\Delta^2$.}
\label{fig:delta2}
\end{figure}

Now, consider an information order for $n=2$. $\bot_2 = (\frac{1}{2},\frac{1}{2})$ is the minimal element and $(1,0)$ and $(0,1)$ are the maximal elements. Call $z(t) = (1-t)\bot_2 + t(1,0)$, then by mixing we have $z(t)\sleq z(t^\prime)$ whenever $t\leq t^\prime$. There are no other comparisons possible because of the above lemma. The entire partial order is determined on $\Delta^2$. We get the structure as seen in Figure \ref{fig:delta2}.

The condition that an information order on $\Delta^n$ restricts to an information order on $\Delta^k\subseteq \Delta^n$ is equivalent to a simpler demand.
\btheor
An information order $\sleq_n$ on $\Delta^n$ induces an information order $\sleq_k$ on $\Delta^k$ iff $\bot_m\sleq \bot_l$ for all $1\leq l \leq m\leq n$.
\etheor
\begin{proof}
Define $\sleq_k$ as follows for any $x,y\in \Delta^k$:
\begin{equation*}
    (x_1,\ldots,x_k)\sleq_k (y_1,\ldots,y_k) \iff (x_1,\ldots,x_k,0,\ldots,0) \sleq_n (y_1,\ldots,y_k,0,\ldots, 0)
\end{equation*}
That is: we interpret $\Delta^k$ as the subset of $\Delta^n$ given by setting the last coordinates to zero. Note that due to permutation invariance it doesn't matter which coordinate we take to be zero: the partial order will be the same. The only if direction follows because an information order on $\Delta^k$ has $\bot_k$ as a minimal element. Since $\bot_l\in\Delta^k$ for $l\leq k$, we have $\bot_k\sleq \bot_l$.

For the other direction we will work by induction. Since $\sleq_k$ is defined as a restriction of $\sleq_n$, it is again a partial order, and it carries over the mixing requirement and the maximal elements. The only property left to check is that $\sleq_k$ has the correct least element. For $k=2$ we have that $\bot_2\sleq \bot_1=\top_1$, so by using permutation invariance and mixing, the entire partial order is determined and we are done.

For the induction hypothesis we will assume that $\bot_{k-1}$ is the least element of $\sleq_{k-1}$. We know that $\bot_k\sleq\bot_{k-1}$, so for any element $y \in \Delta^{k-1}$ we have $\bot_k\sleq y$. Let $x \in \Delta^k$ and define $z(\lambda) = (1-\lambda)\bot_k + \lambda x$. For $0\leq \lambda \leq 1$ this is the line between $\bot_k$ and $x$, but if we take $\lambda\geq 1$ then we extend the line further. We can take this extension until the point where $z(\lambda)$ doesn't lie in $\Delta^n$ anymore. Since normalisation is preserved, at this point one of the coordinates of $z(\lambda)$ must have become less than zero. Take $\lambda$ to be the exact value when $z(\lambda)$ is at the border of $\Delta^n$. At this point $z(\lambda)$ has at least one more zero than $x$, so $z(\lambda)\in \Delta^{k-1}$. But then $\bot_k\sleq \bot_{k-1}\sleq z(\lambda)$ and by the mixing property we get $\bot_k\sleq x=z(1)\sleq z(\lambda)$, so $\bot_k$ is indeed the least element of $\Delta^k$.
\end{proof}

\begin{figure}[htb!]
    \centering
    \includegraphics[width=0.6\textwidth]{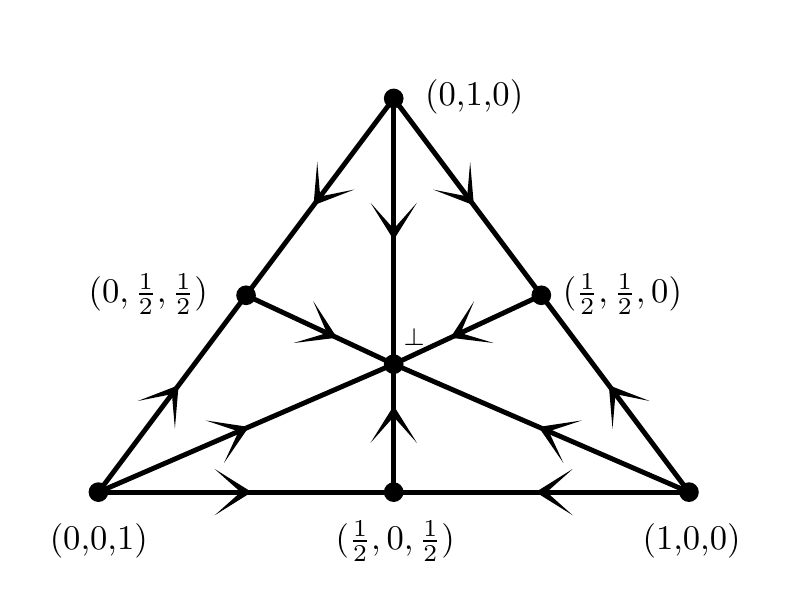}
    \caption[Minimal structure of an information order]{An illustration of the minimal structure of an information order on $\Delta^3$.}
    \label{fig:triangle}
\end{figure}

An illustration of what this structure looks like graphically is given in Figure \ref{fig:triangle}.

The smaller triangles are the sectors $\sigma(\Lambda^n)$, the arrows denote the direction of the comparisons.

\section{Restricted Information Orders}
 We will be looking at a subclass of information orders in detail. We will start by motivating this restriction.
 
\bdefin
    For each $x\in \Delta^n$ there is a unique $y\in \Lambda^n$ such that $y = \sigma(x)$ for some $\sigma\in S^n$. Denote this unique $y$ as $r(x)$. We call $r: \Delta^n\rightarrow \Lambda^n$ the \emph{monotone retraction}.
\edefin
What $r$ does is ordering the coordinates from high to low for any distribution $x$.

\blemma
    If $\sleq$ is an information order on $\Delta^n$, then $r: (\Delta^n,\sleq)\rightarrow (\Lambda^n,\sleq)$ is a strict monotonic map. Furthermore, if $\sleq$ is a closed partial order, than $r$ is Scott-continuous.
\elemma
\begin{proof}
Let $x,y\in \Delta^n$ with $x\sleq y$. Because of permutation invariance, without loss of generality we can take $x$ to be in $\Lambda^n$, so that $r(x)=x$. If $y\in \Lambda^n$, then we are done, so suppose it is not. Let $y$ be in a `neighbouring sector' of $\Lambda^n$: there exists a $\sigma\in S^n$, such that $\sigma(y)\in \Lambda^n$ where $\sigma$ is given by a single permutation of successive coordinates. Name these coordinates $i$ and $i+1$. We then have $x_i\geq x_{i+1}$ and $y_i\leq y_{i+1}$. Let $z(t) = (1-t)x+ty$. For some $t$ we must then have $z(t)_i=z(t)_{i+1}$. But then $\sigma(z(t)) = z(t)$, so  $x\sleq z(t)\sleq y$ translates by permutation invariance to $\sigma(x)\sleq \sigma(z(t)) = z(t) \sleq \sigma(y)$. So we have $x\sleq z(t) \sleq \sigma(y)$, which means that $r(x)\sleq r(y)$. If $y$ doesn't lie in a neighbouring sector than $z(t)$ crosses every sector in between and we can repeat this procedure for a finite amount of $z(t)$ in each of these intermediate sectors. So $r$ is monotone.

Now suppose $x\sleq y$ and $r(x)=r(y)$. Again, without loss of generality we can take $x\in \Lambda^n$ so that $r(x) = x = r(y) = \sigma(y)$ for some $\sigma$. Then $x=\sigma(y)\sleq y$ and from Lemma \ref{lemma:perm} we then get $\sigma(y)=y=x$.

$r$ is a contraction: $d(r(x),r(y))\leq d(x,y)$ so $r$ is continous. Since $\Delta^n$ is a compact subset of $\mathbb{R}^n$, when $\sleq$ is closed, it is upwards small, so that any monotone continuous map is also Scott-continuous (Lemma \ref{lemma:scottcont}).
\end{proof}

This lemma shows us that the behaviour of $\sleq$ on $\Lambda^n$ gives us a lot of information of $\sleq$ on $\Delta^n$: $r$ is a \emph{measurement} of $\sleq$. We can then wonder when $r$ gives us \emph{all} the information of $\sleq$: when do we have $x\sleq y$ if and only if $r(x)\sleq r(y)$? Let $x$ be in $\Lambda^n$. Then for any $\sigma\in S^n$ we have $r(\sigma(x)) = x$ and of course $x=r(x)\sleq x=r(\sigma(x))$, so by the if direction we get $x\sleq \sigma(x)$ for any sigma, which can't happen for an information order.

This procedure then doesn't work, but we can define an information order on $\Delta^n$ that is completely defined by its behaviour on $\Lambda^n$ in a different way:

Set $x\sleq y$ if and only if $r(x)\sleq_{\Lambda^n} r(y)$ \emph{and} there exists a $\sigma$ such that $x,y\in \sigma(\Lambda^n)$, or equivalently, there exists a $\sigma$ such that $\sigma(x),\sigma(y)\in\Lambda^n$.

This extra condition, which makes sure that $x$ and $y$ belong to the same sector, ensures that the situation we described above where we would get $x\sleq \sigma(x)$ can't occur. Now, when does a partial order on $\Lambda^n$ extend to an information order on $\Delta^n$? It is clear that the partial order on $\Lambda^n$, should have $\top_1$ as the maximal element and $\bot_n$ as the minimal element, and furthermore that the partial order should support mixing. There is however one extra condition that the partial order on $\Lambda^n$ should satisfy.

Let $y\in$ int$(\Lambda^n)$ and suppose there is an $x\sleq y$ where $x$ is a border element of $\Lambda^n$. Then it is also in a different sector $\sigma(\Lambda^n)$. Suppose there is a $w\in$ int$(\sigma\Lambda^n)$ such that $w\sleq x$. By transitivity we must have $w\sleq y$ which is a comparison between elements in different sectors which isn't allowed by the form of the partial order.

This problem arises when border elements of $\Lambda^n$ are sometimes smaller and sometimes bigger than an element in the interior of $\Lambda^n$. It is fixed if the partial order has all the border elements below or above the interior elements. Picking them above creates all sorts of problems (not least of which is that $\bot_n$ is a border element, so the partial order will need to have some `discontinuities'), so we will have to choose them below:

\bdefin
    We say that an information order $\sleq$ on $\Delta^n$ has the \emph{degeneracy condition} when for all $x,y\in \Delta^n$, if $x\sleq y$ and for some $i$ and $j$ $y_i = y_j\neq 0$, then $x_i=x_j\neq 0$.
\edefin

Recall that an element $x$ is a border element of $\Lambda^n$ if it has a degenerated spectrum. That is, there is an $i$ such that $x_i=x_{i+1}\neq 0$. It has to be nonzero, because otherwise it wouldn't be a border element. This condition precisely states that elements with a degeneracy on a pair of coordinates $(i,j)$ can't be above elements that don't have a degeneracy on coordinates $(i,j)$.

There is some intuition behind this related to information content. Suppose again we have the situation where Alice and Bob are looking at some boxes of which they know one contains a bar of gold. Their knowledge is represented by probability distributions $p$ and $q$. Suppose $p$ is degenerated on coordinates $i$ and $j$, so $p_i=p_j$, while $q$ isn't. Now someone comes along and \emph{forces} them to choose between boxes $i$ and $j$. Alice wouldn't like this, since she has no preference for any of these boxes so she would have to pick randomly. Bob might also not be too happy about this, but his probability distribution does reflect a difference between these boxes which gives him the opportunity to pick the box which he thinks has the highest probability of containing the gold. Since in this case you would prefer to be Bob, it is not unreasonable to say that we at least know that Alice does not have more information than Bob.

The degeneracy condition turns out to be enough to be able to restrict to a partial order on $\Lambda^n$:
\blemma
Suppose $\sleq$ is an information order with the degeneracy condition on $\Delta^n$, then $\sleq$ can be written as $x\sleq y$ if and only if there exists a $\sigma$ such that $x,y\in \sigma\Lambda^n$ and $r(x)\sleq r(y)$.
\elemma
\begin{proof}
Let $x\in $int$(\Lambda^n)$ and $y$ not in $\Lambda^n$, such that $x\sleq y$. Let $z(t)=(1-t)x+ty$. At some $t$ $z(t)\in \partial \Lambda^n$, so then $x\sleq z(t)$, which breaks the degeneracy condition. So elements are only comparable when they belong to the same sector: $x\sleq y$ implies that there is a $\sigma$ such that $x,y\in \sigma\Lambda^n$. But then $r(x) = \sigma^{-1}(x)$ and $r(y)=\sigma^{-1}(y)$. By permutation invariance, $x\sleq y$ implies $\sigma^{-1}(x) = r(x)\sleq \sigma^{-1}(y) = r(y)$. The other direction works similarly.
\end{proof}

From this we easily get:
\blemma
    There is a one-to-one correspondence between information orders on $\Lambda^n$ with the degeneracy condition and information orders on $\Delta^n$ with the degeneracy condition.
\elemma
So when assuming the degeneracy condition we can interchangably talk about information orders on $\Delta^n$ and on $\Lambda^n$. Let's define a shorthand for this.

\bdefin
    An information order on $\Delta^n$ or equivalently $\Lambda^n$ is called a \emph{restricted} information order (RIO) iff it satisfies the degeneracy condition.
\edefin

We have already seen a RIO: The Bayesian order. The renormalised Löwner orders aren't restricted.

Because $\Lambda^n$ is a simpler space than $\Delta^n$ restricted information orders are a good place to start the study of information orders.

\section{Classification of restricted information orders}
We will give a classification of a limited set of these restricted orders. 

Since a restricted order is completely defined by its behaviour on $\Lambda^n$, we will be working exclusively on $\Lambda^n$ instead of $\Delta^n$ in this section.

The assumption here will be that, since a restricted order encodes information, this information should somehow be encodable in a set of real numbers that represent the information \emph{features} and the partial order would then consist of comparing these features.

Specifically, we would want to define our partial order by an injective map $F: \Lambda^n \rightarrow \mathbb{R}^k$, where we equip $\mathbb{R}^k$ with the product partial order: $v\sleq w$ iff $v_i\sleq w_i$ for all $i\leq k$. $\Lambda^n$ then inherits a partial order via $F$ (it is antisymmetric iff $F$ is injective). Furthermore if $F$ is continuous, then since the order on $\mathbb{R}$ is closed, the partial order on $\Lambda^n$ will be a dcpo.

Note that due to the Urysohn-Carruth Metrization Theorem (Theorem \ref{theor:urysohn}), any closed partial order on $\Lambda^n$ is induced by a continuous injective map $F:\Lambda^n \rightarrow [0,1]^{\mathbb{N}}$, so this is less of a restriction then it might seem.

We will however be carrying on in a slightly different way. Recall that the Bayesian order (which is a restricted information order) was given by a set of inequalities of the form $x_iy_{i+1}\leq y_ix_{i+1}$. Defining $F_i(x) = \frac{x_i}{x_{i+1}}$ we can fit it into this model, but it will go wrong if $x_{i+1}=0$. This could potentially be fixed by allowing $F$ to map into the one-point compacted reals $\mathbb{R}\cup \{\infty\}$, but then what to do when $x_i$ is zero as well? How would you define $\frac{0}{0}$?

We will fix this problem by taking inspiration from the Bayesian order and mapping $\Lambda^n$ into a product of a slightly bigger space.
\bdefin
Define the \emph{extended reals} $E\mathbb{R} = \mathbb{R}^2$ with an order given by: for $x=(x_1,x_2),y=(y_1,y_2)\in \mathbb{R}^2=E\mathbb{R}$, $x\sleq y$ iff $x_1y_2\leq y_1x_2$.
\edefin 
Transitivity and reflexivity of this order are easy enough to check, but it is not antisymmetric. There is a continuous order isomorphic embedding $\mathbb{R}\rightarrow E\mathbb{R}$ given by $r\mapsto (r,1)$, so this is a valid extension of the real numbers. There is also a map defined on the subset of $E\mathbb{R}$ where the second component is nonzero that reflects back: $(x_1,x_2)\mapsto \frac{x_1}{x_2}$ which is monotonic.

The reason we work with this space is because it allows us to represent infinities in a consistent way. These correspond to the elements where the second coordinate is zero. Note also another peculiar property: $\uparrow (0,0) = \downarrow (0,0) = E\mathbb{R}$. The zero is bigger and smaller than any other element. This turns out to be really useful.

So let's look at partial orders defined by an injective map $F: \Lambda^n \rightarrow (E\mathbb{R})^k$. Such a map is given by $k$ pairs of functions $f_i,g_i:\Lambda^n \rightarrow \mathbb{R}$, where we then have $F(x)_i = F_i(x) = (f_i(x),g_i(x))$. $F$ is continuous iff all the $f_i$ and $g_i$ are continuous. The preorder on $E\mathbb{R}$ is closed, so if $F$ is continous than the induced partial order on $\Lambda^n$ will be closed, so that it is a dcpo.

So suppose a partial order on $\Lambda^n$ is given in this way. Then we would have $x\sleq y$ iff $F(x)\sleq F(y)$ iff for all $i\leq k$: $f_i(x)g_i(y)\leq f_i(y)g_i(x)$. Note that $F$ is required to be injective for it to result in a partial order, but it is not the case that \emph{any} injective $F$ will result in a partial order since $E\mathbb{R}$ is only a preorder. Antisymmetry will have to be checked independently.

As the canonical example, for the Bayesian order we have $k=n-1$, and $F_i(x) = (x_i,x_{i+1})$. This approach makes clear what would happen if $x_i = x_{i+1} = 0$. Since this would map $F_i(x)=(0,0)$ in $E\mathbb{R}$, this element would be bigger \emph{and} smaller than any other element. In other words, there is no information to be gained from the $i$th componenet of $F$ for this $x$ in comparison with any other $y\in \Lambda^n$. This makes sense, because if $x\in\Lambda^n$, with its last $k$ coordinates equal to zero, then it is actually an element of $\Lambda^{n-k}$, so that we should be able to use the Bayesian order on $\Lambda^{n-k}$ which is given by an $F$ defined with $n-k-1$ components.

\subsection{Polynomials and affine functions}
If we want to classify these partial orders, we have to know what kind of functions we can choose for $f_i$ and $g_i$. We will be working with continuous functions. This seems like a reasonable enough demand and we will later see an argument for why the only valid information orders are produced by continuous maps. 

Note that $\Lambda^n$ is a compact subset of $\mathbb{R}^n$ so that for a continuous function $f:\Lambda^n\rightarrow\mathbb{R}$ the Stone-Weierstrass theorem states that $f$ can be approximated arbitrarily well by a polynomial. The uppersets and downsets of the partial order are determined by the function values of the $f_i$ and $g_i$. If we can arbitrarily approximate those by polynomials, we can also get arbitrarily close to the correct upper and downsets. So let's look at the situation where we have two $k$th order polynomials $f$ and $g$. 

Any $k$th order polynomial can be written as the product of $k$ affine functions (an affine function is a linear map plus a constant, or equivalently a 1st order polynomial), so we can write $f(x) = \prod_j f_j(x)$ and $g(x) = \prod_j g_j(x)$ where the $f_j$ and $g_j$ are affine functions. The inequality is $f(x)g(y)\leq f(y)g(x)$. Suppose we have $f_j(x)g_j(y)\leq f_j(y)g_j(x)$ for all $j$, then we can take the product of all these inequalities to arrive at $\prod_j f_j(x)g_j(y) = \prod_j f_j(x) \prod_lg_l(y) = f(x)g(y) \leq \prod_j f_j(y)g_j(x) = f(y)g(x)$. So, suppose $(f,g)$ was part of some $F$ to determine a partial order on $\Lambda^n$. If we were to replace these $(f,g)$ in $F$ by the $k$ pairs $(f_j,g_j)$ we would arrive at a more strict partial order.

We can actually do quite a bit more than this. It turns out that any higher order polynomial will not produce a valid information order, because they don't allow mixing. So first, let's assume $f$ and $g$ are first order polynomials. Or in other words: they are affine maps. The defining characteristic of an affine map is that $f((1-t)x+ty) = (1-t)f(x)+tf(y)$ for all $x,y$ and $0\leq t\leq 1$. Now suppose $f(x)g(y)\leq f(y)g(x)$ with $f$ and $g$ affine maps. Let $z(t) = (1-t)x+ty$, then
\begin{align*}
    f(x)g(z(t)) &= f(x)g((1-t)x+ty) = (1-t)f(x)g(x)+ tf(x)g(y) \\
    f(z(t))g(x) &= f((1-t)x+ty)g(x) = (1-t)f(x)g(x) + tf(y)g(x).
\end{align*}
This means that $f(x)g(z(t))\leq f(z(t))g(x)$ if and only if $f(z(t))g(y)\leq f(y)g(z(t))$ if and only if $f(x)g(y)\leq f(y)g(x)$. With affine maps, we get mixing ``for free''. 

For higher order polynomials this doesn't work. We will work with a second order polynomial, and show that it can't have mixing, and then give an argument for why this generalises to higher order polynomials.

Write $f(x) = f_1(x)f_2(x)$ and $g(y)=g_1(y)g_2(y)$ with $f_i$ and $g_i$ affine maps. We can suppose that $f_i(x)$ is nonnegative otherwise we would relabel, the same goes for $g_i$. For the duration of the argument we will also assume that $f(x),g(x),f(y),g(y)\neq 0$, as this would just complicate the situation. Suppose $f(x)g(y)\leq f(y)g(x)$. We'll assume that for the pair of coordinates $x$ and $y$ this inequality was the `deciding factor' that determined that $x\sleq y$ (such a pair must exist, otherwise we could remove this inequality and keep the same partial order), then we must also have $f(x)g(z(t))\leq f(z(t))g(x)$ and $f(z(t))g(y)\leq f(y)g(z(t))$ for $0\leq t \leq 1$. Let's calculate the first expression:

\centerline{\begin{minipage}[c]{\textwidth}
\begin{align*}
    &f(x)g(z(t))\leq f(z(t))g(x) \iff \\
    &f(x)[(1-t)^2g_1(x)g_2(x) + t(1-t)(g_1(x)g_2(y)+g_1(y)g_2(x)) + t^2g_1(y)g_2(y)] \\
    \leq&~ [(1-t)^2g_1(x)g_2(x) + t(1-t)(f_1(x)f_2(y)+f_1(y)f_2(x)) + t^2f_1(y)f_2(y)]g(x) \\ 
    &\iff~t^2[f(x)g(y)-f(y)g(x)] \leq \\ &t(1-t)g(x)(f_1(x)f_2(y)+f_1(y)f_2(x)) -t(1-t)(g_1(x)g_2(y)+g_1(y)g_2(x))f(x) \\
    =&~ t(1-t)f(x)g(x)\left(\frac{f_1(y)}{f_1(x)} + \frac{f_2(y)}{f_2(x)} - \frac{g_1(y)}{g_1(x)} - \frac{g_2(y)}{g_2(x)}\right).
\end{align*}
\vspace{\parindent}
\end{minipage}}

The LHS is negative by assumption, and if $t\rightarrow 0$ it decreases faster to zero than the RHS, because of the quadratic factor so that the RHS must be nonnegative for this to hold. we assumed that $f(x)g(x)>0$, so we must have
\begin{equation*}
    \frac{f_1(y)}{f_1(x)} + \frac{f_2(y)}{f_2(x)} - \frac{g_1(y)}{g_1(x)} - \frac{g_2(y)}{g_2(x)} \geq 0.
\end{equation*}

We can perform the same type of calculation for $f(z(t))g(y)\leq f(y)g(z(t))$ and arrive at 
\begin{equation*}
    \frac{g_1(x)}{g_1(y)} + \frac{g_2(x)}{g_2(y)} -\frac{f_1(x)}{f_1(y)} - \frac{f_2(x)}{f_2(y)}\geq 0.
\end{equation*}
At this point it will prove helpful to define some new variables. Denote $f_i = f_i(x)/f_i(y)$ and $g_i=g_i(x)/g_i(y)$, then we can rewrite these inequalities to
\begin{align*}
    f_1^{-1} + f_2^{-1} - g_1^{-1} - g_2^{-1} &= A \geq 0 \\
    g_1 + g_2 - f_1 - f_2 &= B \geq 0.
\end{align*}
These two inequalities are not independent. Note that $(f_1+f_2)f_1^{-1}f_2^{-1} = f_1^{-1} + f_2^{-1}$ and that $f_1f_2\leq g_1g_2$. Write
\begin{align*}
    A &= f_1^{-1} + f_2^{-1} - g_1^{-1} - g_2^{-1} = (f_1+f_2)f_1^{-1}f_2^{-1} - (g_1+g_2)g_1^{-1}g_2^{-1} \\
    &\iff f_1f_2g_1g_2 A = (f_1+f_2)g_1g_2 - (g_1+g_2)f_1f_2 \\
    &= g_1g_2(f_1+f_2 - g_1-g_2) + (g_1+g_2)(g_1g_2-f_1f_2) \\
    &\iff -B = f_1+f_2 - g_1-g_2 = Af_1f_2 - (g_1^{-1}+g_2^{-1})(g_1g_2-f_1f_2).
\end{align*}
Now suppose $f(x)g(y)=f(y)g(x)$, then $g_1g_2-f_1f_2 = 0$ so that $0\geq -B = Af_1f_2\geq 0$. So $A=B=0$.

Now set $g_1 = f_1 - \epsilon$. Since $B=0$ we then also have $g_2 = f_2+\epsilon$, then
\begin{equation*}
    g_1^{-1} + g_2^{-1} = \frac{1}{f_1-\epsilon} + \frac{1}{f_2+ \epsilon} = \frac{f_1 + f_2}{(f_1-\epsilon)(f_2-\epsilon)}.
\end{equation*}
We also have
\begin{equation*}
    0 = A = g_1^{-1} + g_2^{-1} - f_1^{-1} - f_2^{-1} = (f_1+f_2)\left(f_1^{-1}f_2^{-1} + \frac{1}{(f_1-\epsilon)(f_2-\epsilon)}\right).
\end{equation*}
Since we assumed that $f_1+f_2 > 0$ this is only true when $f_1f_2 = (f_1-\epsilon)(f_2+\epsilon)$, which is the case when $\epsilon = 0$ or when $\epsilon \neq 0$ and $f_1-f_2 = \epsilon$. If $\epsilon \neq 0$ we then get $f_1-g_1=\epsilon = f_1-f_2 = g_2-f_2$, so that $g_1=f_2$ and $f_1=g_2$. If $\epsilon = 0$, we have by the starting assumption $g_1=f_1$ and $g_2=f_2$. By relabelling we can assume that this is the case.

So from $f(x)g(y)=f(y)g(x)$ we get $f_i(x)g_i(y)=f_i(y)g_i(x)$ for $i=1,2$ and otherwise the mixing condition will be violated. If there are no $x$ and $y$ where this equality holds then we must have $f(x)g(y)< f(y)g(x)$ for all $x$ and $y$ so that we can remove the inequality without changing the partial order. Since $f_i$ and $g_i$ are affine maps, the surface where $f_i(x)g_i(y)=f_i(y)g_i(x)$ is given by a plane, while if $f$ and $g$ are true 2nd degree polynomials the equality surface would be some curved space, so this property can in fact only hold when $f$ and $g$ are actually affine maps in disguise for instance with $f_2(x)=g_2(x) = 1$ for all $x$ and $y$. This proves that second order polynomials can't give information orders. If one of the $f_i$ or $g_i$ were zero we could still do the same kind of arguments.

For higher order polynomials we can use that for an arbitrary $y$ there will be $x$ smaller than it arbitarily close (in the usual metric) to $y$ (because of the mixing property). We can then approximate $f$ and $g$ by a second order Taylor series and arrive at the same conclusions. The reason the mixing conditions fails for any higher order polynomial is because mixing `ensures' that the boundaries of uppersets and downsets are straight surfaces. The levelset of a polynomial however is curved. For this reason we also don't have to look at arbitrary (non-polynomial) continuous functions, because at small distances we can also approximate this by a quadratic polynomial.

For discontinuous functions the boundaries of uppersets and downsets will also not be straight, so we can disregard these kinds of functions as well, with a small caveat: the function has to be continuous (and thus affine) on the interior of $\Lambda^n$. On the boundary, so with $x_k=0$ or $x_k=x_{k+1}$ for some $k$ we can in general no longer use these techniques, and in fact some discontinuity when going from the interior to the boundary can still produce a valid information order. See for instance the second renormalised Löwner order $\sleq^-$.

To conclude: to study restricted information orders of the type $x\sleq y$ iff for all $i\leq k$: $f_i(x)g_i(y)\leq f_i(y)g_i(x)$ it suffices to look at $f_i$ and $g_i$ affine. The minimal value for $k$ is $n-1$, because $\Lambda^n$ is $n-1$ dimensional, so $F$ can't be injective otherwise. If we take $k$ to be bigger the partial order is defined by strictly more inequalities so that we get a stricter partial order. So we'll first look at the case for $k=n-1$ first.

\subsection{Classification}
\label{sec:classification}
A long and complete proof of the statement in this section is given in the appendix, here he will present a less extensive proof that relies more on intuition.

Call $H(x,y) = f(x)g(y)-f(y)g(x)$. Then $x\sleq y$ would give $H(x,y)\leq 0$. If $f$ and $g$ are affine, then $H$ is affine as well in both arguments. Furthermore $H$ is antisymmetric. $\Lambda^n$ is a convex space with extremal points $\bot_k = \frac{1}{k}(1,\ldots,1,0,\ldots,0)$. We can write any $x\in \Lambda^n$ as a convex sum of the $\bot_k$ in the following way. Set $a_k = k(x_k-x_{k+1})$, then $x = \sum_k a_k\bot_k$. Also write $y = \sum_k b_k\bot_k$. Then $H(x,y) = H(\sum_k a_k \bot_k,\sum_l b_l \bot_l) = \sum_{k,l}a_kb_l H(\bot_k,\bot_l)$. Using that $H$ is antisymmetric, we can write this as $H(x,y) = \sum_{k<l} (a_kb_l-a_lb_k)H(\bot_k,\bot_l)$. Since we have $\bot_l\sleq \bot_k$ when $l\geq k$, we have $H(\bot_l,\bot_k)\leq 0$, so by antisymmetry $H(\bot_k,\bot_l)\geq 0$. Define $H(\bot_k,\bot_l) = H_{kl}$, so that we can write $H(x,y) = \sum_{k<l} (a_kb_l-a_lb_k)H_{kl}$.

Now, we have $n-1$ such functions $H$. The degeneracy conditions state that when $x\sleq y$ and $y_i=y_{i+1}\neq 0$ we must have $x_i=x_{i+1}\neq 0$. So we actually see that these are $n-1$ seperate conditions. Each of the $H$'s will have to take care of one of these degeneracy conditions. Suppose $H$ has to `enforce' the $k$th degeneracy condition. So when $H(x,y)\leq 0$ and $y_k = y_{k+1}$, then $x_k = x_{k+1}$. From the definition of the $a_i$ and $b_i$ we see that $y_k=y_{k+1}$ precisely when $b_k = 0$. Write out the double sum in $H$ with respect to $k$:
\begin{equation*}
    H(x,y) = \sum_{i<j}(a_ib_j-a_jb_i)H_{ij} = \sum_{i<j,i\neq k, j\neq k}(a_ib_j-a_jb_i)H_{ij} + \sum_{k=i, k<j}a_kb_jH_{ij} \leq 0.
\end{equation*}
Here we have already used that $b_k=0$ in the last double sum. This last term is nonnegative. The first term can be positive and negative depending on how we choose $a_i$ and $b_i$ as long as one of the $H_{ij}$ for $i,j\neq k$ is nonzero. But in that case, there will be a combination of $a_i$ and $b_i$ with $a_k\neq 0$, such that $H(x,y)\leq 0$ which breaks the degeneracy condition. So we must have $H_{ij} = 0$ for all $i$ and $j$ where there is not at least one equal to $k$. The nonzero terms are then $H_{kj}$ for $k<j$ and by antisymmetry also the $H_{ik}$ for $i<k$.

Knowing this, we can write
\begin{align*}
    H(x,y) &= \sum_{i=k<j} (a_kb_j-a_jb_k)H_{kj} = a_k\sum_{j=k+1}^n H_{kj}b_j - b_k\sum_{j=k+1}^n H_{kj}a_j \\
    &= f(x)g(y) - f(y)g(x)
\end{align*}
where $f(x) = a_k$ and $g(y) = \sum_{j=k+1}^n H_{kj}b_j$. We would like to write $f$ and $g$ in terms of the coordinates of the distributions in $\Lambda^n$. Converting back, and rescaling will give us $f(x) = x_k - x_{k+1}$ and $g(y) = \sum_{j=k+1}^n A_{kj} y_j$ for some parameters $A_{kj}$. 

The form of the $f$ ensures that it is always positive, and $f(x)$ is zero exactly when $x$ is degenerated. Let $l\geq k+2$, then $g(\bot_l) = \frac{1}{l}(1+\sum_{j=k+2}^l)$. Suppose $g(\bot_l)\leq 0$. Let $y$ be any nondegenerated element with $y_j=0$ for $j>l$, then we should have $\bot_l\sleq y$ which means $f(\bot_l)g(y) = 0\cdot g(y) = 0 \leq f(y)g(\bot_l)$ where the RHS is a strictly positive number $f(y)$ multiplied by a negative number $g(\bot_l)$ which breaks the inequality. This means that $g$ will also always be nonnegative.

Combining all this, we get the following classification for RIO's defined by $n-1$ pairs of affine maps $f_i$ and $g_i$:
\begin{align*}
    x\sleq_A y \iff f_i(x)g_i(y)&\leq f_i(y)g_i(x) \text{ for all }1\leq i\leq n-1 \\
    \text{where } &\\
    f_i(x) &= x_i-x_{i+1} \\
    g_i(x) &= y_{i+1} + \sum_{j=i+2}^n A^i_j y_j \text{ where } 1+\sum_{j=i+2}^k A^i_j > 0 \\
    \text{for } i+1&<k\leq n
\end{align*}

We denote it $\sleq_A$ because the partial order is determined by a set of parameters $A^i_j$. In total there are $O(n^2)$ free parameters, or specifically, $\frac{(n-2)(n-1)}{2}$. The Bayesian order corresponds to the situation where $A^i_j = 0$ for all the parameters. Note that all the parameters are bounded from below, but not from above. Note also that each inequality has a different amount of free parameters. For $i=n-1$ there are no free parameters, and the inequality is always $(x_{n-1}-x_n)y_n \leq (y_{n-1}-y_n)x_n$ which can be simplified to $x_{n-1}y_n \leq y_{n-1}x_n$. The next inequality $i=n-2$, has 1 free parameter: $(x_{n-2}-x_{n-1})(y_{n-1} + A^{n-2}_ny_n)\leq (y_{n-2}-y_{n-1})(x_{n-1}+A^{n-2}_nx_n)$. The next inequality has 2 free parameters, and so forth.

\subsection{Adding additional inequalities}
The above classification is only for partial orders given by $n-1$ inequalities: the minimal amount. what would happen if we added other inequalities?
So suppose we already have a partial order $\sleq_A$, and that we want to add an inequality $f(x)g(y)\leq f(y)g(x)$. The functions $f$ and $g$ have to be affine. We again write $x,y \in \Lambda^n$ in the convex extremal basis introduced in the previous section: $a_k = k(x_k-x_{k+1})$ and $b_k = k(y_k-y_{k+1})$ so that $x = \sum_k a_k \bot_k$ and $y = \sum_k b_k \bot_k$. In this basis $f$ and $g$ are still affine functions, so we can write $f(x) = \sum_i A_i a_i$ and $g(y) = \sum_j B_j b_j$.

We note that we must have $\bot_k \sleq y$ if $y_{k+1}=0$, so the inequality must satisfy $f(\bot_k)g(y)\leq f(y)g(\bot_k)$. Noting that if we take $x=\bot_k$ then we have $a_k = 1$ and the other $a_i$'s equal to zero, and for $y$ we have $b_j=0$ if $j> k$. This inequality then becomes
\begin{equation*}
    \frac{A_k}{k}\sum_{i\leq k}B_ib_i \leq  \frac{B_k}{k}\sum_{i\leq k}A_ib_i.
\end{equation*}
Which can equally be written as
\begin{equation*}
    \sum_{i<k}b_i (B_iA_k-A_iB_k) \leq 0.
\end{equation*}
Since we can simply take $b_i = 1$ for an arbitrary $i$ (and the other $b_i$'s equal to zero) we must then have
\begin{equation*}
    B_iA_k - A_iB_k \leq 0 \text{ for all } 1\leq i < k\leq n.
\end{equation*}
Now we take $x$ and $y$ arbitrary again.
\begin{align*}
    & f(x)g(y)\leq f(y)g(x) \iff \\
    & \sum_i A_ia_i\sum_j B_jb_j \leq \sum_iA_ib_i\sum_jB_ja_j \iff \\
    & \sum_i\sum_{i<j}A_ia_iB_jb_j + \sum_i\sum_{j\leq i}A_ia_iB_jb_j \leq \sum_i\sum_{i< j}A_ib_iB_ja_j + \sum_i\sum_{j\leq i}A_ib_iB_ja_j.
\end{align*}
Call $f_i(x) = A_ia_i$ and $g_i(y) = \sum_{i<j}B_jb_j$, then we can write this as
\begin{align*}
    \sum_i & \left[f_i(x)g_i(y)-f_i(y)g_i(x)\right]
    \leq \sum_i\sum_{j\leq i} \left(A_ib_iB_ja_j - A_ia_iB_jb_j\right) \\
    = & -\sum_k\sum_{i<k}b_ka_i\left(B_iA_k-B_kA_i\right)
\end{align*}
where in the last line we relabelled the coordinates so that it is easily seen to be always positive (since $B_iA_k - A_iB_k \leq 0$). We then see that instead of adding the inequality $f(x)g(y)\leq f(y)g(x)$ to the partial order, we could make a \emph{stricter} partial order by adding the set of inequalities $f_i(x)g_i(y)\leq f_i(y)g_i(x)$. These inequalities are precisely of the form we have already seen with $f_i(x)=x_i-x_{i+1}$ and $g_i(y)=y_{i+1}+\sum_{j=i+2}^n A^i_jy_j$ (after rescaling). So adding an arbitrary inequality to a partial order $\sleq_A$ will create a partial order that in strictness lies between $\sleq_A$ and an intersection of partial orders $\sleq_{A(i)}$, where $A(i)$ as a set of parameters is the same as $A$, but the parameters relating to the $i$th inequality are taken from $f_i(x)g_i(y)\leq f_i(y)g_i(x)$. 

With this in mind it suffices to study the family of partial orders $\sleq_A$ to ascertain many properties of the restricted information orders.

\subsection{Changing parameters}
\label{sec:parameters}
We have a family of partial orders $\sleq_A$ indexed by the set of parameters $A$ that define the partial order. We will call a partial order $\sleq_A$ \emph{stricter} than a partial order $\sleq_B$ iff for all $x$ and $y$ with $x\sleq_A y$ we have $x\sleq_B y$. That is: the identity map $(\Lambda^n,\sleq_A)\rightarrow (\Lambda^n,\sleq_B)$ is monotone.

Let us fix a set of parameters $A$ and pick $x$ and $y$  such that $x\sleq_A y$. So we have $f_i(x)g_i(y)\leq f_i(y)g_i(x)$ for all $i$. We can rewrite these inequalities to
\begin{equation*}
    \frac{g_i(y)}{g_i(x)} \leq \frac{f_i(y)}{f_i(x)} = \frac{y_i-y_{i+1}}{x_i-x_{i+1}}.
\end{equation*}
Note that only the lefthandside (LHS) depends on the parameters $A$. The righthandside is constant. If we fix $x$ and $y$ we can view $g_i$ as a function of the parameters. If we take the derivative of the LHS with respect to a given parameter and this derivative is nonpositive, than this means that increasing the parameter will decrease the LHS, so that the inequality will still hold. We will show that the derivative to certain parameters is always negative regardless of $x$ and $y$, so that increasing that parameter creates a less strict partial order.

The derivative to $g_i(y)/g_i(x)$ with respect to $A^i_j$ is given by the quotient rule as
\begin{equation*}
    \frac{g_i(x)y_j - g_i(y)x_j}{g_i(x)^2}.
\end{equation*}
The sign of that expression is equal to the sign of
\begin{equation*}
    g_i(x)y_j - g_i(y)x_j = (x_{i+1}y_j - y_{i+1}x_j) + \sum_{k=i+2,k\neq j}^n A_k^i(x_ky_j - y_kx_j).
\end{equation*}

Let's in particular look at the inequality $i=n-2$ which has one free parameter $j=n$. The above expression then simply becomes $x_{n-1}y_n - y_{n-1}x_n$ which is precisely the $i=n-1$ inequality, so we know that this is negative. So increasing $A^{n-2}_n$ gives us a less strict partial order. We can rewrite the $i=n-2$ inequality to the following form:
\begin{equation*}
    x_{n-2}y_{n-1}-y_{n-2}x_{n-1} \leq A_n^{n-2}((y_{n-2}-y_{n-1})x_n - (x_{n-2}-x_{n-1})y_n)
\end{equation*}
Suppose $A^{n-2}_n\geq 0$ and that the righthandside (RHS) term is negative. Increasing $A^{n-2}_n$ would then decrease the RHS, so that at one point the inequality stops holding. So the RHS must be positive. If $A^{n-2}_n < 0$ then it is a stricter partial order than the one with $A^{n-2}_n = 0$, so we must have that the RHS term is positive as well.

So for any $\sleq_A$ and $x$ and $y$, if $x\sleq_A y$, we must have 
\begin{equation*}
    (y_{n-2}-y_{n-1})x_n - (x_{n-2}-x_{n-1})y_n \geq 0.
\end{equation*}
This inequality can be rewritten to
\begin{equation*}
    0 \geq x_{n-1}y_n - y_{n-1}x_n \geq x_{n-2}y_n-y_{n-2}x_n.
\end{equation*}

Now we move on to the next inequality: $i=n-3$, $j=n$. The expression of the sign of the derivative to $A^{n-3}_n$ is
\begin{equation*}
    (x_{n-2}y_n - y_{n-2}x_n) + A_{n-1}^{n-3}(x_{n-1}y_n-y_{n-1}x_n).
\end{equation*}
We know that both the terms involving $x$ and $y$ components are negative so if $A_{n-1}^{n-3}$ is positive this expression is negative. Checking back to the classification we have $1+A_{n-1}^{n-3} > 0$. So, taking $A_{n-1}^{n-3}< 0$ we can write
\begin{align*}
    &(x_{n-2}y_n - y_{n-2}x_n) + A_{n-1}^{n-3}(x_{n-1}y_n-y_{n-1}x_n) \\
    \leq& (x_{n-2}y_n - y_{n-2}x_n) + A_{n-1}^{n-3}(x_{n-2}y_n-y_{n-2}x_n) \\
    =& (A_{n-1}^{n-3}+1)(x_{n-2}y_n - y_{n-2}x_n) \\
    \leq& 0.
\end{align*}

So the derivative to $A^{n-3}_n$ is again always negative. Using the same argument as before we get $(y_{n-3}-y_{n-2})x_n \geq (x_{n-3}-x_{n-2})y_n$ which can be used to form a chain of inequalities:
\begin{equation*}
    0 \geq x_{n-1}y_n - y_{n-1}x_n \geq x_{n-2}y_n-y_{n-2}x_n \geq x_{n-3}y_n-y_{n-3}x_n.
\end{equation*}
This allows us to continue this procedure until we get down to $i=1$. So we know that for each parameter $A^i_n$ where $1\leq i \leq n-1$ increasing it will create a less strict partial order, and for any set of parameters $A$ when $x\sleq y$ we have
\begin{equation*}
    (x_i - x_{i+1})y_n \leq (y_i - y_{i+1})x_n\quad \text{for }1\leq i \leq n-1.
\end{equation*}
Adding these inequalities together we get 
\begin{equation*}
    (x_i - x_j)y_n \leq (y_i - y_j)x_n \quad \text{where }1\leq i < j \leq n.
\end{equation*}
This turns out to be very useful.

\subsection{Antisymmetry and non-contradicting orders}
We can use the previously derived inequalities to derive some nice properties of the partial orders $\sleq_A$.
\blemma
If $x\sleq_A y$ and $x_n\leq y_n\neq 0$ then $x=y$. So if $x\sleq_A y$ then $x_n\geq y_n$.
\elemma
\begin{proof}
Suppose we have $x \sleq_A y$ and $x_n \leq y_n\neq 0$. If $x_n=0$, then the above inequalities directly give $(x_i-x_j)y_n\leq 0$. Taking $j=n$ then immediately gives $y_n=0$, so we must have $x_n\neq 0$. Then the inequalities proven above give 
\begin{equation*}
    x_i - x_j \leq y_i - y_j \quad \text{where } 1\leq i<j\leq n.
\end{equation*}
Specifically, taking $j=n$ gives $x_i - y_i \leq x_j-y_n \leq 0$, so that $x_i\leq y_i$ for all $i$. Because $x$ and $y$ are normalised, this is only possible when $x=y$.
\end{proof}

So we now know that when $x\sleq_A y$ we have $x_n\geq y_n$.

The lemma above also ensures that when $x \sleq_A y$ and $x_n = 0$, then $y_n = 0$. In this case all the parameters $A^i_n$ don't influence $\sleq_A$ and we can prove all the statements about increasing parameters with $n$ replaced by $n-1$. The lemma here then works for $x_{n-1}\geq y_{n-1}$. In fact we get a more general statement.
\blemma
Let $x^-$ denote the smallest nonzero coordinate of $x$ and let $Z(k)=\#\{k\delim x_k=0\}$ the zero counting function, then if $x\sleq_A y$ 
\begin{itemize}
    \item For all $k$ where $x_k = 0$ we have $y_k = 0$, so $Z(y)\geq Z(x)$.
    \item If $Z(x)=Z(y)$, then $x^- \geq y^-$.
    \item If $Z(x)=Z(y)$ and $x^- = y^-$, then $x=y$.
\end{itemize}
\elemma 
\begin{proof}
Note that $x,y\in \Lambda^n$, so $x_k\geq x_{k+1}$, so if $x_k = 0$, then $x_{k+1}=0$. So the first point follows by induction, because we know that when $x_n=0$, $y_n=0$ and then we can reduce the situation to that of $\Lambda^{n-1}$.

If the amount of zeroes in $x$ and $y$ is equal, then there is a unique $k$ such that $x^-=x_k$ and $y^-=y_k$ and $x_{k+1}=0$ and $y_{k+1}=0$. But then $x$ and $y$ are in the situation of the previous lemma, so we get $x^- = x_k \geq y_k = y^-$.

If furthermore we also have $x^-=y^-$, then we have $x^-\leq y^-$, so by the previous lemma we get $x=y$.
\end{proof}

This allows us to define a useful map.
\btheor
Define $\mu^-: \Lambda^n \rightarrow [0,\infty)^*$ as $\mu^-(x) = 2n-3 - 2Z(x) + x^-$. $\mu^-$ is a strict monotonic map for any $\sleq_A$.
\etheor
\begin{proof}
The constant $2n-3$ in $\mu^-$ is chosen so that $\mu^-(\top_1) = 0$, it doesn't affect any of the properties of $\mu^-$. Let $x\sleq_A y$. We know that $Z(y)\geq Z(x)$. If $Z(y)$ is strictly greater than $Z(x)$, than in particular $-2Z(y)+y^- \leq -2Z(y) + 1 < -2Z(x) \leq -2Z(x)+x^-$, so $\mu^-(x)\geq \mu^-(y)$. If $Z(x)=Z(y)$, then we know that $x^-\geq y^-$, so still $\mu^-(x)\geq \mu^-(y)$. This proves the monotonicity of $\mu^-$.

For strictness assume that $x\sleq_A y$, and $\mu^-(x)=\mu^-(y)$. We then must have $Z(x)=Z(y)$, so that from $\mu^-(x)=\mu^-(y)$ we get $x^-=y^-$. Using the previous lemma then gives us $x=y$.
\end{proof}
Note that the map $\mu^-$ is not Scott-continuous: take an arbitrary point $x\in $ int$(\Lambda^n)$, and let $z(t) = (1-t)x + t\top_1$, and construct the sequence $x_i = z(1 - 1/i)$. This sequence is obviously increasing with limit and join $\top_1$. $Z(x_i)=0$ while $Z(\top_1)=n-1$, so there is no way that $\lim \mu^-(x_i) = \mu^-(\lim x_i)$.

The existence of this map has an important consequence. First of all: we haven't actually shown yet that the $\sleq_A$ are partial orders! But now it follows easily, because when $x\sleq_A y$ and $y\sleq_A x$, we have $\mu^-(x)\geq \mu^-(y)$ and $\mu^-(y)\geq \mu^-(x)$, so that $\mu^-(x)=\mu^-(y)$ which gives $x=y$. So the $\sleq_A$ are indeed antisymmetric. 

Furthermore, call two partial orders $\sleq_1$ and $\sleq_2$ on the same set \emph{contradicting} if there exist $x$ and $y\neq x$ such that $x\sleq_1 y$ and $y\sleq_2 x$. We then immediately see that the $\sleq_A$ are not contradicting because of the $\mu^-$ map. The restrictions that produced these partial orders are strict enough to `force' the comparisons in a certain direction. 

\blemma
    The renormalised Löwnner orders also don't contradict any restricted order.
\elemma
\begin{proof}
The order $\sleq^-_L$ also has $\mu^-$ as a strict monotone map, so this follows immediately. For $\sleq^+$ we have to do a little bit more work to get the same result. Suppose $x\sleq_A y$. We have already seen that we then have $(x_i-x_j)y_n \leq (y_i-y_j)x_n$ where $i<j$ (we'll assume $x_n,y_n\neq 0$, the argument still works if these are zero). Take $i=1$, and rewrite the inequalities to $x_1y_n-y_1x_n \leq x_jy_n - y_jx_n$. Now suppose that $y\sleq_L^+ x$. This implies $x_1y_n\geq y_1x_n$ so that $x_jy_n - y_jx_n \geq 0$ for all $j$. We can rewrite this to $x_jy_n\geq y_jx_n$. $x\sleq_A y$ implies $x_n\geq y_n$ so that this inequality can only be satisfied when $x_j\geq y_j$ for all $j$, but then $x=y$.
\end{proof}

Because the restricted information orders don't contradict we could consider the `union' of the partial orders. This turns out to be a restricted information order as well.

\subsection{The maximum restricted order}
Consider the specific case of $n=3$. A partial order $\sleq_A$ is then given by one parameter, and has the form $x\sleq_A y$ iff $x_2y_3\leq y_2x_3$ and $f(x)g(y) = (x_1-x_2)(y_2 + A^1_3y_3)\leq (y_1-y_2)(x_2 + A^1_3y_3) = f(y)g(x)$. We have shown in the previous section that increasing $A^1_3$ gives a more general (less strict) partial order. It would then seem that in the limit of taking this value to infinity we would get the most general partial order. 

Suppose $x_3\neq 0$ and $y_3\neq 0$. Note that we can rescale the inequality $f(x)g(y)\leq f(y)g(x)$ by an arbitrary constant. Divide it by $A^1_3$ so that the expression will remain bounded when $A^1_3\rightarrow \infty$. Then we have $g(y)/A^1_3\rightarrow y_3$. In the limit the inequality would become $(x_1-x_2)y_3 \leq (y_1-y_2)x_3$. Note that the other inequality $x_2y_3\leq y_2x_3$ can also be written as $(x_2-x_3)y_3\leq (y_2-y_3)x_3$ so that the partial order now is
\begin{equation*}
    x\sleq y \iff (x_k-x_{k+1})y_n \leq (y_k-y_{k+1})x_n \text{ for } 1\leq k \leq n-1.
\end{equation*}
This is the case for $y_3\neq 0$ and $x_3\neq 0$. If we have $x_3=0$, then by necessity also $y_3=0$ in which case the partial order simplifies to $x\sleq y$ iff $x_1y_2\leq y_1x_2$ which is the unique information order on $\Lambda^2$. 

There is one case left: $y_3=0$ and $x_3\neq 0$. In this case for very large $A^1_3$ the inequality gets close to
\begin{equation*}
    (x_1-x_2)y_2 \leq (y_1-y_2)A^1_3x_3.
\end{equation*}
We must again distinguish two cases. If $y_1\neq y_2$, then the RHS blows up while the LHS stays bounded, so in this case the inequality would be trivial. The other inequality $(x_2-x_3)y_3 = 0 \leq (y_2-y_3)x_3 = y_2x_3$ is also trivially satisfied. If $y_1=y_2$ then the RHS is zero, so that this inequality is only satisfied when $x_1=x_2$ (this is simply the degeneracy condition).

This motivates the definition of the maximal order.
\bdefin
For $x,y \in \Lambda^n$ we set $x \sleq_{max}^n y$ if and only if one of the following mutually exclusive options hold.
\begin{itemize}
    \item $y_n=x_n=0$ and $x \sleq_{max}^{n-1} y$.
    \item $x_n,y_n\neq 0$ and for all $1\leq k\leq n-1$ we have $(x_k-x_{k+1})y_n \leq (y_k-y_{k+1})x_n$.
    \item $y_n=0$ and $x_n\neq 0$ and for all $1\leq k\leq n-1$ such that $y_k=y_{k+1}$ we have $x_k=x_{k+1}$.
\end{itemize}

The base case $\sleq_{max}^2$ is defined as $x\sleq_{max}^2 y$ iff $x_1\geq y_1$.

We call $\sleq_{max}^n$ the \emph{maximal restricted order} on $\Lambda^n$.
\edefin

\btheor
$\sleq_{max}^n$ is indeed the maximal restricted order:
\begin{itemize}
    \item $\sleq_{max}^n$ is a restricted information order.
    \item If $x\sleq_A y$ for any $\sleq_A$ and $x,y\in\Lambda^n$, then $x\sleq_{max}^n y$.
\end{itemize}
\etheor
\begin{proof}
We prove by induction. We know that $\sleq_{max}^2$ is a restricted information order and for $n=2$ any $\sleq_A$ is equal to the unique information order on $n=2$. Suppose it is true for $\sleq_{max}^{n-1}$. Reflexivity is trivial. For transitivity we have to distinguish cases. Let $x\sleq_{max}^n y$ and $y\sleq_{max}^n z$. If $x_n=y_n=z_n=0$ we reduce to $\sleq_{max}^{n-1}$. So suppose $x_n\neq 0$. If $y_n =0$ then $z_n=0$, and for all $k$ where $z_k=z_{k+1}$ we have $y_k=y_{k+1}$ so that $x_k=x_{k+1}$, so $x\sleq_{max}^n z$. When $y_n\neq 0$ and $z_n=0$ we can do it similarly. When $x_n,y_n,z_n\neq 0$, we rewrite the inequalities to $\frac{x_k-x_{k+1}}{x_n}\leq \frac{y_k-y_{k+1}}{y_n}$ in which case transitivity is clear.

For antisymmetry we can assume that $x$ and $y$ contain no zero coordinates since otherwise we would simplify to $\sleq_{max}^{n-1}$. So $x\sleq_{max}^n y$ and $y \sleq_{max}^n x$ would give $(x_k-x_{k+1})y_n=(y_k-y_{k+1})x_n$. We can rewrite this to $x_ky_n - y_kx_n = x_{k+1}y_n - y_{k+1}x_n$. For $k=n-1$ we get $0 = x_{n-1}y_n-y_{n-1}x_n$. So we in fact get $\frac{y_k}{x_k} = \frac{y_n}{x_n}$ for all $k$. If we suppose that $x_n\neq y_n$, then we see that either $y_k>x_k$ for all $k$ or $y_k<x_k$ for all $k$, both can't happen. So we must have $x_n=y_n$, which then gives $y_k=x_k$ for all $k$, so $x=y$. 

The proof that this partial order allows mixing is similar to the proof that the renormalised Löwner orders allow mixing. That $\bot_n$ is the minimal element and that $\top_1$ is the maximal element can simply be checked directly.

Now suppose $x\sleq_A y$. $\sleq_A$ satisfies the degeneracy condition so we know that when $y_k=y_{k+1}\neq 0$ then $x_k=x_{k+1}\neq 0$ and when $x_n=0$ then $y_n=0$. So if $y_n=0$ and $x_n\neq 0$, we immediately have $x \sleq_{max}^n y$. If $x_n,y_n\neq 0$ we have already seen that we can derive that $(x_i-x_j)y_n\leq (y_i-y_j)x_n$ for all $1\leq i<j \leq n-1$, so that we also have $x \sleq_{max}^n y$. If $x_n=y_n=0$ we use the induction hypothesis.
\end{proof}

$\sleq_{max}^n$ is less well behaved than the $\sleq_A$. In particular, it is not closed and it is not a dcpo. We will show this explicitly:
Take an arbitrary element $x=(x_1,x_2,x_3,0,\ldots,0) \in \Lambda^n$ with $x_1>x_2>x_3$ and let $\bot_2 = (\frac{1}{2},\frac{1}{2},0,\ldots,0)\in \Lambda^n$. We know that $x \sleq_{max}^n \bot_2$ is \emph{not} the case because $(\bot_2)_1 = (\bot_2)_2$ while $x_1\neq x_2$. Define $z(t) = (1-t)x + t\bot_2$, and let $x_i = z(1-1/i)$ for $i\geq 1$, so $x_1=x$ and all the $x_i$ have their first three coordinates nonzero and the rest zero, so to derive if $x_i \sleq_{max}^n x_j$ we can restrict to $n=3$. We have $(x_1-x_2)z(t)_3 = (x_1-x_2)(1-t)x_3 \leq (z(t)_1-z(t)_2)x_3 = (1-t)(x_1-x_2)x_3$ and $x_2z(t)_3 = (1-t)x_2x_3 \leq z(t)_2x_3 = (1-t)x_2x_3 + t\frac{1}{2}x_3$, so $x \sleq_{max}^n x_i$ for all $i$. We can use the same argument to show that $x_i$ is an increasing sequence. Any element $y=(p,1-p,0,\ldots,0)$ with $p>\frac{1}{2}$ is an upper bound of this sequence. In fact, we can set $y_k =  \frac{1}{2}(1+1/k,1-1/k,0,\ldots,0)$ which is a decreasing sequence of upperbounds of $(x_i)$. This sequence has a highest lower bound: $\bot_2$. Since $\bot_2$ is not a upper bound of $(x_i)$ we see that $(x_i)$ has no least upper bound.

\begin{figure}[htb!]
    \centering
    \begin{subfigure}[b]{0.48\textwidth}
    \centering
        \includegraphics[width=0.8\textwidth]{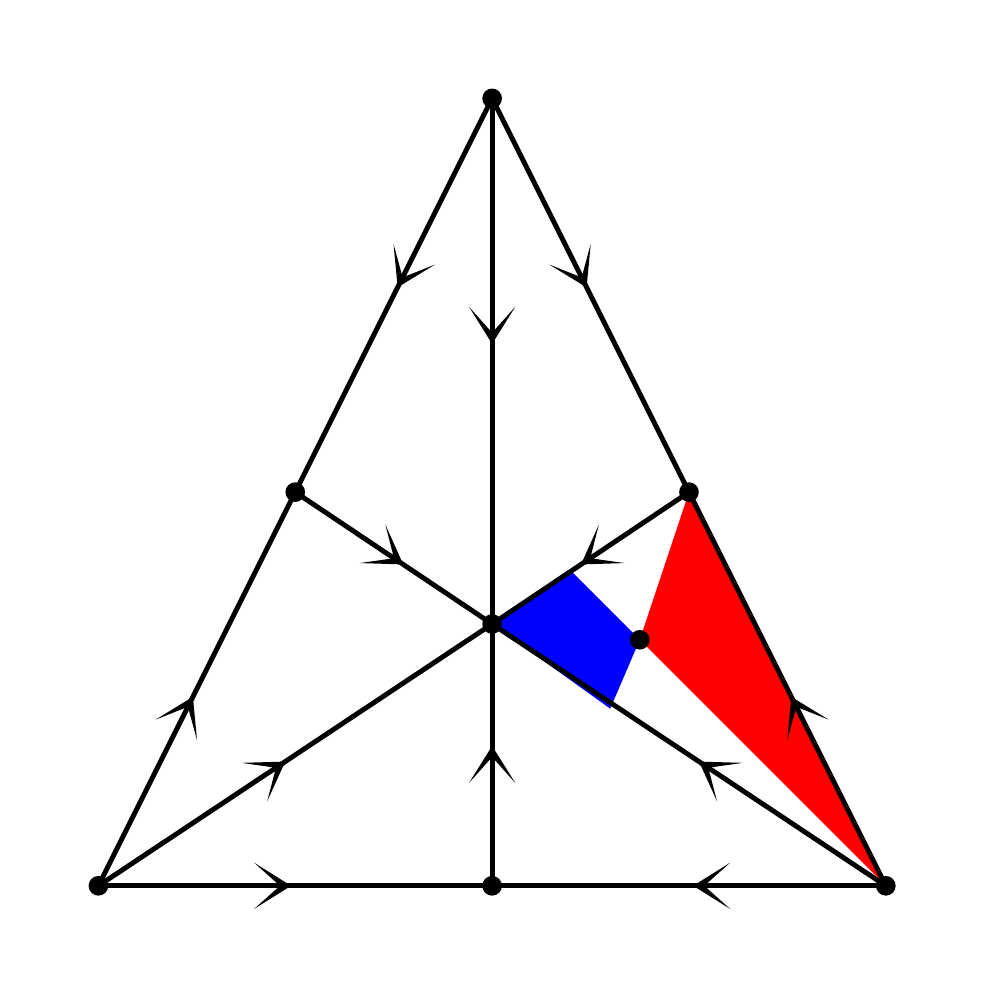}
        \caption{$\sleq_{max}^3$.}
    \end{subfigure}
    ~
    \begin{subfigure}[b]{0.48\textwidth}
    \centering
        \includegraphics[width=0.8\textwidth]{figures/figure_L2}
        \caption{$\sleq_L^-$.}
    \end{subfigure}
    \caption[Comparison of $\sleq_{max}^3$ and $\sleq_L^-$]{Upperset (red) and downset (blue) of the distribution $y=\frac{1}{30}(15,10,5)$ with respect to $\sleq_{max}^3$ and $\sleq_L^-$. Note that $\bot_2$ is not in the upperset in the case of $\sleq_{max}^3$.}
    \label{fig:compmax}
\end{figure}

The reason $\sleq_{max}^n$ is not a dcpo is because we had to exclude elements like $\bot_2$ from being above any nondegenerated element to preserve the degeneracy condition. If we were to include these elements in the uppersets we would have a dcpo, but in that case by transitivity $x\sleq \bot_2\sleq (0,1,0)$, so if we want to preserve mixing we would also have to include the line between $x\in \Lambda^3$ and $(0,1,0)$. In fact: the smallest information order that is a dcpo that covers $\sleq_{max}^n$ is $\sleq_L^-$. This is best seen when comparing the uppersets in a picture. See Figure \ref{fig:compmax}.

\subsection{Entropy}
\label{sec:entropy}
The Shannon entropy $\mu_S(x) = - \sum_i x_i\log x_i$ is a strict monotone Scott-continuous map for the Bayesian order \cite{Coecke2010book}. We can wonder when this is the case for the other information orders as well.

We turn back to the method of taking the derivatives of the parameters to prove inequalities about the partial orders. Suppose $x\sleq_A y$. Recall that the sign of the derivative with respect to $A^k_i$ of the LHS of
\begin{equation*}
    \frac{g_k(y)}{g_k(x)} \leq \frac{f_k(y)}{f_k(x)}
\end{equation*}
is equal to the sign of
\begin{equation*}
    g_k(x)y_i - g_k(y)x_i = x_{k+1}y_i - y_{k+1}x_i + \sum_{j=k+2}^{i-1}A_j(x_jy_i-y_jx_i) - \sum_{j=i+1}^n A_j(x_iy_j - x_jy_i).
\end{equation*}
In Section \ref{sec:parameters} we showed that for $i=n$ this expression is always negative, which also produced the inequalities 
\begin{equation*}
    0\geq x_ky_n-y_kx_n \geq x_{k-1}y_n - y_{k-1}x_n
\end{equation*}
for $1\leq k \leq n-1$. Now set $k=n-3$ and $i=n-1$, then the sign equation is
\begin{equation*}
    x_{n-2}y_{n-1} - y_{n-2}x_{n-1} - A^{n-3}_n(x_{n-1}y_n - y_{n-1}x_n).
\end{equation*}
Recall that the $(n-2)$th inequality is
\begin{equation*}
    (x_{n-2}-x_{n-1})(y_{n-1}+A^{n-2}_ny_n) \leq (y_{n-2}-y_{n-1})(x_{n-1} + A^{n-2}_nx_n)
\end{equation*}
which can be rewritten to
\begin{equation*}
    x_{n-2}y_{n-1}-y_{n-2}x_{n-1} \leq A^{n-2}_n[(y_{n-2}-y_{n-1})x_n - (x_{n-2}-x_{n-1})y_n].
\end{equation*}
We have already seen that $(y_{n-2}-y_{n-1})x_n - (x_{n-2}-x_{n-1})y_n \geq 0$, so if $A^{n-2}_n\leq 0$, then the RHS is negative, which means that $x_{n-2}y_{n-1}-y_{n-2}x_{n-1} \leq 0$. Looking back at the sign equation for $k=n-3$ and $i=n-1$ we can see that if we also take $A^{n-3}_n\leq 0$, the expression will always be negative. We will then get some new inequalities we can use in our derivations of the sign of $A^{n-4}_{n-1}$, and we can continue this procedure. The inequalities resulting from this procedure are
\begin{equation*}
    (x_i-x_j)y_k\leq (y_i-y_j)x_k \text{for all } 1\leq i<j\leq k\leq n.
\end{equation*}

The clue here is that we took the parameters to be negative. If some parameters are positive then we can't say anything in general about the sign of the derivative of $A^k_j$. For the partial orders given by small (negative) parameters we can say something general.
\blemma
    Let $\sleq_A$ and $\sleq_B$ be restricted information orders given by sets of parameters $A$ and $B$, with $A^k_j\leq 0$ for all $k$ and $j$. If $A^k_j\leq B^k_j$ for all $k$ and $j$, then $\sleq_A$ is stricter than $\sleq_B$: $x\sleq_A y$ implies $x\sleq_B y$.
\elemma
\begin{proof}
For all components where $B^k_j$ is negative this follows directly from the sign equations as defined above. If some of the components of $B$ are positive it follows because we can start from the lowest values of $j$ and work our way up. So let $C$ be the set of parameters given by $C^k_j = \min(0,\max(A^k_j,B^k_j))$, then it should be clear that $\sleq_C$ is less strict than $\sleq_A$. Then we can pick the lowest $j$ and $k$ such that $B^k_j \geq 0$. Increasing this parameter on $C$ still gives a more general order, since all the relevant parameters in $C$ for this to hold are nonpositive, so let $C^\prime$ be $C$ but with $C^k_j = B^k_j$, then $\sleq_{C^\prime}$ lies above $\sleq_C$. We can then carry on to the next positive parameter which only depends on the parameters with a higher value for $k$ and $j$. In the end we indeed have $\sleq_A$ stricter than $\sleq_B$.
\end{proof}

Recall that the Bayesian order is given by $\sleq_A$ with $A^k_j=0$ for all $k$ and $j$. This lemma shows that the Bayesian order is the maximal order with respect to these negative parameter orders. It then follows that these orders also allow Shannon entropy as a strict monotone Scott-continuous map (Scott-continuity folllows from the continuity of the entropy function and the closedness of the orders). 

We can now also consider a \emph{minimal} information order as the order given by the intersection of all the $\sleq_A$'s. Since the RIO's with negative parameters are already stricter than the ones with positive parameters, we only have to consider the intersection of the negative parameter orders. For illustration, consider the $n=3$ case. Then there is one free parameter $A^1_3$, and we have the condition $1+A^1_3>0$, so the intersection would consist of taking the limit $A^1_3 \rightarrow -1$. Let's look at the order given by $A^1_3=-1$. It looks like
\begin{equation*}
    x \sleq y \iff (x_1-x_2)(y_2-y_3) \leq (y_1-y_2)(x_2-x_3)\quad \& \quad x_2y_3\leq y_2x_3.
\end{equation*}
This works completely fine as an information order, except when we have a pair $x$ and $y$ with $x_1=x_2$ and $x_2=x_3$ and $y_2=y_3$. In this case both inequalities reduce to $0\leq 0$, so that $x\sleq y$ and $y\sleq x$. The problem is that $x$ has a double degeneracy that the first inequality can't deal with. Fortunately, this is only the case when $x_1=x_2=x_3$ in which case $x=\bot_3$. So this problem is easily fixed by adding the extra requirement to the above order that $y\neq \bot_3$.

For $n>3$ there are multiple parameters that we could take the limit of. In general these all produce different partial orders, where one is not stricter than the order. The minimal order is then the intersection of all these limit orders. This order is the following:
\begin{equation*}
    x\sleq_{min}^n y \iff y\neq \bot_n \text{ and } (x_k-x_{k+1})(y_{k+1}-y_j)\leq (y_k-y_{k+1})(x_{k+1}-x_j)
\end{equation*}
for all $1\leq k < j \leq n$.

The Bayesian order is not the maximal restricted order having Shannon entropy as a measurement. As a counter example let
\begin{equation*}
    x\sleq_E y \iff (x_1-x_2)(1-y_1)\leq (y_1-y_2)(1-x_1)~~\&~~ x_ky_{k+1}\leq y_ky_{k+1}
\end{equation*}
for all $2\leq k \leq n-1$.
This partial order looks almost the same as the Bayesian order except for the first inequality. Note that this partial order is given by $A^j_k = 0$ for $2\leq j < k\leq n$, and $A^1_k = 1$ for $3 \leq k \leq n$ where we use that $y_2 + \sum_{k=3}^n A^1_k y_k = y_2 + \sum_{k=3}^n y_k = 1 - y_1$ due to the normalisation of $y$. Because these parameters are all bigger than (or equal to) zero this order is indeed less strict than the Bayesian order.
\blemma
    The order $\sleq_E$ has the following properties.
    \begin{itemize}
        \item The map $\mu^+: \Lambda^n \rightarrow [0,\infty)^*$, $\mu^+(x) = 1-x_1$ is a measurement.
        \item If $x\sleq_E y$ then there is a $k$ such that $y_i\geq x_i$ for $i\leq k$ and $y_i\leq x_i$ for $i>k$.
        \item $\sleq_E$ allows Shannon Entropy as a measurement.
    \end{itemize}
\elemma
\begin{proof}
Let $x\sleq_E y$, and suppose $x_1\geq y_1$. Then $1-x_1\leq 1-y_1$. Since $(x_1-x_2)(1-y_1)\leq (y_1-y_2)(1-x_1)$ we must then have $x_1-x_2\leq y_1-y_2$. Rewriting then gives $0\leq x_1-y_1 \leq x_2-y_2$, so that $x_2\geq y_2$. The second inequality is $x_2y_3\leq y_2x_3$ which then gives $y_3\leq x_3$. We repeat so that we get $x_k\geq y_k$ for all $k$. This is only possible when $x=y$. So if $x\sleq_E y$ we have $y_1\leq x_1$, and if $x_1=y_1$ then $x=y$. This proves that $\mu^+$ is strict monotone. Scott-continuity follows from the continuity of the map and the closedenss of $\sleq_E$.

Let $x\sleq_E y$. Suppose that $x_k\geq y_k$ for a $k\geq 2$. Then $x_ky_{k+1}\leq y_kx_{k+1}$, so then also $x_{k+1}\geq y_{k+1}$. So if for some $k$, $x_k\geq y_k$, then also $x_j\geq y_j$ for all $j\geq k$. Since $x_n\geq y_n$ and $x_1\leq y_1$, there is a minimal such $k$. The proof that $\sleq_E$ has Shannon Entropy as a measurement then follows completely analogous to the proof that the Bayesian order allows Shannon Entropy as a measurement in \cite{Coecke2010book}.
\end{proof}

At the moment it is not clear if this is the maximal restricted order that is compatible with Shannon entropy. 

The \emph{majorization} preorder is a partial order when restricted to $\Lambda^n$ and is given by
\begin{equation*}
    x\sleq_M y \iff \sum_{i=1}^k x_i \leq \sum_{i=1}^k y_i \text{ for all } 1\leq k \leq n.
\end{equation*}
It is well known that the majorization order is monotone over all Schur-convex functions, an example of which is Shannon Entropy \cite[Chapter 12]{convex1992}. This means that an order that contradicts the majorization order on a pair of points can't be monotone over Shannon Entropy. Note that $\mu^-$ is monotone over $\sleq_M$. This means that any restricted information order is compatible with majorization. Majorization is also monotone over $\mu^+$. Note that majorization does not satisfy the degeneracy conditions, so it doesn't extend to an information order on $\Delta^n$. The maximum order $\sleq_{max}^n$ does not agree with entropy, but it still allows $\mu^-$ as a strict monotone map, so it doesn't contradict majorization. This shows that non-contradiction with majorization is not enough to prove that it is compatible with entropy.

$\sleq_E$ is not the least restrictive information order that has $\mu^+$ as a measurement. 
\begin{equation*}
    x \sleq_1 y \iff (x_k-x_{k+1})(1-\sum_{i=1}^k y_i) \leq (y_k-y_{k+1})(1-\sum_{i=1}^k x_i) \text{ for } 1\leq k \leq n-1.
\end{equation*}
We call it $\sleq_1$, because this is the order where $A^k_j = 1$ for all $k$ and $j$. Suppose again that $x\sleq_1 y$ and $x_1\geq y_1$, then as in the other lemma we get $x_2\geq y_2$. We then have $x_1+x_2 \geq y_1+y_2$, so using the second inequality we must have $x_2-x_3 \leq y_2-y_3$, which rewrites to $0\leq x_2-y_2 \leq x_3-y_3$ so that $x_3\geq y_3$. Rinse and repeat: $x_k\geq y_k$ for all $k$, so $x=y$. This order does not have the property that $x_k\leq y_k$ for small $k$ and $x_k\geq y_k$ for large $k$, so the proof that $\sleq_E$ has Shannon entropy as a measurement can't be applied here. It is not clear if this partial orders is compatible with entropy. Some new proof method would be needed. Note that for $n\leq 3$, $\sleq_1$ and $\sleq_E$ are the same.

\section{Domains}
\btheor
$\Lambda^n$ is a domain (continuous dcpo) when equipped with any $\sleq_A$.
\label{theor:domain}
\etheor
\begin{proof}
This proof is an adapted more general version of the proof that the Bayesian order is a domain as given in \cite{Coecke2010book}. We have already seen that $\sleq_A$ is a dcpo, so we only have to show that $\sleq_A$ is continuous: that for each $y\in \Lambda^n$ Approx$(y)$ is a directed set with least upper bound $y$. Since $\sleq_A$ is an upwards small partial order, it suffices to find an increasing sequence $(x_i)$ with $\vee x_i = y$ such that $x_i \ll y$. We will show that $x(t)\ll y$ for $t<1$ with $x(t) = (1-t)\bot_n + ty$ which proves the statement.

Let $a_n\rightarrow a$ with join/limit $a$ such that $y\sleq a$. We note that we then have $f_k(y)g_k(a) - f_k(a)g_k(y)\leq 0$ for all $k$. Since all the $f_k$ and $g_k$ are continuous and the $a_n$ are converging, for any $\epsilon>0$ there will be an $N$ such that for all $n> N$ we have $f_k(y)g_k(a_n)-f_k(a_n)g_k(y)\leq \epsilon$. We note further that if $f_k(a_n) = 0$ then either $f_k(a)=0$ in which case $f_k(y) = 0$ so that we can ignore this $k$th inequality, or $f_k(a)> 0$, in which case there is an $N$ such that $f_k(a_n)>\frac{1}{2}f_k(a)$ for all $n>N$. Now choose 
\begin{equation*}
    \epsilon = \frac{1}{4}\frac{1-t}{t}\min\left\{f_k(a)g_k(\bot_n)\delim \forall k: f_k(a)\neq 0\right\}.
\end{equation*}
Since $g_k(\bot_n)>0$ for all $k$ we have $\epsilon>0$, then take $N$ such that for all $n> N$ we have $f_k(y)g_k(a_n)-f_k(a_n)g_k(y)\leq \epsilon$ and $f_k(a_n)>\frac{1}{2}f_k(a)$. Pick a $n > N$. We can now calculate:
\begin{align*}
    &f_k(x(t))g_k(a_n) - f_k(a_n)g_n(x(t)) \\
    =& tf_k(y)g_k(a_n) - f_k(a_n)[(1-t)g_k(\bot_n) + tg_k(y)] \\
    =& t(f_k(y)g_k(a_n) - f_k(a_n)g_k(y)] - (1-t)f_k(a_n)g_k(\bot_n) \\
    \leq& t\epsilon - (1-t)f_k(a_n)g_k(\bot_n) \\
    \leq& \frac{1}{4}(1-t)f_k(a)g_k(\bot_n) - (1-t)f_k(a_n)g_k(\bot_n) \\
    \leq& 0
\end{align*}
for all $k$ where $f_k(a)\neq 0$. If $f_k(a)=0$ then $f_k(y)=0$ so because $x(t)\leq y$ we also have $f_k(x(t))=0$ so that the $k$th inequality trivially holds. So $x(t)\sleq a_n$. Since the sequence $(a_n)$ was arbitrary we have $x(t)\ll y$ which proves that $(\Lambda^n, \sleq_A)$ is a domain.
\end{proof}

We can use a similar argument for $\sleq_L^+$: recall that it was defined as $x\sleq_L^+ y$ iff $x^+y_k \sleq y^+x_k$. Now if for a certain $k$ $y_k=y^+$, then we must also have $x_k=x^+$. This means that when $a_n\rightarrow a$ increasing and $x\sleq y\sleq a$, there is a $k$ such that $a_k=a^+$, $y_k=y^+$, $x_k=x^+$ and $a_n^+ = (a_n)_k$. So writing $f_i(x) = x_k = x^+$ and $g_i(x) = x_i$ for all $i$ we can use exactly the same argument as above. So $\sleq_L^+$ is also a domain.

When restricted to $\Lambda^n$, the second renormalised Löwner order is also a domain: when we only consider $x,y\in\Lambda^n$ we can write
\begin{equation*}
    x\sleq_L^- y \iff Z(x)<Z(y)\text{ or } x_ky^-\leq y_kx^- \text{for all $k$ with }x_k,y_k\neq 0
\end{equation*}
where $Z(x)$ is the zero counting function. Again let $x(t) = (1-t)\bot_n + ty$ for $0\leq t<1$ then $Z(x(t)) = 0$. Let $a_m\rightarrow a$ be an increasing sequence. Suppose $(a_m)$ contains an element with $Z(a_m)> 0$, then we have $x(t)\sleq_L^- a_m$ and we are done, so assume $Z(a_m)=0$. Suppose $Z(a)> 0$, then $(a_m)_n\rightarrow 0$. But then $x(t)_k(a_m)_n\rightarrow 0 \leq (a_m)_kx_n$, so at some point $x(t)\sleq_L^- a_m$. So finally suppose $Z(a)=0$, then if $y\sleq_L^- a$ we also have $Z(y)=0$. But in this case we can just set $f_i(x)=x_i$ and $g_i(x)=x_n$ and use the same arguments as above.

\begin{figure}[htb!]
\centering
\includegraphics[width=0.4\textwidth]{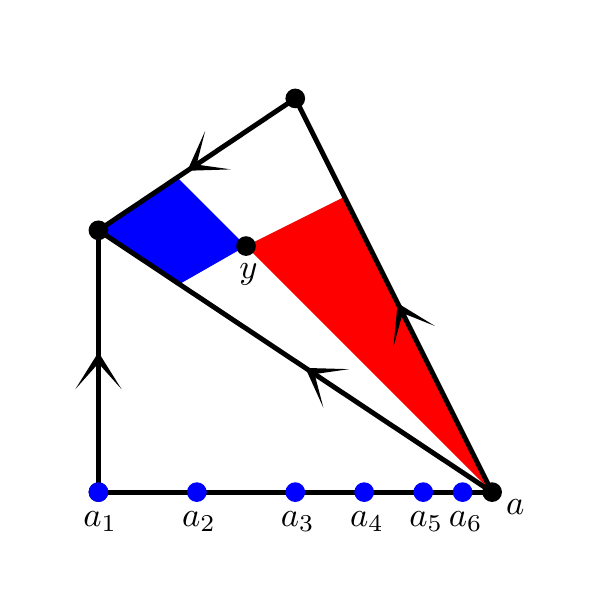}
\caption[Increasing sequence with no approximation]{An increasing sequence $a_n\rightarrow a=(1,0,0)$. Upperset and downset of $y$ are denoted in red and respectively blue.}
\label{fig:increasing}
\end{figure}

We have now shown that all these partial orders are domains when restricted to $\Lambda^n$. However they do \emph{not} produce domains on $\Delta^n$ in general. This is best illustrated with a picture: see Figure \ref{fig:increasing}.

Let $y=(y_1,y_2,y_3)$ with $y_1>y_2>y_3>0$ and $a=\top_1=(1,0,0)$, then $y\sleq a$ and construct $a_n=(1-\frac{1}{n},0,\frac{1}{n})$ for $n\geq 3$, then $a_n$ is an increasing sequence with join $a$. For any $\sleq_A$ and $\sleq_L^-$ there is not any element in int$(\Lambda^n)$ that is bigger then any of these $(a_n)$, so there can't possibly be an increasing sequence that converges to $y$ (which is in the interior of $\Lambda^n$).

The maximum eigenvalue Löwner order \emph{is} still a domain on $\Delta^n$. The argument used in Theorem \ref{theor:domain} also works for $\sleq^+$ in $\Delta^n$. The reason why it doesn't fail for this specific $a_n$ is because $\top_1$ is in the \emph{interior} of the upperset of any element, so that any sequence converging to the top element will at some point lie in the upperset itself.

\section{Summary and conclusions}
In this chapter we defined a new kind of structure called an information order. These are partial orders on $\Delta^n$ that carry a certain kind of information-like structure. We looked in detail at a certain class of these orders called restricted information orders (RIO's) and found that these can be classified by a group of real parameters. See Figure \ref{fig:properties} for a short summary of the different properties of a couple of information orders encountered.

\begin{figure}[htb!]
\centerline{
\begin{tabular}{l l || c | c | c | c | c}
    Name & Symbol & Measurements & Restricted? & dcpo? & closed? & domain? \\
    \hline
    Bayesian order & $\sleq_B$ & $\mu^-,\mu^+,\mu_S$ &  \checkmark & \checkmark & \checkmark & X  \\
    RIO parametrized by $A$ & $\sleq_A$ & $\mu^-$ & \checkmark & \checkmark & \checkmark & X \\
    Maximal order & $\sleq_{max}^n$ & $\mu^-$ & \checkmark & X & X & X \\
    Minimum eigenvalue & $\sleq^-_L$ & $\mu^-$ & X & \checkmark & X & X \\
    Maximum eigenvalue & $\sleq^+_L$ & $\mu^+$ & X & \checkmark & \checkmark & \checkmark
\end{tabular}
}
\caption[Summary of properties of information orders]{Table of the various information orders encountered and their properties.}
\label{fig:properties}
\end{figure}

Note that all the orders in the table that are dcpo's are also domains when restricted to $\Lambda^n$, the monotone sector. The only order found so far that is also a domain on the entirety of $\Delta^n$ is $\sleq^+_L$. With `measurements' we mean maps that are strict monotone in that order. $\mu^+(x) = 1-x^+$ and $\mu_S$ the Shannon entropy are continuous so that they are also Scott-continuous for closed partial orders. $\mu^-$ is not continuous and is also not Scott-continuous.

We've seen that some restricted information orders have Shannon entropy as a measurement and some have $\mu^+$ as a measurement (Section \ref{sec:entropy}), but is not yet clear whether these are the maximal orders with these properties. It is also not yet clear if $\sleq_L^+$ has $\mu_S$ as a measurement. 

Other open questions relate to what kind of information orders exist that are not of the restricted kind like the renormalised Löwner orders. $\sleq_L^+$ still satisfies the degeneracy condition on the first coordinate, while $\sleq_L^-$ satisfies it on the last coordinate, perhaps there are other information orders that only satisfy the degeneracy condition on the $k$th coordinate? 

It is clear that for an information order to be a domain on $\Delta^n$ any element $x\in \Lambda^n$ must have $\top_1$ in the interior of its upperset. This means that any information order that is also a domain can't have the degeneracy condition on the $k$th coordinate for $k\geq 2$. Information orders that are domains will in that sense be `similar' to $\sleq_L^+$. We'll investigate this in more detail in the following chapter.

We'll also look at how the orders studied here can be extended to the space of density operators. In particular, we'll take a closer look at $\sleq_L^+$ and $\sleq_L^-$ and see that they have some more interesting properties.
\chapter{Ordering Density Operators}
In the previous chapter we looked at information orders on the space of probability distributions. In this chapter we will extend this idea to the space of density operators. We will also define some extra properties and hypothesise that the maximum eigenvalue renormalised Löwner order is the unique order satisfying these properties.

\section{Preliminaries}

\bdefin
    A linear operator\footnote{We will use the terms `operator' and `matrix' interchangeably in this chapter.} $A: \mathbb{C}^n \rightarrow \mathbb{C}^n$ is called \emph{positive} when for all $v\in\mathbb{C}^n$, $v^\dagger A v \geq 0$. We denote the space of positive operators on $\mathbb{C}^n$ as $PO(n)$. If $A$ is positive we also write $A\geq 0$.
\edefin

A positive operator is necessarily Hermitian, so it has a spectrum of eigenvalues. It is not too hard to see that a Hermitian operator is positive iff all its eigenvalues are nonnegative. Note that if $A,B \in PO(n)$ and $r\in \mathbb{R}_{\geq 0}$, then $A+B\in PO(n)$ and $tA\in PO(n)$, so $PO(n)$ is a \emph{cone} (a special type of convex subspace) in the space of linear operators on $\mathbb{C}^n$.

\bdefin
    A matrix $U$ is called \emph{unitary} if $U^{-1} = U^\dagger$. If $A$ as a matrix is written in a certain orthonormal basis, then for any basis transformation there is a unitary $U$ such that $A$ written in that basis is $UAU^\dagger$. We denote the space of unitary matrices as $U(n)$.
\edefin

\bdefin
    For a given orthonormal basis $(e_i)$ we call a linear operator $M$ \emph{diagonal} with respect to this basis iff $Me_i = \lambda_i e_i$ for a set of eigenvalues $\lambda_i$. Denote the space of Hermitian diagonal matrices on $\mathbb{C}^n$ as Diag$(n)$. For any Hermitian matrix $A$ we can find an unitary $U$ and a diagonal $D$ such that $A = UDU^\dagger$.
\edefin

Note that Diag$(n) \cong \mathbb{R}^n$ since for a Hermitian matrix all the eigenvalues are real and $PO(n)\cap $Diag$(n) \cong \mathbb{R}_{\geq 0}^n$.

\bdefin
    The $\emph{trace} $ of a matrix $M$ is given by Tr$(M) = \sum_i e_i^\dagger M e_i$ where $(e_i)$ is an orthonormal basis.
\edefin

\blemma
    Let $M$, $N$ and $K$ be matrices, $U$ unitary, $A$ a Hermitian matrix with $A = UDU^\dagger$ where $D=$ diag$(\lambda_1,\ldots,\lambda_n)$ is diagonal. The trace has the following properties.
    \begin{itemize}
        \item The trace of a matrix is linear and independent of the chosen orthonormal basis.
        \item The trace is permutation invariant: Tr$(MNK)=$ Tr$(KMN)=$ Tr$(NKM)$.
        \item The trace is invariant under basis change: Tr$(UMU^\dagger) = $ Tr$(M)$.
        \item The trace of a Hermitian matrix is equal to the sum of its eigenvalues: Tr$(A) = $Tr$(UDU^\dagger)=$ Tr$(D) = \sum_i e_i^\dagger D e_i = \sum_i \lambda_i$.
        \item The trace of a positive matrix $B$ is positive and is zero iff $B=0$.
    \end{itemize}
\elemma

\bdefin
 We call a positive operator $A$ \emph{normalised} iff Tr$(A) = 1$. A normalised operator is also called a \emph{density operator}. The space of density operators is denoted as $DO(n)$. Note that $DO(n) = \{A: \mathbb{C}^n\rightarrow \mathbb{C}^n \text{ linear}\delim A^\dagger = A, A\geq 0, \text{Tr}(A) = 1\}$. We will from now on denote density operators by greek letters (such as $\rho$ and $\pi$).
\edefin

All the density operators are positive and thus Hermitian. This means that for $\rho \in DO(n)$ we can find a unitary matrix $U$ and diagonal matrix $D$ such that $\rho = UDU^\dagger$. We have 1 =  Tr$(\rho) = $ Tr$(D) = \sum_i \lambda_i$. Since $\lambda_i\geq 0$ and $\sum_i \lambda_i = 1$, the eigenvalues of $\rho$ form a probability distribution. So in fact we have Diag$(n)\cap DO(n) \cong \Delta^n$. Because of this we will simply denote the space of diagonal density operators as $\Delta^n$ and identify them with probability distributions.

This is no coincidence. A probability distribution is a classical notion of a state: a configuration that a classical system can be in. A density matrix on the other hand represents a quantum state. Such a state can be more complex because there are observables that don't commute with each other (such as position and momentum). This is represented in $DO(n)$ by the fact that not all density matrices can be diagonalised simultaneously. The diagonal density operators can then be seen as the subspace where all the observables do commute, so this is again a purely classical state.

In the previous chapter we were discussing orders that capture the idea of information content on classical states (probability distributions). Can we extend these to these quantum states (density matrices). This is in fact the case.

\bdefin
    Define the \emph{uniform distribution} of $DO(n)$ as $\bot_n = \frac{1}{n}I_n$ where $I_n$ is the identity operator on $\mathbb{C}^n$. $\bot_n$ is obviously diagonal and when interpreted as en element of $\Delta^n$ it is precisely the uniform distribution in $\Delta^n$.The \emph{pure states} of $DO(n)$ are precisely the rank-1 projections, or in other words: $\rho\in DO(n)$ is a pure state if there is a normalised vector $v\in\mathbb{C}^n$ such that $\rho v = v$ (so $v$ is an eigenvector of $\rho$ with eigenvalue 1. By normalisation all the other eigenvalues are then equal to zero). 
\edefin

\btheor
Let $\sleq$ be an information order as defined in Definition \ref{def:informationorder} on $\Delta^n$. This extends to a partial order on $DO(n)$ that allows mixing with least element $\bot_n$ and the pure states as the maximal elements.
\etheor
\begin{proof}
For $\rho,\pi \in DO(n)$ define $\rho \sleq_{DO(n)} \pi$ iff there is a $U\in U(n)$ such that $U\rho U^\dagger, U\pi U^\dagger \in \Delta^n$ (that is: $\pi$ and $\rho$ can be diagonalised at the same time) and $U\rho U^\dagger \sleq U\pi U^\dagger$.

Simultaneous diagonalisability defines an equivalence relation on $DO(n)$ so the transitivity, reflexivity and antisymmetry of $\sleq$ carries over to $\sleq_{DO(n)}$. Note that the unitary matrix $U$ is not uniquely defined: we can permute the coordinates in the basis and still be left with diagonal matrices, but this doesn't matter for the partial order precisely because $\sleq$ is permutation invariant. 

For any $U\in U(n)$ we have $U\bot_n U^\dagger = \frac{1}{n} UI_n U^\dagger = \bot_n UU^\dagger = \bot_n$, so $\bot_n$ is simultaneously diagonalisable with any density matrix, so it is indeed the least element. For any $\rho$ we have the $n$ $p_i$ that project to one of the eigenspaces of $\rho$. When simultaneously diagonalised, these $p_i$ have a 1 somewhere on the diagonal and zeroes everywhere else, so they are represented as the $\top_i$ on $\Delta^n$, which indeed makes them maximal elements by the properties of $\sleq$. That $\sleq_{DO(n)}$ also allows mixing follows from the mixing property on $\sleq$ and the fact that if $\rho$ and $\pi$ are simultaneously diagonalisable, then any convex combination of them will also be simultaneously diagonalisable.
\end{proof}

It is also true that if $\sleq$ is closed or a dcpo on $\Delta^n$ that this extension will also be closed or a dcpo on $DO(n)$. None of these extensions will be a domain. This can be demonstrated using the same argument as is demonstrated in Figure \ref{fig:increasing}: for any $\pi$ we can make an increasing sequence converging to the projection operator of the highest eigenvalue of $\pi$ such that this increasing sequence is everywhere written in a different basis than $\pi$ is. 

Although these extensions give information-like orderings on $DO(n)$ they might be considered too restrictive: the requirement that operators be simultaneously diagonisable for them to be comparable is very strong and trows away most of what makes $DO(n)$ interesting. We might hope there is a better set of information orders on $DO(n)$.

On the space of positive operators $PO(n)$ there is a natural choice of partial order \cite{lowner1934monotone}.
\bdefin
    Define the \emph{Löwner order} $\sleq_L$ on $PO(n)$ as follows.
    \begin{equation*}
        A \sleq_L B \iff B-A\geq 0.
    \end{equation*}
\edefin

\blemma
Some properties of the Löwner order.
\begin{itemize}
    \item The Löwner order is invariant under basis transformations (unitary invariant), allows mixing and has zero as the unique least element.
    \item The Löwner order is downwards small (so the dual is a dcpo).
    \item The Löwner order has kernel inclusion: if $A\sleq_L B$ and $Bv = 0$ for some $v$, then $Av=0$. In other words ker$(B)\subseteq $ ker$(A)$.
    \item The trace is strict monotone. For the dual order the trace is also Scott-continuous when we consider Tr$: PO(n) \rightarrow [0,\infty)^*$.
\end{itemize}
\elemma
\begin{proof}
Let $U$ be a unitary operator and suppose $A\sleq B$, so for all $v$: $v^\dagger (B-A) v \geq 0$. Since this holds for all $v$. This also holds for all $Uv$, as $U$ is a bijection of $\mathbb{C}^n$. Then for all $v$: $(Uv)^\dagger(B-A)(Uv) \geq 0$ which can be rewritten to $v^\dagger(U^\dagger B U - U^\dagger A U)v \geq 0$, so $U^\dagger A U \sleq U^\dagger B U$.

Define $C = (1-t)A+tB$ for $0\leq t \leq 1$. Then $C - A = t(B-A)$ and $B-C = (1-t)(B-A)$, so it is clear that when $A\sleq B$ then also $A\sleq C \sleq B$ for all $t$.

Suppose $A\sleq B$, then $B-A\in PO(n)$, so then Tr$(B-A)\geq 0$, and by linearity of the trace: Tr$(B)\geq $ Tr$(A)$. So the trace is monotone over the Löwner order. Furthermore if Tr$(B)= $ Tr$(A)$ then Tr$(B-A)=0$ and since $B-A$ is a positive operator, we must then have $B-A = 0$, so $B=A$, which proves strictness. The monotonicity of the trace means that the downset of a positive operator $B$ is contained in $\{A \in PO(n) \delim \text{Tr}(A)\leq \text{Tr}(B)\}$ which is a bounded set. Now downwards smallness follows if we prove that the Löwner order is closed. To do this note that $f_v(A) = v^\dagger A v$ is a linear (and thus continuous) map to the real numbers. So suppose $C_i \rightarrow C$ converging and that $B\sleq C_i$ for all $i$. Then $C_i - B \rightarrow C-B$ and for a given $v$ $f_v(C_i-B)\geq 0$. $f_v$ is continuous so it preserves limits, so that we then also have $f_v(C-B)\geq 0$ for any $v$ which proves that $B\sleq C$. The same holds for downsets of $B$. The Scott-continuousness of the trace with respect to the dual order then follows by noting that the trace is a continuous map.
\end{proof}

\btheor
The dual of the Löwner order is a domain.
\etheor
\begin{proof}
Since the Löwner order is downwards small, the dual is upwards small, so it suffices to find for each $B\in PO(n)$ an increasing sequence $A(t)$ such that $A(t)\rightarrow B$ and $A(t)\ll B$, and to prove this condition we only need to work with converging increasing sequences. 

Define $A(t) = tI_n + B$, then $A(t)-A(t^\prime) = (t-t^\prime)I_n$ so it is clear that $A(t)$ is increasing for $t\rightarrow 0$ and that $A(t)\rightarrow B$ as $t\rightarrow 0$. Note that $PO(n)$ is a subset of the Euclidean space $\mathbb{C}^{n^2}$, so it is a metric space. There are multiple equivalent metrics but for this proof it is easiest if we work with the sup-norm induced metric: 
\begin{equation*}
    d(A,B) = \sup_{v,w}\left\{\left\lvert \frac{v^\dagger (A-B) w}{\lVert v \rVert \lVert w \rVert} \right\rvert \right\}.
\end{equation*}
Let $C_i\rightarrow C$ be an increasing sequence such that $B\sleq^* C$, so $B-C\geq 0$. Because of upwards smallness the sequence is converging in the metric. In particular from some $i$ onward we have $d(C_i,C)\leq t$, which means that for all normalised $v$: $v^\dagger (C - C_i) v \geq -t$. Then we have for all normalised $v$:
\begin{align*}
    &v^\dagger(A(t)-C_i)v = v^\dagger(B-C_i) v + tv^\dagger I_n v = v^\dagger(B-C + C-C_i)v + t \\
    =& v^\dagger(B-C)v + v^\dagger(C-C_i)v + t \geq v^\dagger(B-C)v -t + t \geq 0.
\end{align*}
So $A(t)\sleq^* C_i$ and since this increasing sequence was arbitrary $A(t)\ll B$.
\end{proof}

While the dual of the Löwner order is a domain, the Löwner order isn't even directed complete. This is because any sequence $A_n = n\cdot A$, is increasing and obviously does not have any least upper bound. The restriction of $PO(n)$ to any compact subset is directed complete with the Löwner order.

The Löwner order might be seen as a good candidate for an information order on $DO(n)$, but unfortunately as the strict monotonicity of the trace demonstrates, when restricted to $DO(n)$ the Löwner order is trivial: $\rho \sleq \pi$ if and only if $\rho=\pi$. Fortunately there are some other candidates.

\section{Renormalising the Löwner order}
We revisit the two renormalised Löwner orders that we looked at in Chapter 2, but now we define them in the bigger context of the density operators. So first of all, some definitions.

\bdefin
    Let $\rho \in DO(n)$ with its set of eigenvalues in descending order (with degeneracies) $(\lambda_1,\ldots,\lambda_k,0,\ldots,0)$ with corresponding eigenvectors $v_i$, then $\rho^+ = 
    \lambda_1$ denotes the maximum eigenvalue of $\rho$. $\rho^- = \lambda_k$ denotes the smallest nonzero eigenvalue. Define $L^{\lambda_j}(\rho)= \{v\in \mathbb{C}^n \delim \rho v = \lambda_j v\}$ the linear span of the $j$th eigenvalue, and specifically $L^+(\rho) = L^{\rho^+}(\rho)$ and similarly for $\rho^-$. The \emph{kernel} of $\rho$ is given by ker$(\rho) = \{v\in\mathbb{C}^n \delim \rho v = 0\}$.
\edefin

We are now ready to define the renormalised Löwner orders.
\bdefin
    For $\rho, \pi \in DO(n)$ define
    \begin{equation*}
        \rho \sleq^+ \pi \iff \frac{\rho}{\rho^+} - \frac{\pi}{\pi^+} \geq 0 \iff \pi^+\rho - \rho^+ \pi \geq 0.
    \end{equation*}
    $\sleq^+$ is called the \emph{maximum eigenvalue order}.
\edefin

\blemma
    $\sleq^+$ has the following properties.
    \begin{itemize}
        \item $\sleq^+$ is closed and a dcpo.
        \item If $\rho \sleq^+ \pi$ then ker$(\rho)\subseteq$ ker$(\pi)$.
        \item If $\rho \sleq^+ \pi$ then $L^+(\pi)\subseteq L^+(\rho)$.
        \item $\sleq^+$ allows mixing.
        \item $\sleq^+$ is unitary conjugation invariant: for all $U\in U(n)$: $\rho \sleq^+ \pi$ iff $U\rho U^\dagger \sleq^+ U\pi U^\dagger$.
        \item The uniform distribution $\bot_n$ is the least element and the pure states are the maximal elements.
    \end{itemize}
\elemma
\begin{proof}
We can view $\sleq^+$ as being implemented by the map $F:DO(n)\rightarrow (PO(n),\sleq^*_L)$ where $F(\rho)=\rho/\rho^+$ in the sense of Theorem \ref{theor:inducedpo}: $F$ is continuous and injective and $\sleq^*_L$ is a closed partial order. Furthermore $DO(n)$ is a compact metric space, so $F$ induces a small partial order on $DO(n)$, which is $\sleq^+$. So $\sleq^+$ is a dcpo and closed.

Let $v\in\mathbb{C}^n$ be a normalised vector. If $\rho v = 0$, then $-\rho^+ v^\dagger \pi v \geq 0$ which is only the case when $\pi v = 0$. So indeed ker$(\rho)\subseteq $ ker$(\pi)$. Suppose $\pi v = \pi^+ v$. Then $\pi^+ v^\dagger \rho v - \rho^+ \pi^+ \geq 0$, so $v^\dagger \rho v \geq \rho^+$. The LHS is a convex combination of the eigenvalues of $\rho$ while the RHS is the highest eigenvalue of $\rho$, so this equality can only hold when we have equality: $\rho v = \rho^+ v$, so $L^+(\pi)\subseteq L^+(\rho)$.

Suppose $\rho \sleq^+ \pi$. Let $z(t) = (1-t)\rho + t \pi$. In general $z(t)^+ \leq (1-t)\rho^+ + t\pi^+$, but because of the above, there is a $v$ such that $\rho v = \rho^+$ and $\pi v = \pi^+$, so $z(t) v = (1-t)\rho^+ + t \pi^+$, which means that $z(t)^+ = (1-t)\rho^+ + t\pi^+$. Now we calculate $z(t)^+ \rho - \rho^+ z(t) = (1-t)\rho^+\rho + t\pi^+\rho - \rho^+(1-t)\rho + \rho^+t\pi = t(\pi^+ \rho - \rho^+ \pi) \geq 0$, so $\rho \sleq z(t)$ and we can use the same argument to get $z(t)\sleq \pi$.

For unitary conjugation invariance we note that $(U\rho U^\dagger)^+ = \rho^+$ since a basis transformation doesn't change the eigenvalues. We then note that
\begin{equation*}
    v^\dagger((U\pi U^\dagger)^+ U\rho U^\dagger - (U\rho U^\dagger)^+ U\pi U^\dagger) v = (U^\dagger v)^\dagger (\pi^+ \rho - \rho^+ \pi) (U^\dagger v).
\end{equation*}
Since $U$ is a bijection it maps the $v$ one to one, which means it preserves the positivity structure.

That the uniform distribution is minimal and the pure states are maximal can be checked directly.
\end{proof}

\bdefin
For $\rho,\pi \in DO(n)$ define $\rho \sleq^- \pi$ iff one of the following mutual exclusive options holds.
\begin{itemize}
    \item ker$(\rho) = $ ker$(\pi)$ and $\rho^-\pi - \pi^-\rho\geq 0$.
    \item ker$(\rho)\subset $ ker$(\pi)$ and $L^-(\rho)\cap $ ker$(\pi) \neq \{0\}$.
\end{itemize}
$\sleq^-$ is called the \emph{minimal eigenvalue order}.
\edefin
\blemma
    $\sleq^-$ has the following properties.
    \begin{itemize}
        \item $\sleq^-$ is a dcpo (but is not closed).
        \item If $\rho \sleq^+ \pi$ then ker$(\rho)\subseteq$ ker$(\pi)$.
        \item $\sleq^-$ allows mixing.
        \item $\sleq^-$ is unitary conjugation invariant.
        \item The uniform distribution $\bot_n$ is the least element and the pure states are the maximal elements.
    \end{itemize}
\elemma
\begin{proof}
That $\sleq^-$ is a dcpo (and not closed) is done in the same way as for $\sleq^-$ on $\Delta^n$. The clue is that for a given $\rho$ its upperset is a closed convex space if $L^-(\rho)$ is $1$-dimensional (so if its lowest nonzero eigenvalue is nondegenerated). Otherwise the upperset will be a finite union of closed convex spaces. We can then use Theorem \ref{theor:convex}.

Suppose $\rho \sleq^- \pi$. If ker$(\rho)=$ ker$(\pi)$, then just as in the proof for $\sleq^+$ we see that $L^-(\pi)\subseteq L^-(\rho)$ so that $z(t)^- = (1-t)\rho^- + t\pi^-$ where $z(t)=(1-t)\rho + t \pi$. Mixing then follows easily. If ker$(\rho)\subset $ ker$(\pi)$ then because $L^-(\rho)\cap $ ker$(\pi)\neq \{0\}$ there is a $v$ such that $\rho v = \rho^- v$ and $\pi v = 0$. Then $z(t)v = (1-t)\rho^- v = z(t)^- v$ and we proceed in the same way.

For unitary conjugation invariance we note that ker$(U\rho U^\dagger) =$ $U$ker$(\rho)$ and the same for $L^-(\rho)$. The invariance then follows easily.

That $\bot_n$ is the minimal element follows because $L^-(\bot_n)$ is equal to the entire space $\mathbb{C}^n$. Similarly let $p$ be a pure state and suppose $p\sleq^- \pi$, then either ker$(p)\subseteq$ ker$(\pi)$ in which case $\pi$ is also a pure state which means that $\pi=p$. So the pure states are indeed maximal.
\end{proof}

Note that the condition $L^+(\pi)\subseteq L^+(\rho)$ (or the same with $L^-$) is essentially a degeneracy condition of the highest (or lowest) eigenvalue as we say in the previous chapter, because it implies dim$(L^+(\pi))\leq $ dim$(L^+(\rho))$.

There is another property the partial orders share relating to conjugation invariance.
\blemma
    Let $(DO(n),\sleq)$ be a poset with a conjugation invariant closed $\sleq$. Then if $\rho \sleq U\rho U^\dagger$ we must have $\rho = U\rho U^\dagger$.
\elemma
\begin{proof}
    Denote $U(\rho) = U\rho U^\dagger$. Assume $\rho\sleq U(\rho)$. By conjugation invariance then also $U(\rho)\sleq U^2(\rho)$, and so on. So $(U^n(\rho))$ is an increasing sequence. Because $\sleq$ is closed it is a dcpo so that this sequence has a join, and moreover this sequence is convergent, but that is only possible if $U(\rho)=\rho$.
\end{proof}
Note that although $\sleq^-$ is not closed, it still has this property which can be checked directly. This property is the density operator analog of Lemma \ref{lemma:perm}, that for probability distributions states that $x\sleq \sigma(x)$ implies $x=\sigma(x)$.

\section{Composing systems}
The renormalised Löwner orders also share another property with the normal Löwner order: the ability to compose systems. We'll define what we mean by that, but first we need some preliminaries.

\bdefin
    Let $V$ be a $n$-dimensional vector space with basis $(v_i)_{i=1}^n$ and $W$ a $m$-dimensional vector space with basis $(w_j)_{j=1}^m$. We define their \emph{tensor product} $V\otimes W$ to be the $nm$-dimensional vector space with basis $(v_i\otimes w_j)$ with $1\leq i \leq n$ and $1\leq j \leq m$. An arbitrary vector in $V\otimes W$ can then be written as $\sum_{i,j}c_{i,j}v_i\otimes w_j$.
    Given linear operators $A:V\rightarrow V$ and $B:W\rightarrow W$ we define their tensor product $A\otimes B:V\otimes W \rightarrow V\otimes W$ as the unique linear map with $(A\otimes B)(v_i\otimes w_j) = (Av_i)\otimes (Bw_j)$.
\edefin.

\blemma
    Let $A$ and $B$ be Hermitian linear operators on some finite dimensional vector spaces $V$ and $W$.
    \begin{itemize}
        \item As a map $\otimes: V\times W \rightarrow V\otimes W$, $\otimes $ is bilinear.
        \item If $A v = \lambda v$ and $B w = \mu w$, then $(A\otimes B)(v \otimes w) = \lambda\mu (v\otimes w)$.
        \item Tr$(A\otimes B)$ = Tr$(A)$Tr$(B)$.
        \item $(A\otimes B)^+ = A^+B^+$.
        \item $(A\otimes B)^- = A^-B^-$.
        \item if $A$ and $B$ are positive/Hermitian, then $A\otimes B$ is also positive/Hermitian.
    \end{itemize}
\elemma

Note also that if we have two finite dimensional complex vector spaces $V$ and $W$ and dim$(V)=$ dim$(W)$, then for any choice of basis $(v_i)$ of $V$ and $(w_j)$ of $W$ we have an isomorphism $v_i \mapsto w_i$ so that $V\cong W$. Because of this we can identify $\mathbb{C}^n\otimes \mathbb{C}^m \cong \mathbb{C}^{nm}$ for a given choice of basis. In fact if we pick the standard basis of $\mathbb{C}^n$: $e_i = (0,\ldots,0,1,0,\ldots,0)$, then we can identify $\mathbb{C}^n\subseteq \mathbb{C}^m$ when $n\leq m$. In this case we can construct $\mathbb{C}^\infty = \bigcup_n \mathbb{C}^n$ and view the tensor product as $\otimes: \mathbb{C}^\infty \times \mathbb{C}^\infty \rightarrow \mathbb{C}^\infty$.

This sort of construction also works for $PO(n)$ and $DO(n)$: since the tensor product of $A\in PO(n)$ and $B\in PO(m)$ is again positive, we have $(A\otimes B)\in PO(nm)$ for some choice of basis, and if Tr$(A)=1$ and Tr$(B)=1$, then Tr$(A\otimes B) = $Tr$(A)$Tr$(B) = 1$, so if $A\in DO(n)$ and $B\in DO(m)$ then $(A\otimes B)\in DO(nm)$ for some choice of basis.

Now why is this useful? A finite quantum state is decribed by a density matrix. For instance, the state of $n$ qbits is described by a density matrix in $DO(2^n)$. Now suppose we have two quantum systems that don't interact with each other with the first one described by the state $\rho_1\in DO(n)$ and the second one described by $\rho_2\in DO(m)$. The fact that the systems don't interact precisely means that we can describe the composite system as the tensor product of the states: $\rho_1\otimes \rho_2$. If they had some sort of interaction then the composite system is more complex than the sum of its parts and we wouldn't be able to describe it as a pure tensor of the individual states. 

Taking this idea a bit further. Suppose we had some measure of information content on the density operators that then also describes the information content in the particular quantum states. So supposing that we have quantum systems 1 and 2 that are either described by states $\rho_i$ or states $\pi_i$ for $i=1,2$ and $\rho_i \sleq \pi_i$, so that the states $\rho_i$ contain less information than the states $\pi_i$.  If we suppose that system 1 and 2 are noninteracting then we can describe the composite system as either $\rho_1\otimes \rho_2$ or $\pi_1\otimes \pi_2$. Now of course since the $\rho_i$ contain less information then the $\pi_i$ we would assume that composing these noninteracting systems together would not change the information content. In fact we might assume that \emph{information in composite system = information in system 1 + information in system 2}, so this information order should also have $\rho_1\otimes \rho_2 \sleq \pi_1 \otimes \pi_2$.

\bdefin
    If we have a family of posets $(DO(n),\sleq^n)$ we'll say this family allows \emph{composing} when $\rho_i\sleq \pi_i$ for $i=1,2$ implies $\rho_1\otimes \rho_2 \sleq \pi_1 \otimes \pi_2$.
\edefin

\bdefin
    Let Let $F^n:DO(n)\rightarrow PO(n)$ for $n\geq 1$ be a family of maps. We say that the $F_n$ are \emph{compatible} when for any $n\leq m$, $F^n = F^m_{\lvert DO(n)}$. So if $\rho \in DO(n)$ then $F^n(\rho) = F^m(\rho)$. We say that this family \emph{splits under tensor products} when for $\rho_1 \in DO(n)$ and $\rho_2 \in DO(m)$ we have $F^{nm}(\rho_1\otimes \rho_2) = F^n(\rho_1) \otimes F^m(\rho_2)$. Instead of $DO(n)$, we can also take $PO(n)$ to be the domain in this definition.
\edefin

Examples of a compatible family of maps that splits under tensor products is given by the set of identities $id^n: PO(n)\rightarrow PO(n)$, or $F^n: DO(n) \rightarrow PO(n)$ where $F^n(\rho) = \rho/\rho^+$ or similarly $F^n(\rho)=\rho/\rho^-$.

\btheor
    Let $F^n:DO(n)\rightarrow PO(n)$ be a compatible family that splits under tensor products. For $\rho,\pi \in DO(n)$ define $\rho \sleq_n \pi$ iff $F^n(\pi) - F^n(\rho) \geq 0$. Then this family of partial orders allows composing.
\etheor
\begin{proof}
Let $\rho_1,\pi_1 \in DO(n)$ and $\rho_2,\pi_2 \in DO(m)$ and $\rho_1 \sleq_n \pi_1$ and $\rho_2 \sleq_m \pi_2$ then we need to show that $\rho_1 \otimes \rho_2 \sleq_{nm} \pi_1 \otimes \pi_2$.

\centerline{\begin{minipage}[c]{\textwidth}
\begin{align*}
    &F^{nm}(\pi_1\otimes \pi_2) - F^{nm}(\rho_1\otimes \rho_2) \\
    =& F^n(\pi_1)\otimes F^m(\pi_2) - F^n(\rho_1)\otimes F^m(\rho_2) \\
    =&  F^n(\pi_1)\otimes F^m(\pi_2) - F^n(\rho_1)\otimes F^m(\pi_2) + F^n(\rho_1)\otimes F^m(\pi_2) - F^n(\rho_1)\otimes F^m(\rho_2) \\
    =& (F^n(\pi_1)- F^n(\rho_1))\otimes F^m(\pi_2) + F^n(\rho_1)\otimes (F^m(\pi_2) - F^m(\rho_2)) \\
    \geq& 0
\end{align*} 
\vspace{\baselineskip}
\end{minipage}}
\noindent because by assumption $F^n(\pi_1)-F^n(\rho_1)\geq 0$ and the same for $\pi_2$ and $\rho_2$, so indeed $\rho_1 \otimes \rho_2 \sleq_{nm} \pi_1 \otimes \pi_2$.
\end{proof}

This proves that the Löwner order and the maximum eigenvalue order allow composing. For the minimal eigenvalue order we need to some extra work. Supposing that $\rho_i\sleq^- \pi_i$, if ker$(\rho_i) = $ ker$(\pi_i)$ for $i=1,2$ we can use the theorem above. If however ker$(\rho_1)\subset $ ker$(\pi_1)$, so that $L^-(\rho_1)\cap $ ker$(\pi_1)\neq \{0\}$. Let $v$ be such that $\rho_1 v = \rho_1^- v$ and $\pi_1 v = 0$. We have
\begin{equation*}
    \text{ker}(\rho_1\otimes \rho_2) = \text{ker}(\rho_1)\otimes \text{ker}(\rho_2)\subset \text{ker}(\pi_1)\otimes \text{ker}(\pi_2) =  \text{ker}(\pi_1\otimes \pi_2),
\end{equation*} 
and for any $w\in L^-(\rho_2)$ we have $(\rho_1\otimes \rho_2)(v\otimes w) = \rho_1^-\rho_2^- (v\otimes w) = (\rho_1\otimes \rho_2)^- (v\otimes w)$ and $(\pi_1 \otimes \pi_2)(v\otimes w) = (\pi_1 v)\otimes (\pi_2 w) = 0 \otimes (pi_2 w) = 0$. So indeed $\rho_1\otimes \rho_2 \sleq^- \pi_1\otimes \pi_2$.

We also have the following.
\blemma
    Let $\sleq^n$ be a family of partial orders on $DO(n)$ induced by a family of compatible maps that split under tensors products. Let $\kappa \in DO(m)$ be such that $F^m(\kappa)\neq 0$. The \emph{right tensor map} $R_\kappa: DO(n) \rightarrow DO(nm)$ defined as $R_\kappa(\rho) = \rho \otimes \kappa$ is a strict monotone continuous map and $DO(n)$ is order isomorphic to $R_\kappa(DO(n))$. The same holds for the left tensor map $L_\kappa$ given by $L_\kappa(\rho) = \kappa \otimes \rho$. This also works when working with $PO(n)$ instead of $DO(n)$.
\elemma
\begin{proof}
Monotonicity follows by the previous theorem. Strictness follows from injectivity of $R_\kappa$. Continuity follows from the linearity of the tensor product. To prove that this map is an order isomorphy we only have to show that if $R_\kappa(\rho)\sleq R_\kappa(\pi)$ then $\rho \sleq \pi$. If $R_\kappa(\rho)\sleq R_\kappa(\pi)$ then $F(\pi\otimes \kappa) - F(\rho\otimes \kappa) = F(\pi)\otimes F(\kappa) - F(\rho)\otimes F(\kappa) = (F(\pi)-F(\rho))\otimes F(\kappa) \geq 0$. Since $F(\kappa)\neq 0$, there is a $w$ such that $F(\kappa)w \neq 0$. Pick this $w$, then we must have, for all $v$: 
\begin{equation*}
    (v\otimes w)^\dagger (F(\pi)-F(\rho))\otimes F(\kappa) (v\otimes w) = v^\dagger (F(\pi)-F(\rho)) v \cdot w^\dagger F(\kappa) w \geq 0.
\end{equation*}
In particular for all $v$: $v^\dagger (F(\pi)-F(\rho)) v \geq 0$, so indeed $\rho \sleq \pi$.
\end{proof}

Although $\sleq^-$ is not completely described by a family of maps spliiting under tensor producs you can still check that this property holds for $\sleq^-$ for any $\kappa \in DO(m)$.

This property has an intuitive explanation: $\kappa$ can be seen as a sort of `external environment' that does not interact with the system we are interested in. Adding such an environment to the description of your system should not change any of the fundamental properties of the system. In this case this translates to $DO(n)$ being order isomorphic to $DO(n)\otimes \kappa$: the relative information content of states in $DO(n)$ isn't in any way affected by adding an external state.

It should be noted that any of the restricted information orders on $\Delta^n$ extended to $DO(n)$ does not in any way allow composing of systems in a natural way. The reason for this is that elements in $\Delta^n$ are comparable when they belong to the same sector, for instance $\Lambda^n$. The problem is that there is no obvious way in which the tensor product of $x_1\in \Lambda^n$ and $x_2 \in \Lambda^m$ is an element of $\Lambda^{nm}$. We'd have to choose some kind of basis that maps the coordinates of $x_1\otimes x_2$ in a descending order so that it belongs to $\Lambda^{nm}$, but this mapping would depend on the $x_1$ and $x_2$ you started with. The only properties that are preserved by this tensor product are the highest values of $x_1$ and $x_2$ and their lowest values, which is why the renormalised Löwner orders \emph{do} allow composing.

So far we have two partial orders on $DO(n)$, the renormalised Löwner orders, that seem to behave really nicely: they are directed complete, they have mixing, they allow composing, they both have reasonable behaviour on kernels, the least element is the uniform distribution and the maximal elements are the pure states. The maximum eigenvalue order however seems to work better then the minimal eigenvalue order. We have already seen that the minimal eigenvalue order is not closed, while the maximum eigenvalue is. There is however another way in which they are different. We have already seen that $\sleq^-$ when restricted to $\Delta^n$ is not a domain so it is also not a domain on $DO(n)$. However:

\btheor
    $(DO(n),\sleq^+)$ is a domain:
    \begin{itemize} 
    \item For all $\pi \in DO(n)$ and $0\leq t <1$: $(1-t)\bot_n + t\pi \ll \pi$.
    \item If $\rho \ll \pi$ then ker$(\rho) = \{0\}$.
    \item If $\rho \ll \pi$ then for $0<t<1$: $\rho \ll (1-t)\rho + t\pi \ll \pi$.
    \end{itemize}
\etheor
\begin{proof}
    This proof is an adaption of the proof in Theorem \ref{theor:domain} and is also similar to the proof that the dual of the Löwner order is a domain. $\sleq^+$ is upwards small so we only have to work with converging increasing sequences and it suffices to find for any $\pi\in DO(n)$ an increasing sequence $(\rho(t))$ such that $\rho(t)\ll \pi$ and $\rho(t)\rightarrow \pi$.
    
    Let $a_i\rightarrow a$ be an increasing sequence in $DO(n)$ with limit/join $a$. We have $L^+(a)\subseteq L^+(a_j)\subseteq L^+(a_i)$ for all $i\leq j$. From some $N$ we must have equality for all $i\geq N$: suppose this is not the case, then there is a normalised $v$ such that $a_i v = a_i^+ v$ for all $i$ while $a v \neq a^+ v$. Then $\lVert a_i v \rVert = a_i^+$. Since we have $a_i\rightarrow a$ we also have $a_i^+ \rightarrow a^+$, so $\lVert a_i v \rVert \rightarrow a^+$, but $\lVert a v \rVert \leq a^+ - \delta$ for some $\delta$. $(a_i)$ converges in the matrix norm so this is a contradiction. We will now assume that $L^+(a)=L^+(a_j)$, otherwise we could just take the tail of the sequence where this is the case.
    
    Let $\rho(t) = (1-t)\bot_n + t\pi$. Assume $\pi \sleq^+ a$. Then $L^+(a_j)=L^+(a) \subseteq L^+(\pi)$. The goal is to find a $i$ such that $\rho(t)\sleq^+ a_i$. This is the case when for all normalised $v$:
    \begin{align*}
        &v^\dagger(a_i^+\rho(t) - \rho(t)^+ a_i)v \\
        =&~tv^\dagger(a_i^+\pi - \pi^+ a_i)v + (1-t)v^\dagger(a_i^+ \bot_n - \bot_n^+a_i)v \\
        =&~tv^\dagger[(a_i^+-a^+)\pi - \pi^+(a_i-a)]v + tv^\dagger(a^+\pi - \pi^+ a)v + \frac{1-t}{n}(a_i^+-v^\dagger a_i v) \\
        \geq&~0.
    \end{align*}
    Suppose $a_i v = a_i^+ v$, then $\pi v = \pi^+ v$ as well, which implies $\rho(t) v = \rho(t)^+ v$ so that the expression above is equal to zero. We can therefore restrict to the $v$'s where $a_i v \neq a_i^+ v$. Then the last 2 terms in the expression above are strictly positive, say bigger than some $\epsilon>0$. Now since $a_i \rightarrow a$ in the sup-norm, we can find an $N$ such that for all $i\geq N$ $\lvert a_i^+- a^+\rvert \leq \frac{\epsilon}{2t\pi^+}$ and also $\lvert v^\dagger(a_i-a)v\rvert\leq \frac{\epsilon}{2t\pi^+}$ for all $v$. We can then write
    \begin{equation*}
        \left \lvert tv^\dagger[(a_i^+-a^+)\pi - \pi^+(a_i-a)]v \right \rvert \leq 2t\pi^+ \frac{\epsilon}{2t\pi^+} = \epsilon.
    \end{equation*}
    This means that $(a_i^+\rho(t)-\rho(t)^+a_i)v \geq 0$ for all $v$, so indeed $\rho(t)\sleq^+ a_i$. We therefore have $\rho(t)\ll \pi$ which proves that $(DO(n),\sleq^+)$ is indeed a domain.
    
    Now suppose $\rho \ll \pi$. Then for some $0\leq t<1$ we have $\rho \sleq \rho(t)$. We then have ker$(\rho)\subseteq $ ker$(\rho(t)) = \{0\}$.
    
    Define $z(t) = (1-t)\rho + t\pi$ for $0<t<1$. In the same way as above we write
    \begin{align*}
        &v^\dagger(a_j^+z(t) - z(t)^+a_j)v \\
        =&~ (1-t)v^\dagger(a_j^\dagger \rho - \rho^+a_j)v + tv^\dagger(a_j^+\pi - \pi^+ a_j)v.
    \end{align*}
    Just as before we can ignore the $v$ where $a_j v = a_j^+$. Since $\rho \ll \pi$ we can pick $j$ high enough such that the first term is then strictly positive. The second term goes to zero just as we saw earlier in this proof. So at some point this expression is positive. So indeed $z(t)\ll \pi$. That $\rho \ll z(t)$ works similarly.
\end{proof}

\section{Unicity of the maximum eigenvalue order}
We'll now try to establish some ways in which the maximum eigenvalue order is unique. Unique in what way?

\bdefin
    Let $(DO(n),\sleq^n)$ be a family of posets. We'll call the sequence of partial orders $\sleq = (\sleq^n)$ a \emph{quantum information order} iff
    \begin{itemize}
        \item The uniform distribution $\bot_n$ is the minimal element in $DO(n)$.
        \item The pure states are the maximal elements in each $DO(n)$.
        \item The partial orders are compatible: if we view $DO(n)\subseteq DO(m)$ for $m\geq n$, then for $\rho,\pi \in DO(n)$ we have $\rho\sleq^n \pi$ iff $\rho\sleq^m \pi$.
        \item The partial orders are unitary invariant: For all $U\in U(n)$: $\rho \sleq^n \pi$ iff $U\rho U^\dagger \sleq^n U \pi U^\dagger$.
        \item The partial orders allow mixing: for $\rho,\pi \in DO(n)$, if $\rho \sleq^n \pi$ then $\rho \sleq^n (1-t)\rho + t \pi \sleq^n \pi$.
        \item The partial orders allow composing: for $\rho_i,\pi_i \in DO(n_i)$ with $\rho_i\sleq^{n_i} \pi_i$ for $i=1,2$ then $\rho_1\otimes \rho_2 \sleq^{n_1n_2} \pi_1\otimes \pi_2$.
        \item Items increase in specificity: If $\rho \sleq^n \pi$ then ker$(\rho)\subseteq$ ker$(\pi)$.
    \end{itemize}
\edefin

Note that these properties are not all independent (for instance, the increase in specificity ensures that the pure states are maximal). The identification of $DO(n)\subseteq DO(m)$ is done in a chosen basis: if we have $\rho\in DO(n)$ then we can add extra columns are rows o the matrix representation of $\rho$ that are filled with zeroes. Because of the unitary invariance, it doesn't matter in which basis we do this, the resulting order structure is the same. The last property in this definition ensures that for $\rho \in DO(n)$ $\uparrow \rho \subseteq DO(n)$. This definition can also be used for an order structure on $PO(n)$, in which case we require that the maximal element is the zero element, and that there are no minimal elements.

We have already seen that both $\sleq^+$ and $\sleq^-$ satisfy these conditions (and the dual Löwner order satisfies the definition on $PO(n)$), so obviously these properties do not define a unique order on $DO(n)$. We \emph{hypothesize} (not prove) the following:
\begin{quote}
    \textsc{Hypothesis}: The only quantum information order that is also a domain is the maximum eigenvalue order.
\end{quote}

It might be that the question we should be trying to answer is that the maximum eigenvalue order is the only closed quantum information order, or even the only closed order which is also a domain, or it might be that all these coincide.

We will show a specific construction that uniquely produces the maximum eigenvalue order, but so far there is no proof that this is the only possible construction.

Specifically, we will consider partial orders on $DO(n)$ that are induced by a continuous map $F:DO(n)\rightarrow (PO(n),\sleq_L^*)$. In fact, we will consider a compatible family of continuous injective maps that split under tensor products given by $F^n:DO(n)\rightarrow PO(n)$. It should be clear that the family of partial orders induced by such a family of maps is closed (so they are dcpo's), compatible and allow composing. Note that this construction is similar to how we constructed partial orders on $\Lambda^n$. In that case we looked at an embedding $F:\Lambda^n \rightarrow \mathbb{R}^n$, which can be seen as the diagonal restriction of a map $F:DO(n)\rightarrow PO(n)$.

Pick a given $n$ and for brevity denote $F^n=F$. Note that if $F(\pi)=0$, then $\pi$ will be the unique maximal element of $DO(n)$, which can't be the case. So $F(\pi)\neq 0$ for all $\pi\in DO(n)$. Define $f: DO(n) \rightarrow \mathbb{R}_{>0}$ as $f(\pi) = 1/$Tr$(F(\pi))$ and define $K:DO(n)\rightarrow DO(n)$ as $K(\pi) = F(\pi)/$Tr$(\pi)$. Then we can write $F(\pi)=K(\pi)/f(\pi)$ where $K$ is some self map of $DO(n)$ and $f$ is a map from $DO(n)$ to the positive reals. Note that since Tr$(A\otimes B) = $ Tr$(A)$Tr$(B)$, both $K$ and $f$ split under tensor products. Both $K$ and $f$ are also continuous. Furthermore $f$ is a strict monotonic map to $\mathbb{R}_{>0}$ with the regular ordering. Note that although $F$ is injective, $K$ doesn't have to be.

A partial order is not uniquely defined by a $F$:  if we were to scale $F$ by a positive constant, this would not change the induced order so there is some gauge freedom in choosing $F$. What kind of freedom do we have? Suppose we have some linear matrix valued map $H:M_n\rightarrow M_n$ and that we transform $F$ to $H\circ F$. Then we should have $\rho \sleq_F \pi$ iff $\rho \sleq_{H\circ F} \pi$ iff $H\circ F(\rho) - H\circ F(\pi) = H(F(\rho)-F(\pi))\geq 0$ iff $F(\rho) - F(\pi)\geq 0$. $H$ should be injective to preserve antisymmetry (and so is a bijection), and it should be clear that this holds only if $H$ maps all positive operators to positive operators and doesn't map any nonpositive operator to a positive operator. That is: $H$ is a positive map and its inverse is positive as well. It turns out these maps have been classified \cite{cariello2012}: $H$ is either $H(X) = AXA^\dagger$ for all $X$ or $H(X) = AX^TA^\dagger$ where $A$ is some invertible matrix and $X^T$ denotes the transpose of $X$. So in particular, we can transform $F$ by rescaling, taking the transpose or conjugate it with a unitary operator and it wouldn't change the partial order.

Note that the restriction of $\sleq_F$ to the diagonal operators gives an information order on $\Delta^n$. As we saw in the previous chapter, such a partial order has to be defined by affine maps. So we will take $K$ to be an affine map. This makes additional sense as the primary structure of $DO(n)$ is that of a convex space, so the affine maps are the structure preserving maps. We'll now proceed with a short proof of a well known fact.

\blemma
    There is a one-to-one correspondence between affine maps $K: DO(n)\rightarrow DO(n)$ and linear positive trace preserving maps $K_m: M_n\rightarrow M_n$, where $M_n$ denotes the space of all complex valued $n\times n$ matrices.
\elemma
\begin{proof}
The restriction of a linear positive trace preserving map to $DO(n)$ is clearly an affine self map. We'll look at the other direction.

Start with an affine map $K:DO(n)\rightarrow DO(n)$. We extend this to a map on the positive operators $K_p:PO(n)\rightarrow PO(n)$ by setting $K_p(0)=0$ and $K_p(A) = $Tr$(A)K(\frac{A}{\text{Tr}(A)})$ for $A\neq 0$. It should be clear that then $K_p(rA) = r K_p(A)$ for all $r\geq 0$ and by using the affineness of $K$ we can prove $K_p(A+B) = k_p(A)+K_p(B)$.

Now we extend to Hermitian matrices. For any Hermitian $A$ we can uniquely write $A = A^+ - A^-$ where $A^+$ and $A^-$ are positive operators. We then define $K_h(A) = K_p(A^+) - K_p(A^-)$. That $K_h$ is linear is routine to check.

And finally to extend to all matrices we note that any matrix $A$ can be written as the sum of a Hermitian matrix and an anti-Hermitian matrix: $A = A^h + iA^a$ where $A^h$ and $A^a$ are both Hermitian. Again we define $K_m(A) = K_h(A^h) + i K_h(A^a)$.

The resulting map $K_m:M_n\rightarrow M_n$ is linear and sends positive matrices to positive matrices. It is furthermore easy to check per step that it preserves the trace. We also have that $K$ is injective iff $K_m$ is injective, and if $K$ is surjective then $K_m$ is surjective. If $K$ splits under tensor products then $K_m$ does so as well.
\end{proof}

\subsection{The rank of operators}
The rank of an operator is the dimension of the subspace covered by its image: rnk$(A) = $dim(Im$(A))$. For a hermitian operator it is the sum of the amount of nonzero eigenvalues (counting multiplicities). So for instance rnk$(\bot_n)=n$, and for a pure state $p$ rnk$(p) = 1$. Indeed another way to define pure states is as the rank 1 projections. The rank, just like the trace, distributes over tensor products: rnk$(A\otimes B) =  $rnk$(A)$rnk$(B)$.

Choose an orthonormal basis $(e_i)$ and fix it for all $DO(n)$. Let $P_1$ be the projection to the first basis element $e_1$. So $P_1e_1 = e_1$ and $P_1e_j = 0$ for all $j\geq 2$. Then we can view $P_1\in DO(n)$ for \emph{all} $n$ since the matrix representation of $P_1$ is just a $1$ in the left upper corner followed by zeroes everywhere else. The tensor product $P_1\otimes P_1$ is represented in the same way (although in a bigger space), so we can identify $P_1\otimes P_1 = P_1$. Recalling that $F$ splits under tensors we get $F(P_1) = F(P_1\otimes P_1) = F(P_1)\otimes F(P_1)$. Calculating the rank on both sides gives rnk$(F(P_1)) = $ rnk$(F(P_1))^2$, so rnk$(F(P_1)) = 1$. This is not a completely rigorous argument since we are abusing notation here, but it should make intuitive sense. If the rank of $F(P_1)$ would be bigger than $1$, then taking the repeated tensor product $F(\otimes^n P_1) = \otimes^n F(P_1)$ would result in an operator with an arbitrarily high rank, while the underlying operator $\otimes^n P_1$ would still have rank 1.

The basis we chose was arbitrary, so $P_1$ was also arbitrary. Any other rank $1$ projection can be found by taking a unitary conjugation of $P_1$ which corresponds to a basis change. If we want the partial order to be unitary conjugation invariant then it makes sense to require that all rank $1$ projections preserve their rank. So we'll assume that if rnk$(P) = 1$ then rnk$(F(P)) = 1$.

Any rank $1$ operator $R$ can be written as $R = cP$ where $c$ is some nonzero complex number and $P$ is a rank $1$ projection. For $K$ an affine map on $DO(n)$ we then have $K_m(R) = K_m(cP) = c K(P)$, so if $K$ preserves rank $1$ operators, then $K_m$ does so as well. Since $F = \frac{K}{f}$ preserves rank 1 operators, $K$ does so as well, so we are left with a linear rank $1$ preserving operator $K_m$ on $M_n$. It just so happens that \cite{marcus1959} has classified these operators:

\blemma
Let  $T:M_n\rightarrow M_n$ be a linear map preserving the rank of all rank 1 operators, then there exist invertible matrices $A,B\in M_n$ such that for all $M\in M_n$, $T(M) = AMB$ or $T(M) = AM^TB$.
\elemma

As we've already seen, we have gauge freedom to apply transpose to $F$, so we can assume that we are dealing with the first case. This means we have $K_m(M) = AMB$. Noting that $K_m$ is trace preserving we get Tr$(M) = $ Tr$(K_m(M))=$ Tr$(AMB)$ = Tr$(BAM)$ for all $M$. This is only possible if $BA = I_n$, so $B = A^{-1}$. $K_m$ is a positive linear map. In \cite{cariello2012} they showed that for any such map we have $K_m(M)\dagger = K_m(M^\dagger)$. This is now only possible when $A^{-1} = A^\dagger$, so $A$ is a unitary matrix. We then have $K_m(M) = UMU^\dagger$ for some $U\in U(n)$, and by using our gauge freedom we can simply set $U=I_n$. This means we can take $K$ to be the identity.

\subsection{Alternative derivations}
\label{sec:altder}
Since the argument from the rank 1 operators is not completely rigorous we offer some other assumptions on $K$ that would produce the same result.

Once we get $K_m(M) = AMB$ we can simplify $K_m$ to be the identity. There are other ways to prove $K_m$ must have this form. \cite{marcus1959} lists a few possibilities:
If a linear map $T:M_n \rightarrow M_n$ preserves one of the following:
\begin{itemize}
    \item the determinant of every matrix or
    \item the rank of all rank $n$ matrices or
    \item the rank of rank 2 matrices
\end{itemize}
then it must be of the form $T(M) = AMB$ or $T(M) = AM^TB$. In \cite{cariello2012} they show that a positive linear bijection $T:M_n \rightarrow M_n$ with positive inverse has to be of the form $T(M) = AMA^\dagger$ or $T(M) = AM^TA^\dagger$. In particular, if we assume that $K:DO(n)\rightarrow DO(n)$ is surjective and affine, then the inverse of the extension $K_m$ is positive, and this holds. In fact, because $K$ is affine, it is enough to require surjectivity to the pure states.

There are also results for bijective (not necessarily affine) maps $K: DO(n) \rightarrow DO(n)$.
If $K$ preserves one of the following:
\begin{itemize}
    \item the fidelity between all matrices,
    \item the transition probabilities, 
    \item the Bures metric,
    \item the trace distance metric,
    \item the relative entropy between all matrices, or
    \item the Jensen-Shannon divergence between all matrices
\end{itemize}
then $K$ has to be of the form $K(A) = UAU^\dagger$ where $U$ is a unitary or anti-unitary operator \cite{molnar2001,molnar2002,molnar2008a,molnar2008b}.

We will not use any of these results, but it should be clear that if $K$ is to preserve any kind of structure on $DO(n)$ it has to be a very \emph{simple} map.

\subsection{Last steps}
Now that we've established (or at least made it very plausible) that we can set $K(\rho) = \rho$ for all $\rho\in DO(n)$ we are left with $F(\rho) = \rho/f(\rho)$. The partial order is then given by
\begin{equation*}
    \rho \sleq \pi \iff \frac{\rho}{f(\rho)} - \frac{\pi}{f(\pi)}  \geq 0 \iff \forall v: v^\dagger (f(\pi)\rho - f(\rho)\pi)v \geq 0.
\end{equation*}
So we simply modify the Löwner order by dividing all the elements by a scalar.

We can reuse the argument about rank $1$ operators here. We know that $f$ splits over tensors, so let $P_1$ denote the projection to the first basis element, so that we can identify $P_1\otimes p_1 = P_1$.Then we should also have $f(P_1) = f(P_1\otimes P_1) = f(P_1)^2$, so $f(P_1)=1$. Again, the choice of basis is arbitrary, so set $f(P) = 1$ for all rank 1 projections $P$.

The rank $1$ projections are the maximal elements. Suppose we have $\rho \sleq P$ for $\rho \in DO(n)$ and $P$ some rank 1 projection. Let $v$ be the unique normalised vector such that $P v = v$. Then we must have 
\begin{equation*}
    v^\dagger(f(P)\rho - f(\rho)P)v = v^\dagger \rho v - f(\rho) \geq 0,
\end{equation*}
or slightly rewritten: $v^\dagger \rho v \geq f(\rho)$. The rank 1 projections are the only maximal elements, so there must be a projection where this inequality holds. the LHS is a convex sum of $\rho$'s eigenvalues. In particular, it is smaller than $\rho^+$. So we must at least have $f(\rho)\leq \rho^+$.

\blemma
    $f$ is unitary conjugation invariant: $f(U\rho U^\dagger) = f(\rho)$ for all $U\in U(n)$ and $\rho\in DO(n)$.
\elemma
\begin{proof}
The only maximal elements are the rank 1 projections, so for each $\rho\in DO(n)$ there must exist a projection $P$ such that $\rho \sleq P$. Because of unitary conjugation invariance, we must then for all $U\in U(n)$ have
\begin{align*}
    U\rho U^\dagger \sleq UP U^\dagger \iff& U\left(f(UPU^\dagger)\rho - f(U\rho U^\dagger)P\right) U^\dagger \geq 0 \\
    \iff& \rho - f(U\rho U^\dagger) P \geq 0.
\end{align*} 
So we must have $\rho - f(\rho) P \geq 0$ if and only if $\rho - f(U\rho U^\dagger) P \geq 0$. This is only possible when $f(\rho) = f(U\rho U^\dagger)$ for all $U\in U(n)$ and $\rho \in DO(n)$. 
\end{proof}

This has important consequences: any $\rho \in DO(n)$ can be written as $\rho = UDU^\dagger$ where $D\in \Lambda^n$ is the diagonal matrix with on the diagonal the eigenvalues of $\rho$ in decreasing order. $f(\rho) = f(UDU^\dagger) = f(D)$, so $f$ only depends on the \emph{ordered} eigenvalues of an operator.

As was discussed earlier in this chapter, the space $\Lambda^n$ doesn't behave well under tensor products: there is not a single choice of basis such that when $x\in \Lambda^n$ and $y\in \Lambda^m$ then $x\otimes y \in \Lambda^{nm}$ since in general it is not possible to preserve the order of the eigenvalues. In fact, there are 3 preserved quantities when tensoring: the highest eigenvalue, the lowest eigenvalue, and the lowest nonzero eigenvalue. Since $f$ splits under tensor products, it can only depend on these quantities. The lowest nonzero eigenvalue is noncontinuous, and since $f$ is continuous it can't depend on that. That leaves two variables $f$ can depend on: $f(\rho) = h(\rho^+,\rho^-)$ where in this case $\rho^-$ denotes the \emph{lowest} eigenvalue, not the lowest \emph{nonzero} eigenvalue. We then have $f(\rho_1\otimes \rho_2) = h(\rho_1^+\rho_2^+,\rho_1^-\rho_2^-) = f(\rho_1)f(\rho_2) = h(\rho_1^+,\rho_1^-)h(\rho_2^+,\rho_2^-)$. So $h$ is a function that satisfies $h(x_1x_2,y_1y_2)=h(x_1,y_1)h(x_2,y_2)$. We furthermore know that $f(P_1)=1$, and $P_1^+ = 1$, $P_1^- = 0$, so $f(\rho\otimes P_1)=f(\rho)\cdot 1 = h(\rho^+\cdot 1,\rho^-\cdot 0) = h(\rho^+, 0) = h(\rho^+,\rho^-)\cdot h(1,0) = h(\rho^+, \rho^-)$, so we have $h(x,y) = h(x,0)$ which means that $h$ is only a function of its first argument: $h(x,y) = h(x)$. Then $h:[0,1]\rightarrow \mathbb{R}_{>0}$ with $h(xy)=h(x)h(y)$. In other words: $h$ is a group homomorphism from a subgroup of $(\mathbb{R}_{>0},\cdot)$ to itself. These have all been classified: such a group homomorphism is always of the form $h(x) = x^r$ for some $r \in \mathbb{R}$. Since $f(\rho) = h(\rho^+)$ is a strict monotone function we must have $r > 0$. We have already seen that $f(\rho)\leq \rho^+$, which is only the case when $r \geq 1$.

\blemma
    The partial order induced by $f(\rho) = (\rho^+)^r$ for $r\geq 1$ only allows mixing when $r=1$.
\elemma
\begin{proof}
As we've seen in Chapter 2, an information order follows from an affine map, so mixing only holds when $r=1$, but in this case we can show this explicitly. Suppose we have the partial order given by $f(\rho) = (\rho^+)^r$ with $r>1$. Let $\rho$ be a diagonal matrix, and let $T_i$ be the diagonal matrix $(0,\ldots,0,1,0,\ldots,0)$ with the 1 at the $i$th position. We have $\rho \sleq T_i$ iff $(\rho^+)^r (T_i)_j \leq (T_i^+)^r\rho_j$ for all the diagonal components indexed by $j$. The LHS is zero except for $j=i$. So we have $\rho \sleq T_i$ iff $(\rho^+)^r \leq \rho_i$. Now, since $r$ is strictly bigger then 1, there exist $\rho$ such that $\rho\sleq T_i$ with $\rho_i \neq \rho^+ = \rho_k$ where $i\neq k$. Pick such a $\rho$. Let $z(t) = t\rho + (1-t)T_i$. The mixing condition then says that $\rho\sleq z(t)$ for all $t$. Note that $z(t)_i = t\rho_i + (1-t)$ and $z(t)_k = t\rho^+$. For some $0<t<1$ we will have $z(t)_i = z(t)_k = z(t)^+$. Since we have $\rho \sleq z(t)$ we must have $(\rho^+)^rz(t)_k\leq (z(t)^+)^r\rho_k$. Filling in the values of $z(t)_k$ we get $t(\rho^+)^{r+1}\leq t^r(\rho^+)^{r+1}$ so that $t\leq t^r$, which is never the case for any $0<t<1$ and $r>1$. We conclude that for $r>1$ there exists a diagonal $\rho$ such that $\rho \sleq T_i$ while there is a $0<t<1$ such that $\rho \sleq z(t)$ does \emph{not} hold. So the partial order doesn't satisfy the mixing condition. That the partial order does satisfy the mixing requirement for $r=1$ has already been shown.
\end{proof}

The proof of the unicity of the maximum eigenvalue order above is far from rigorous at some points. We will list some of the problems in the proof.

\begin{itemize}
    \item Why would a general quantum information order on $DO(n)$ have to be induced by a map $F: DO(n)\rightarrow PO(n)$?
    \item Can the affineness of $K$ be derived explicitly instead of implicitly?  Can it perhaps be shown that it has to satisfy some other property mentioned in Section \ref{sec:altder}?
    \item Can the argument using the rank $1$ operators be made more rigorous?
    \item Can the argument about $f$ only depending on ``quantities preserved by the tensor product'' be made more rigorous?
\end{itemize}

The first point might be solved by applying something similar to the Urysohn-Carruth Metrization Theorem, but that instead of the partial order being induced by a map to $[0,1]^\infty$, it will be induced by a map to $PO(n)$.

For the third point it might be possible to use the fact that the family of $K^n$'s which extend to $K^n_m: M_n\rightarrow M_n$ can be combined into a single linear positive trace preserving map that splits under tensor products $K: M \rightarrow M$, where $M=\bigcup_{n=1}^\infty M_n$. The preservation of rank $1$ operators might then be cast in terms of the continuity of $K$, or that we want to describe $K$ in a coordinate free way.

That $f$ has to depend solely on the maximum eigenvalue can possibly also be derived in another way: the reason $f$ is affine between operators with $\rho\sleq^+ \pi$ is because $L^+(\pi)\subseteq L^+(\rho)$, if $f(\rho)$ were to depend on other eigenvalues of $\rho$ as well, then these eigenspaces must also be preserved in some way by $\rho\sleq_f \pi$, because otherwise affineness would fail. But in general, when mixing the operators together, the ordered eigenvalues might change in order, so it is not clear how a partial order would work that also depended on, say the second highest eigenvalue.

\subsection{Graphical intuition}
\begin{figure}[htb!]
    \centering
    \includegraphics{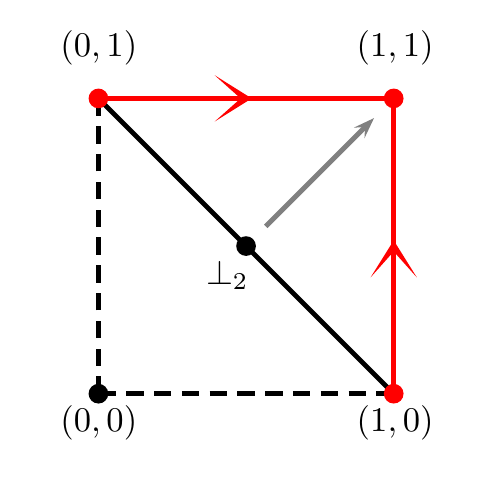}
    \caption[diagonal matrices in $DO(2)$ mapped by $F^+$]{An illustration of the area of diagonal matrices in $PO(2)$ with highest eigenvalue less than 1.}
    \label{fig:2dmod}
    \vspace{-1cm}
\end{figure}

There is a nice graphical intuition behind the operation of the map $F^+: \rho \mapsto \rho/\rho^+$. Consider $n=2$, and pick the subset of diagonal matrices in $DO(2)$ and $PO(2)$. $PO(2)$ can then be seen as $[0,\infty)^2$ and $DO(2)$ is a diagonal line between the points $(1,0)$ and $(0,1)$. See Figure \ref{fig:2dmod}. 

The subspace of $PO(2)$ with highest eigenvalue less than $1$ is then a square with with the sides given by $(0,0),(1,0),(0,1),(1,1)$. The red area illustrates the image of $F^+$. The comparable elements belong to the same red line segment. These lines correspond to different eigenspaces $L^+(\rho)$.

\begin{figure}[htb!]
    \centering
    \includegraphics{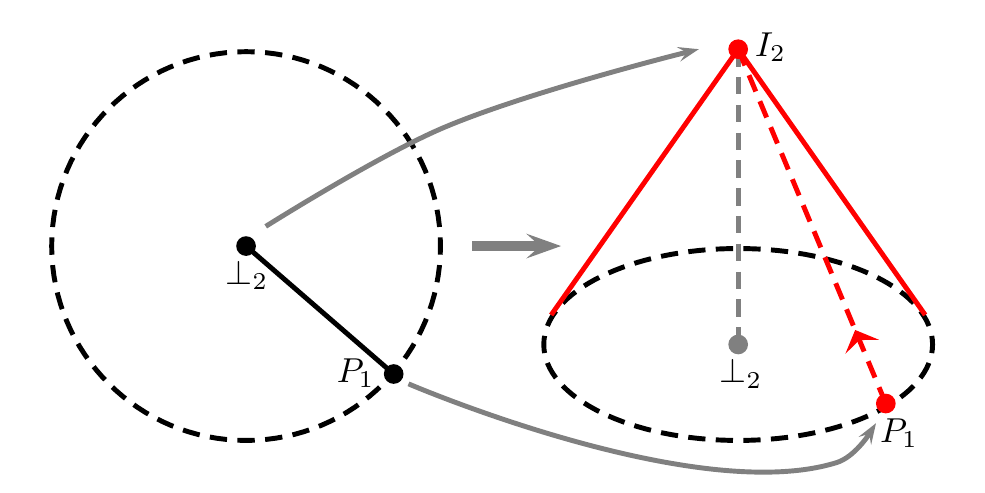}
    \caption[$DO(2)$ mapped into a cone]{An illustration of $DO(2)$ being transformed by $F^+$ into a cone in $PO(2)$.}
    \label{fig:cone}
\end{figure}

The entirety of $DO(2)$ can be seen as $U(1)\times \Lambda^2 \cong S(1) \times [\frac{1}{2},1]$, or in other words, it is the filled circle with $\bot_2$ in the middle and the pure states on the outside. See Figure \ref{fig:cone}. 

The map $F^+$ `pulls' the middle up so that the circle becomes a cone. The points of $F^+(DO(2))$ lie on the outside of the cone. Two points in $DO(2)$ are comparable if they belong to the same line piece from $I_2$ to a pure state.

\begin{figure}[htb!]
    \centering
    \includegraphics{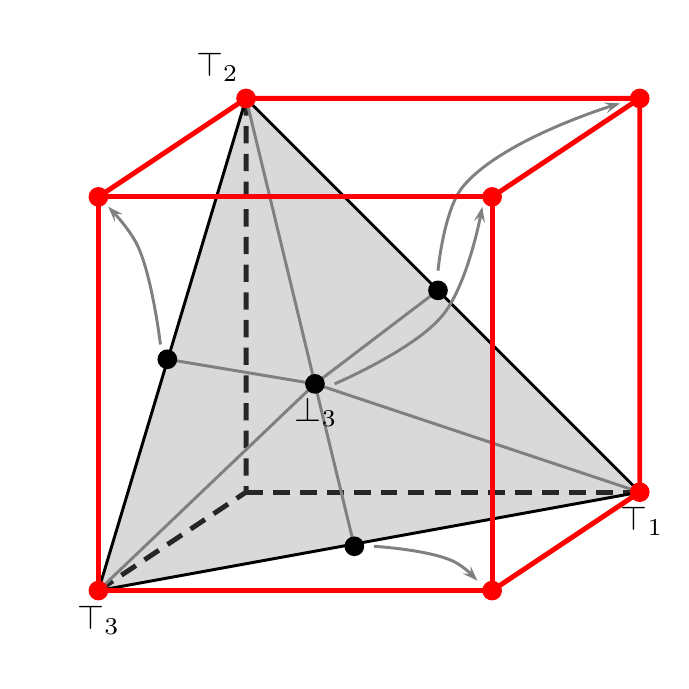}
    \caption[$\Delta^3$ mapped by $F^+$]{$\Delta^3$ as embedded in the space of positive operators with highest eigenvalue less than 1. The grey arrows denote where certain points are mapped to.}
    \label{fig:cube}
\end{figure}

We can draw a similar picture for $DO(3)$. Because humans are bad at visualising high dimensional spaces we'll only consider the diagonal matrices. See Figure \ref{fig:cube}.

The triangle $\Delta^3$ that is embedded into this cube is mapped into three of the faces of the cube. These faces correspond to the eigenspace of the highest eigenvalue $L^+(\rho)$. Points mapped to different faces of the cube are not comparable in $\sleq^+$. The reason $\sleq^+$ allows mixing is then simply that the map $F^+$ is affine as long as points belong to the same `face'.

\section{Extending the order to positive operators}
The Löwner order gives an order structure on the positive operators, that allows comparisons between operators that have a \emph{different} trace: If $A,B \in PO(n)$ and Tr$(A)=$ Tr$(B)$ then either $A\sleq_L B$ or $B\sleq_L A$ implies $A=B$ (in other words, the trace is a strict monotone map). The `distance' between operators $A\sleq_L B$ can be seen as measured by the difference in the trace: Tr$(B-A)$.

The maximum eigenvalue order as defined on $DO(n)$ allows nontrivial comparisons and is not bound by the trace of the operators (as they all have the same trace). Instead the `distance' between two comparable items can be seen as the difference between the maximum eigenvalues. Seeing as $DO(n)$ is a subset of $PO(n)$, one can wonder if we can extend this partial order in a natural way to the entirety of $PO(n)$. We'll show a couple of ways that it can be extended, and the different properties that each of these extensions have.

\bdefin
    The minimal extension: For $A,B \in PO(n)$ let $A\sleq_{min}^+ B$ iff Tr$(A)=$ Tr$(B)$ and $A/$Tr$(A)\sleq^+B/$Tr$(B)$.
\edefin
This extension has $(PO(n),\sleq_{min}^+) \cong (DO(n)\times \mathbb{R}_{> 0} \cup {(0,0)}, \sleq^+\times =)$ order isomorphically, where $\mathbb{R}_{>0}$ is equipped with the equality order: $s\sleq t \iff s=t$. This is a product of two domains with an additional element $(0,0)$, so it is still a domain, and it obviously preserves all the other properties that $\sleq^+$ has, but this extension is not very interesting. It merely glues together copies of $DO(n)$ onto $PO(n)$ and doesn't use any of the extra structure that $PO(n)$ has.

Any map $F: DO(n) \rightarrow PO(n)$ can be extended to a map $F_p: PO(n)\rightarrow PO(n)$ by setting $F_p(0)=0$ and $F_p(A) = $ Tr$(A)F(A/$Tr$(A))$. If we apply this procedure to $F(\rho) = \rho/\rho^+$ we get
\bdefin
    The intuitive extension: For $A,B \in PO(n)$ let $A\sleq_{int}^+ B$ iff Tr$(B)B/B^+ \sleq_L $ Tr$(A)A/A^+$, and we set $0$ as the unique maximal element.
\edefin
This extension works pretty well. In particular it allows composing, and it is a domain, but it doesn't allow mixing: Let $A=$ diag$(2,3)$ and $B= $ diag$(2,1)$, then $A\sleq_{int}^+ B$, but $C= \frac{1}{2}A + \frac{1}{2} B = $ diag$(2,2)$, then not $A\sleq_{int}^+ C$.

We can also define a sort of `maximal extension':
\bdefin 
    The maximal extension: For $A,B \in PO(n)$ let $A\sleq_{max}^+ B$ iff Tr$(A)= $ Tr$(B)$ and $B/B^+\sleq_L A/A^+$ or Tr$(A) > $ Tr$(B)$ and ker$(A) \subseteq $ ker$(B)$.
\edefin
This extension is a quantum information order and includes the Löwner order (if $A\sleq_L^* B$ then $A\sleq_{max}^+ B$), but it is no longer closed or even a dcpo.

\bdefin
    The natural extension: For $A,B \in PO(n)$ let $A\sleq_{nat}^+ B$ iff Tr$(A)\geq $ Tr$(B)$ and $B/B^+ \sleq_L A/A^+$.
\edefin
The order $\sleq_{nat}^+$ has all the properties we require of a quantum information order. Furthermore it is closed, so it is a dcpo, and for any $A\in PO(n)$ let $Z(t) = A + t\bot_n$. Then $Z(t)\ll A$ is an increasing sequence for $t\rightarrow 0$, so $\sleq_{nat}^+$ is also a domain. We also have: if $A\sleq_{nat}^+ B$ and $B\sleq_L^* A$, then Tr$(A)= $ Tr$(B)$ so that $A=B$. So $\sleq_{nat}^+$ and $\sleq_L^*$, the dual Löwner order, are non-contradicting. Furthermore we have $Z(t)\ll A$ in $\sleq_{int}^+$ \emph{and} in $\sleq_L^*$, so the intersection of the orders will again be a domain and this domain also preserves all the properties of a quantum information order. Note that this is indeed a stricter order: if we take $A=$ diag$(3,1$) and $B= $ diag$(2,1)$, then $A\sleq_L^* B$ but not $A\sleq_{nat}^+ B$ and if we let $C = $ diag$(1,0)$ then $\bot_2\sleq_{nat}^+ C$ but not $\bot_2\sleq_L^* C$, so it is not the case that one of the orders is contained in the other.

The dual Löwner order gives information on operators with different traces, while the natural extension of the maximal eigenvalue order also gives information on operators with the same trace. Since they are non-contradictory, a natural question to ask next is wether there is an order that includes the both of them. The answer to that is yes, because the maximal extension does precisely that. However, this extension is not a dcpo and doesn't really use any information on the content of operators with different traces. So the question then becomes, ``is there a quantum information order that encapsulates both $\sleq_L^*$ and $\sleq_{nat}^+$ and is itself also a domain?'' If such an order were to exist we would have a very rich information content structure on the space of positive operators that allows comparisons between operators with the same, or with different traces. Such a partial order would have to include at least the transitive closure of the following relation:
\begin{quote}
    $A \sleq B$ iff Tr$(A) = $ Tr$(B)$ and $A\sleq^+ B$ or $\exists B^\prime\in PO(n)$ such that Tr$(B^\prime) = $ Tr$(B)$ and $B^\prime \sleq^+ B$ and $A\sleq_L^* B^\prime$.
\end{quote}
Since this relation contains an existential quantifier on the downset of $B$ the transitive closure would be very big indeed. It is not clear at the moment what a partial order containing this relation would look like. It could very well be that $\sleq_{max}^+$ is also the \emph{minimal} order containing $\sleq_L^*$ and $\sleq_{nat}^+$.  If this is the case then there is a sort of trade-off in having a partial order based on the structure of $PO(n)$: either you can compare elements with different traces in detail, or you can compare elements with the same trace in detail.
\chapter{Entailment in Distributional Semantics}
In this chapter we will outline a potential application of the structures studied in Chapter 2 and 3. This application is the study of entailment and disambiguation in distributional natural language models. We'll also take a short look at the question of how we can apply entailment at the sentence level by composing words.

\section{Distributional natural language models}
The problem of making computers deal with natural language when you are interested in the semantic content of sentences and words is often tackled by using a distributional natural language model. This is a model based on the distributional hypothesis: \emph{the meaning of a word is defined by the context in which it is used}. What this means is that if you have a large database of written text (a \emph{corpus}), then by merely looking at the context in which each word occurs, that is, which words surround it, you can infer the meaning of the word. Or rather: the meaning \emph{is} the context in which it occurs.

How this works in practice usually is that you take a large corpus of text, for instance the British National Corpus, and you pick a few thousand basis words, usually the most occurring words. Then for every unique word in the corpus you count how many times it \emph{co-occurres} with each of the basis words, say within a distance of five words of the basis word. The result is that for each word in the corpus, you have a vector representing how the word is distributed trough it. This vector is then often normalised in some way to ensure that the vectors are sufficiently comparable.

You can do all kinds of things with this model. For instance, a measure of how similar two words are is the cosine distance between the distributions of the words: if word $a$ and $b$ are represented by vectors $v_a$ and $v_b$, then the similarity is given by
\begin{equation*}
    \text{Sim}(a,b) = \frac{<v_a,v_b>}{\lVert v_a \rVert\lVert v_b\rVert}.
\end{equation*}

We will be looking in more detail at another application.

\section{Entailment and disambiguation}
A common relation between words that often occurs is that of \emph{entailment}: word $a$ entails word $b$ if in a true sentence containing the word $a$, $a$ can usually be replaced by $b$ and still produce a true sentence. An example of such a pair is \emph{dog} and \emph{animal}. If we know that \emph{the dog bites the cat} is true, then \emph{the animal bites the cat} is also true since every dog is an animal. Note that we said that in an entailment pair, the word can \emph{usually} be replaced. It can for instance go wrong when using universal quantifiers: \emph{all dogs eat meat} does not entail \emph{all animals eat meat}. This has to do with the \emph{positivity} of words. In a positive sentence, if we have entailment on each of the words, then we have entailment on the sentence. An existential quantifier is positive, but universal quantifiers and negations are negative. 

Another important thing happening in language is \emph{disambiguation}. This is when an ambiguous word is disambiguated by the context in which it occurs. Consider for instance the word \emph{bank}. Without a suitable context you can't say whether the speaker meant \emph{river bank} or \emph{financial institution}.

While disambiguation and entailment might be seen as two disparate structures in language, distributionally they have something in common. The word \emph{dog} is more narrowly used then the word \emph{animal} since it is more specific. We would expect this to be reflected in the distributional properties of the words. Namely, that \emph{dog} is used in less contexts than \emph{animal}. We expect the same thing to happen in the case of disambiguation. \emph{River bank} disambiguates \emph{bank} because it is more specific, so again we would expect to see it in less contexts.

Another way of viewing these problems is trough the lens of information content. \emph{Dog} contains more information than \emph{animal}, precisely because it is more specific. \emph{Dog bites cat} gives you more information about the situation than \emph{Animal bites cat}. The same goes for \emph{I went to the bank} and \emph{I went to the river bank} (although in this case you would probably expect the first sentence to be referring to the financial institution. Technically you can't know for certain). This view of looking at these problems points towards an application for the partial orders studied in this thesis so far. But first, an overview of what has been done so far in this field.

\section{Known methods}
A very comprehensive paper on different kinds of measures of entailment and other assymetric texual relations was written by Kotlerman et al. \cite{kotlerman2010}. In it they distinguish three important properties that a measure of entailment should have:
\begin{itemize}
    \item Promoting the similarity scores if included features are highly relevant for the narrower term; the estimation of feature relevance may be better based on feature ranks rather than on feature weights.
    \item Promoting the similarity scores when included features are placed higher in the vector of the broader term as well.
    \item Demoting similarities for short feature vectors
\end{itemize}
Features here refer to the components of the word vector, and feature relevance is the relative size of the component with respect to the other components. A \emph{short} feature vector is a vector that contains many zero components. This is a word that doesn't occur much in the corpus, and thus there is more uncertainty about the properties of the word. In the paper they produce a similarity measure based on the Average Precision measure taking these points into account, which does well on standard tests.

Another approach that focuses more on the relation between entailment and compositionality (how entailment at the word level transitions to entailment at the sentence level) is taken in \cite{balkir2016}. Instead of using vectors to represent words, they use density matrices, the logic behind this being that density matrices are more suitable to represent correlations between features which is important for these kinds of problems. The function they based their similarity measure on is the Kullback-Leibler (KL) divergence:
\begin{equation*}
    KL(\rho,\pi) = \text{Tr}(\rho(\ln \rho - \ln \pi))
\end{equation*}
which is the density matrix analog of the \emph{relative entropy} between probability distributions. The normalised version of KL-divergence which they call the \emph{representativeness} is given by
\begin{equation*}
    R(\rho,\pi) = \frac{1}{1+ KL(\rho,\pi)}.
\end{equation*}
This is a number between zero and one. It is only 1 when $\rho=\pi$. The model based on the representativeness performed well when presented with simple sentences.

Another approach was taken in \cite{bankova2016} where they looked at \emph{graded} entailment: a pair like \emph{dog} and \emph{pet} can be considered a partial entailment pair. A lot of dogs are pets, but not all of them. This can be presented by a probability specifying with what chance the entailment holds. This is what is meant by grading. They implement this by representing words by density matrices and then using a graded version of the Löwner order:
\begin{equation*}
    \rho \leq_k \pi \iff \pi - k\rho \geq 0
\end{equation*}
where $k$ is the value of the grading.  We will not go into detail here how the compositionality in distributional models is achieved. For us it suffices to say that the composition of words is achieved by tensoring together the representations of words and then performing some linear map on the resulting tensor that reduces it to a simpler object. For the details see for instance \cite{clark2008, coecke2010}. This graded entailment preserves its structure when words are composed together.

\section{Applications of Information Orders}
 Entailment and disambiguation relations are related to the information content in words or similarly in the distributions that represent those words. Since the information orders on $\Delta^n$ and $DO(n)$ studied in Chapter 2 and 3 were expressly designed to incorporate this idea of information content they might prove suitable for the task of entailment and disambiguation.

Let's first consider the case of restricted information orders (RIO). In empirical natural language models the vector spaces used often have at least a few thousand dimensions. As we saw in Chapter 2, the amount of free parameters that can be chosen for a RIO scales with the square of $n$, so there is a lot of freedom in choosing which particular order you'll use for a specific linguistic application. The downside however is that RIO's can only compare elements that belong to the same monotone sector. The amount of sectors in $\Delta^n$ scales with the factorial of $n$. When we have $n$ in the thousands this would probably mean that each words belongs to its own sector, so that no words are comparable at all. This is a problem for the other information orders as well, so let's discuss some possible remedies.

\subsection{Smoothing}
A technique often used in information retrieval tasks is to use some form of \emph{smoothing} to deal with zeroes in distributions that would otherwise cause problems (by causing singularities on some points for instance). Such smoothing often takes the form of adding a small constant term somewhere or mixing together different distributions in some way to produce a more homogeneous distribution. 

In the case of information orders we will consider the following form of smoothing:
\begin{equation*}
    x \sleq_\alpha y \iff x\sleq \alpha x + (1-\alpha) y.
\end{equation*}

\begin{figure}[htb!]
    \centering
    \includegraphics{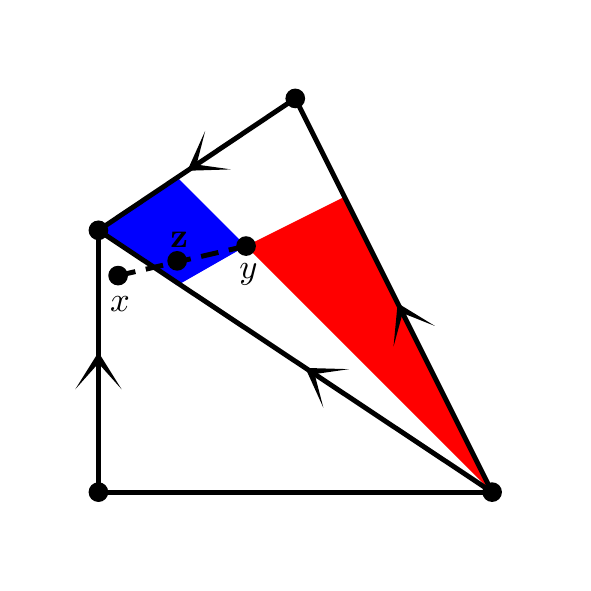}
    \caption[Smoothing in information orders]{An example of how smoothing can produce a new comparison.}
    \label{fig:smoothing}
\end{figure}

The advantage of this is that it leaves the original partial order structure intact, because the orders allowed mixing. So when we have $x\sleq y$ then also $x\sleq \alpha x + (1-\alpha) y$. What this smoothing does, is that while $x$ and $y$ might be in incomparable regions of the distributional space (for instance, for RIO's they could belong to different sectors), if we take a mix of them we might cross into the right sector which makes the distributions comparable. See Figure \ref{fig:smoothing}. Here $z$ denotes the smoothed element between $x$ and $y$. While $x$ would originally not be comparable to $y$, with this smoothing we do have a comparison.

It should be noted that this kind of smoothing breaks transitivity, so it is in general no longer a partial order. The other similarity measures used to study entailment are not partial orders either, so this is not necessarily a problem.

\subsection{Forms of grading}
There are multiple forms of grading we can consider. We'll start with the most intuitive form, based on the grading of the Löwner order considered in \cite{bankova2016}.

The orders on $\Delta^n$ studied in this thesis have the form $x\sleq y$ iff for all $i$: $f_i(x)g_i(y)\leq f_i(y)g_i(x)$ (except for the maximum RIO and $\sleq^-$, there the behaviour is slightly different when presented with elements with different amounts of zeroes, but the general idea still holds). Grading on this partial order can be easily implemented by saying that $k$-graded entailment is given by $x\sleq_k y$ iff for all $i$: $k\cdot f_i(x)g_i(y) \leq f_i(y)g_i(x)$. In the same vain, if we have an order on $DO(n)$ given by $\rho \sleq \pi$ iff $F(\rho)- F(\pi)\geq 0$, then grading is implemented by $\rho \sleq_k \pi$ iff $F(\rho) - kF(\pi) \geq 0$. This is a straightforward generalisation of the grading in \cite{bankova2016} where they took $F=id$.

Another form of grading can be considered based on the three important properties that an asymmetric similarity measure should have as outlined in Kotlerman et al. For a RIO the partial order consists of $n-1$ comparisons and if all these comparisons hold then we say that the elements are comparable.  Recall that the $j$th inequality only makes reference to coordinates $j$ and higher (where the coordinates are ordered from high to low). If we take the advice from Kotlerman et al. then these last inequalities should count less towards our understanding of the entailment relations. Instead of letting the partial order return a binary 1 or 0 determining whether there is entailment or not we can count the amount of inequalities that point the right way, and scale the points awarded based on the exact inequality. So if the partial order is given by $x\sleq y$ iff for all $i$: $F_i(x,y)\leq 0$ (for a RIO $F_i(x,y) = f_i(x)g_i(y)-f_i(y)g_i(x)$), then we could make an assymetric similarity measure in the following way:
\begin{equation*}
    \text{Sim}(x,y) = \sum_i A_i \text{sign}(F_i(x,y)).
\end{equation*}
The $A_i$ are weights that determine how important a certain inequality is. Following the advice of Kotlerman et al. these should be decreasing with increasing $i$. The $A_i$'s should be positive and sum up to 1, so that when $x\sleq y$ we have Sim$(x,y)=1$.\footnote{These conditions actually ensure that $(A_i)\in \Lambda^{n-1}$.} If the $F_i$ is antisymmetric in its arguments (as it the case for a RIO), then we also have Sim$(x,y) = -$ Sim$(y,x)$.

Something similar can be done for the maximum eigenvalue order on $DO(n)$. Let $\rho,\pi \in DO(n)$, and let $(v_i)$ and $(w_i)$ denote orthonormal bases of eigenvectors for $\rho$ respectively $\pi$. Call $s_i = v_i^\dagger(\pi^+\rho - \rho^+\pi)v_i$ and $t_i = w_i^\dagger(\pi^+\rho - \rho^+\pi)w_i$, then we can define
\begin{equation*}
    \text{Sim}(\rho,\pi) = \sum_i A_i(\text{sign}(s_i)+\text{sign}(t_i)).
\end{equation*}
Since again, the spaces corresponding to the higher eigenvalues correspond to the features occurring most of the time they should be assigned higher weights.

It might be that slightly different forms of similarity measures prove more useful in practice. This is merely to show that similarity measures \emph{can} be constructed from these partial orders. Empirical research is needed to assess the practical usefulness of these partial orders, and the specific combination of smoothing and grading of the first and second type that works best.

\section{Compositionality}
The graded entailment of \cite{bankova2016} uses the Löwner order as the structure that defines the grading. Since the Löwner order is the trivial order on $DO(n)$ you only have $\rho\sleq_k \pi$ with $k=1$ when $\rho=\pi$. This might not be what you want since $k$ determines the degree of entailment. It is reasonable to assume that there are pairs of distinct terms that have a `perfect' entailment relation, but this can't fit into this model. In this model when $\rho_1\sleq_k \pi_1$ and $\rho_2\sleq_l \pi_2$ then $\rho_1\otimes \rho_2 \sleq_{kl} \pi_1\otimes \pi_2$. If the $\rho_i$ and $\pi_i$ are distinct then $k$ and $l$ will always be smaller than 1, so we see that tensoring terms together (composing words) gives a smaller level of entailment: $kl<k,l$. 

As they already touched on in their paper, this problem is alleviated by normalising the positive operators in some other way. They reference \cite{dhondt2006} as using the space of positive operators with the maximum eigenvalue bounded by 1. Call this space $MO(n)$ for \emph{maximum eigenvalue normalised} operators. We then see that the map used in Chapter 3 to describe the maximum eigenvalue order $F(\rho) = \rho/\rho^+$ is actually a map $F: DO(n) \rightarrow MO(n)$. So this idea is actually using the maximum eigenvalue order in a disguised way.

Using the maximum eigenvalue order instead of the normal Löwner order doesn't change much. If we have $\rho \sleq^+_k \pi \iff \pi^+\rho - k\rho^+\pi \geq 0$ then we have $\pi \sleq_{k^\prime} \rho$ with $k^\prime = k\rho^+/\pi^+$. The only difference is in what the exact value of $k$ is. The grading with the maximum eigenvalue order behaves better with respect to composing however, since the values for $k$ will in general be higher, so that the entailment strength doesn't artificially drop off as you compose words together. It therefore seems that the maximum eigenvalue order is a more natural choice for doing this kind of entailment.
\chapter*{Conclusion}
\addcontentsline{toc}{chapter}{\tocEntry{Conclusion}}
\pdfbookmark[1]{Conclusion}{Conclusion}
\markboth{Conclusion}{Conclusion}

In this thesis we set out to define some properties that a partial order on $\Delta^n$ or $DO(n)$ should satisfy to qualify as an information order. Starting from a minimal set of properties we saw that there were many examples of such partial orders on $\Delta^n$. When given another restriction, the degeneracy condition, the resulting class of partial orders could be classified, and we saw that this uniquely defines a direction for the partial orders. This class generalises the Bayesian order, which is a specific example of a restricted information order. These restricted information orders can be seen as being defined on the monotone sector $\Lambda^n$ and when restricted to $\Lambda^n$ they are continuous dcpo's (domains). This seems to imply that $\Lambda^n$ is a more \emph{simple} space than $\Delta^n$: the information orders on $\Lambda^n$ have a unique direction, and they are all domains. For information orders on $\Delta^n$ this is no longer the case. There are for instance the two renormalised Löwner orders that contradict each other in some points. The only information order that has been found to be a domain on $\Delta^n$ is the maximum eigenvalue Löwner order. It was further put forward (and hopefully made likely) that this might be the only domain structure on $\Delta^n$ that is also an information order.

When looking for information orders on $DO(n)$ we saw that those coming from $\Delta^n$ are too restrictive. We could reformulate the renormalised Löwner orders to get information orders on $DO(n)$, and they satisfied a new property: the order structure is preserved when composing systems together using the tensor product. It was again seen that only the maximum eigenvalue Löwner order was a domain, and it was shown that a certain construction had this order as the unique solution.

In the introduction it was stated that the question of combining an order and the idea of information content is a rather fundamental one. The only direct work in the direction of this question was done with regard to the Bayesian order, which served as starting point for this thesis. It is the hope of the author that this thesis might serve as a starting point for looking into this question in a bit more depth.

Some interesting open questions that have arisen from this thesis:
\begin{itemize}
    \item What is the least restrictive information order on $\Delta^n$ having $\mu_S$ (Shannon entropy) as a measurement? What about $\mu^+$? Does $\sleq_L^+$ have Shannon entropy as a measurement?
    \item What kind of nonrestricted information orders are there on $\Delta^n$ next to the renormalized Löwner orders? Specifically, are there information orders that satisfy the $k$th degeneracy condition, but not any of the others?
    \item What kind of orders are there on $DO(n)$ that allow the composing of systems? Are there more quantum information orders on $DO(n)$ next to $\sleq_L^+$ and $\sleq_L^-$? Is $\sleq_L^+$ the unique quantum information order that is also a domain?
    \item Is there a way to combine $\sleq_L^+$ and the dual Löwner order into a single order on $PO(n)$? If this is not possible is there some deeper reason behind this?
\end{itemize}

Next to these more formal mathematical questions it might also be interesting to see if the orders seen in this thesis would be practically applicable to computational linguistics.

\cleardoublepage
%********************************************************************
% Bibliography
%*******************************************************
% work-around to have small caps also here in the headline
\manualmark
\markboth{\spacedlowsmallcaps{\bibname}}{\spacedlowsmallcaps{\bibname}} % work-around to have small caps also
%\phantomsection 
\refstepcounter{dummy}
%\addtocontents{toc}{\protect\vspace{\beforebibskip}} % to have the bib a bit from the rest in the toc
\addcontentsline{toc}{chapter}{\tocEntry{\bibname}}
\bibliographystyle{plain}
\label{app:bibliography} 
\bibliography{cite}
\nocite{*}

\appendix
\cleardoublepage
%\part{Appendix}
\chapter{Classification of restricted orders}
Here we will show a complete proof that restricted information orders on $\Lambda^n$ given by $n-1$ pairs of affine functions have to be of the form as seen in Section \ref{sec:classification}.

Let $f(x)g(y)\leq f(y)g(x)$ be the inequality we want to study. We write
\begin{equation*}
    f(x) = \sum_{i=1}^n a_ix_i + c,\quad
    g(x) = \sum_{i=1}^n b_ix_i + d.
\end{equation*}

Writing out the inequality then gives us
\begin{equation*}
    \sum_i da_ix_i + cb_iy_i + \sum_{i,j} a_ib_jx_iy_j + cd \leq
    \sum_i cb_ix_i + da_iy_i + \sum_{i,j} a_jb_ix_iy_j + cd
\end{equation*}
which by grouping terms together can be written as
\begin{equation*}
    \sum_{i=1}^{n-1}\sum_{j=i+1}^nA_{ij} (x_iy_j - x_jy_i) \leq \sum_{i=1}^nB_i (y_i - x_i)
\end{equation*}
or more succinctly as $x^T A y \leq B^T(y-x)$ for some antisymmetric matrix $A$ and a vector $B$. 

The condition $\bot_n\sleq y$ then translates into the inequality 
\begin{equation*}
    \sum_{i=1}^{n-1}\sum_{j=i+1}^n A_{ij}\frac{1}{n}(y_j-y_i) \leq \sum_{k=1}^nB_k(y_k - \frac{1}{n}).
\end{equation*}

Setting $y_1=\ldots =y_{n-1}$ then gives
\begin{equation*}
    0\leq \sum_{k=1}^n B_k(y_1-\frac{1}{n})
\end{equation*}
where the last term is not negative since $y_1\geq \frac{1}{n}$ so that we get
\begin{equation*}
    \sum_{k=1}^nB_k \geq 0.
\end{equation*}

Now comes the main problem. We want to show what restriction on $A$ and $B$ we get when the inequality must satisfy one of the degeneracy conditions. We'll show here the proof for the first degeneracy condition (so when $x\sleq y$ and $y_1=y_2$), the others follow analogously. So what we want to show is that when $f(x)g(y)\leq f(y)g(x)$ with $y_1=y_2$, then we \emph{must}  have $x_1=x_2$. Write $y = (p,p,q_3,\ldots,q_{n-1},q_n)$ and $x=(x_1,\ldots, x_n)$. Expanding the inequality gives 
\begin{align*}
    &A_{12}p(x_1-x_2) + \sum_{j=3}^n(A_{1j}(x_1q_j - px_j) + A_{2j}(x_2q_j-px_j)) \\
    +& \sum_{i=3}^{n-1}\sum_{j=i+1}^nA_{ij}(x_iq_j - q_ix_j)
    \leq B_1(p-x_1)+B_2(p-x_2) + \sum_{k=3}^nB_k(q_k-x_k).
\end{align*}
Since it is our goal to prove $x_1=x_2$, we will write this inequality in terms of the difference between $x_1$ and $x_2$: Write $x_2 = x_1 - \epsilon$ and $x_i = x_1 - \epsilon - \delta_i$.  We then get the following relations:
\begin{align*}
    x_1q_j-px_j &= p(x_1-x_j) - (p-q_j)x_1 = p(\epsilon+\delta_j) - (p-q_j)x_1, \\
    x_2q_j-px_j &= p\delta_j - (p-q_j)(x_1-\epsilon), \\
    x_iq_j-q_ix_j &= x_1(q_j-q_i) - \epsilon(q_j-q_i) - (\delta_iq_j - \delta_jq_i), \\
    B_1(p-x_1) + B_2(p-x_2) &= -B_1\epsilon + (B_1+B_2)(p-x_2), \\
    q_k - x_k &= p - x_k - (p-q_k).
\end{align*}

We can then write the lefthandside (LHS) after dividing by $p$ as
\begin{align*}
    &A_{12}\epsilon + \sum_{j=3}^n\left(A_{1j}(\epsilon + \delta_j) + A_{2j}\delta_j - A_{1j}\frac{p-q_j}{p}x_1 - A_{2j}\frac{p-q_j}{p}(x_1-\epsilon)\right) \\
    &+\sum_{i=3}^{n-1}\sum_{j=i+1}^nA_{ij}\left((x_1-\epsilon)\frac{q_j-q_i}{p} - \frac{\delta_iq_j - \delta_jq_i}{p}\right) \\
    =& \\
    &\epsilon\left(\sum_{j=2}^nA_{1j} + \sum_{j=3}^nA_{2j}\frac{p-q_j}{p} - \sum_{i=3}^{n-1}\sum_{j=i+1}^nA_{ij}\frac{q_j-q_i}{p}\right) \\
    +&\sum_{l=3}^n\delta_l\left(A_{1l}+A_{2l} - \sum_{j=l+1}^nA_{lj}\frac{q_j}{p} + \sum_{i=3}^{l-1}A_{il}\frac{q_i}{p}\right) \\
    +&x_1\left(\sum_{j=3}^n(A_{1j}+A_{2j})\frac{p-q_j}{p} + \sum_{i=3}^{n-1}\sum_{j=i+1}^nA_{ij}\frac{q_j-q_i}{p}\right).
\end{align*}
And the righthandside (RHS) after dividing by $p$ as
\begin{align*}
    &-\frac{B_1}{p}\epsilon + \frac{B_1+B_2}{p}(p-x_1+\epsilon) + \sum_{k=3}^nB_k\left(\frac{p-x_k}{p} - \frac{p-q_k}{p}\right) \\
    =& \frac{B_2}{p}\epsilon + \sum_{k=3}^n\frac{B_k}{p}\epsilon + \sum_{l=3}^n\delta_l \frac{B_l}{p} + \left(\sum_{k=1}^nB_k\right)(1-\frac{x_1}{p}) - \sum_{k=3}^nB_k\frac{p-q_k}{p}.
\end{align*}

In the original inequality we bring every $\epsilon$ or $\delta_l$ term to the left and the rest to the right:
\begin{align*}
    &\epsilon\left(\sum_{j=2}^nA_{1j} + \sum_{j=3}^nA_{2j}\frac{p-q_j}{p} - \sum_{i=3}^{n-1}\sum_{j=i+1}^nA_{ij}\frac{q_j-q_i}{p} - \sum_{k=2}^{n-1}\frac{B_k}{p}\right) \\
    +& \sum_{l=3}^n\delta_l\left(A_{1l}+A_{2l} - \sum_{j=l+1}^nA_{lj}\frac{q_j}{p} + \sum_{i=3}^{l-1}A_{il}\frac{q_i}{p} - \frac{B_l}{p}\right) \\
    \leq& \\
    &\left(\sum_{k=1}^nB_k\right)(1-\frac{x_1}{p}) - \sum_{k=3}^n\frac{B_k}{p}(p-q_k)\\
    +& \frac{x_1}{p}\left(\sum_{j=3}^n(A_{1j}+A_{2j})(p-q_j) + \sum_{i=3}^{n-1}\sum_{j=i+1}^nA_{ij}(q_j-q_i)\right).
\end{align*}

Now let $x_1=p$ and $\delta_l = p-q_l$, then the $\delta_l$ tems on the left cancel out against all the terms on the right so that we are left with 
\begin{equation*}
    \epsilon\left(\sum_{j=2}^nA_{1j} + \sum_{j=3}^nA_{2j}\frac{p-q_j}{p} - \sum_{i=3}^{n-1}\sum_{j=i+1}^nA_{ij}\frac{q_j-q_i}{p} - \sum_{k=2}^n\frac{B_k}{p}\right) \leq 0
\end{equation*}
This inequality holds whenever $\epsilon=0$ (which is fine) or when $\epsilon>0$ and the term after it is negative or zero. Therefore, for any allowable values of $p$ and $q_j$ this term must be strictly positive. Write $C$ for this value. So we have $C>0$. For brevity we will also write $B=\sum_{k=1}^nB_k$. Note that the $\bot_n\sleq y$ condition tells us that $B\geq 0$. 

Now let $x_1 < p$, but keep the $\delta_l$ at the same values: $\delta_l = p-q_l$. By taking the $\delta_l$ terms on the LHS to the RHS we can combine the terms:
\begin{align*}
    C\epsilon &\leq B(1-\frac{x_1}{p}) + (\frac{x_1}{p}-1)\left(\sum_{j=3}^n(p-q_j)(A_{1j}+A_{2j})+\sum_{i=3}^{n-1}\sum_{j=i+1}^nA_{ij}(q_i-q_j)\right) \\
    &= \left(1-\frac{x_1}{p}\right)\left(B - \left(\sum_{j=3}^n(p-q_j)(A_{1j}+A_{2j})+\sum_{i=3}^{n-1}\sum_{j=i+1}^nA_{ij}(q_i-q_j)\right)\right)
\end{align*}

If the RHS is positive then we can find a strictly positive $\epsilon$ that satisfies this inequality. So the RHS must be negative for any allowable values of $p$ and $q_i$. Since $x_1<p$ we have $1-\frac{x_1}{p}>0$. Take $p=q_3=\ldots=q_{n-1}$. Then all the $A_{ij}$ terms cancel and we we are left with
\begin{equation*}
    C\epsilon \leq (1-\frac{x_1}{p})B
\end{equation*}
so that we must have $B\leq 0$, which gives $B=0$.

Now let all the $\delta_l$ terms be free and take them to the RHS. Then the inequality becomes
\begin{align*}
    C\epsilon \leq& \sum_{k=3}^n\frac{B_k}{p}(\delta_k - (p-q_k)) + \sum_{j=3}^n(A_{1j}+A_{2j})((p-q_j)\frac{x_1}{p} - \delta_j)\\ +&\sum_{i=3}^{n-1}\sum_{j=i+1}^nA_{ij}((1-\frac{\delta_j}{p})q_i - (1-\frac{\delta_i}{p})q_j)^.
\end{align*}

Define a new quantity $\phi_k = \delta_k - (p-q_k)$. This value can be positive and negative. Rewriting the inequality with this new quantity gives
\begin{equation*}
    C\epsilon \leq \sum_{k=3}^n\frac{B_k}{p}\phi_k + \sum_{j=3}^n(A_{1j}+A_{2j})((p-q_j)(\frac{x_1}{p}-1)-\phi_j) + \sum_{i=3}^{n-1}\sum_{j=i+1}^nA_{ij}\left(\frac{\phi_i}{p}q_j - \frac{\phi_j}{p}q_i\right)
\end{equation*}

Let $\phi_i=0$ for all $i<n$ and set $p=q_1=q_2=\ldots=q_{n-1}$, then again many terms cancel and we are left with
\begin{equation*}
    C\epsilon \leq \frac{1}{p}B_n\phi_n + (p-q_n)(\frac{x_1}{p}-1)(A_{1n}+A_{2n}) - \phi_n\sum_{i=1}^{n-1}A_{in}
\end{equation*}

By setting $x_1=p$ we then must have $\phi_n(\frac{B_n}{p} - \sum_{i=1}^{n-1}A_{in}) \leq 0$ for any allowable $p$ and $\phi_n$. Since $\phi_n$ can change sign the only way to let this hold is to have the term be equal to zero. But since this must hold for multiple values of $p$ this can only be the case when $B_n = 0 = \sum_{i=1}^{n-1}A_{in}$. 

We can repeat this procedure, by keeping $x_1=p$ and setting all the $\phi_i=0$ except for $\phi_{n-1}$ and letting $p=q_1=q_2=\ldots=q_{n-1}$ The inequality then becomes
\begin{equation*}
    C\epsilon \leq \phi_{n-1}\left(\frac{B_{n-1}}{p} - \sum_{i=1}^{n-2}A_{i(n-1)} + A_{(n-1)n}\frac{q_n}{p}\right)
\end{equation*}
The term must again be zero because $\phi_{n-1}$ can change sign and each individual term must also be zero since we can independently change $p$ and $q_n$. So $B_{n-1} = \sum_{i=1}^{n-2}A_{i(n-1)} = A_{n-1)n} = 0$.

We can continue this procedure $k$ times where $n-k\geq 3$. The $k$th term will look like
\begin{equation*}
    \phi_{n-k}\left(\frac{B_{n-k}}{p} - \sum_{i=1}^{n-k-1}A_{i(n-k)} + \sum_{j=n-k+1}^{n-1}A_{(n-k)j} \frac{q_j}{p}\right).
\end{equation*}
Therefore we will have $B_i=0$ for $i>2$ and $B_1+B_2=0$ and $A_{1j}+A_{2j} = 0$ and $A_{ij} = 0$ when $i>2$ and $j>2$. This procedure gives no condition on $A_{12}$. 

When we do all this for the $k$th degeneracy condition where $y_k=y_{k+1}$, the results are
\begin{align*}
    B_k + B_{k+1} &= 0 \\
    B_i &= 0\quad\forall i\neq k,k+1 \\
    A_{kj} + A_{(k+1)j} &= 0\quad \forall j\neq k,k+1 \\
    A_{ij} &= 0\quad \forall i,j\neq k,k+1
\end{align*}
where we define $A_{ji} = - A_{ij}$. Note that we have no condition on $A_{k(k+1)}$.

Now writing out the original inequalities again in terms of $A$ and $B$ with these new conditions we get
\begin{align*}
    &\sum_{j=1}^n A_{kj}(x_jy_{k+1}-y_jx_{k+1}) - A_{kj}(x_jy_k-y_jx_k) \\
    =& \sum_{j=1}^n A_{kj}[y_j(x_k-x_{k+1})-x_i(y_k-y_{k+1})] \\
    \leq & B_k(y_k-x_k) - B_k(y_{k+1}-x_{k+1}) = B_k((y_k-y_{k+1}) - (x_k-x_{k+1})
\end{align*}
which after grouping terms together becomes
\begin{equation*}
    (x_k-x_{k+1})g(y) \leq (y_k-y_{k+1})g(x)
\end{equation*}
where $g(x) = \sum_{i=1}^n A_{ki}x_i + B_k$. Since $\sum_i x_i = 1$, we can absorb this $B_k$ term into the $A_{ki}$ constants, so that we can write $g(x) = \sum_i A_i x_i$. 

We require that if $y_{k+1}=0$, then $\bot_k \sleq y$, which translates into
\begin{equation*}
    \frac{1}{k}g(y) \leq y_k g(\bot_k) = y_k\sum_{i=1}^k \frac{A_i}{k}
\end{equation*}
which can be rewritten as
\begin{equation*}
    \sum_{i=1}^{k-1} (y_i-y_k)A_i \leq 0
\end{equation*}
for all combinations of $y_i$, so in fact $A_i\leq 0$ for $i<k$. We can absorb the parameter $A_k$ into $A_{k+1}$ because the parameter $A_k$ gives the term $(x_k-x_{k+1})A_ky_k - (y_k-y_{k+1})A_kx_k = -A_k(x_ky_{k+1} - y_kx_{k+1})$, so that we can just as well set $A_{k+1}^\prime = A_{k+1}-A_k$. So we take $A_k=0$. 

Now note that if we take $x=\bot_{k+1}$ then the inequality becomes
\begin{equation*}
    0 \leq (y_k-y_{k+1})\frac{1}{k+1}(\sum_{i=1}^{k-1}A_i + A_{k+1})
\end{equation*}
so that $A_{k+1}\geq 0$ to offset the negative $\sum_{i=1}^{k-1}A_i$. 

Now let $k=n-1$ and pick $x$ with $x_{n-1} = x_n$. The inequality then has the form
\begin{equation*}
    0 \leq (y_{n-1}-y_n)g(x)
\end{equation*}
Suppose there is an $A_i<0$ with $i<k=n-1$. Then if we take $x_{n-1}=x_n\rightarrow 0$, at some point the RHS becomes strictly negative which prevents any $y$ from being bigger than $x$. So $A_i=0$ for $i<n-1$. As we saw we can take $A_k=0$ and we have $A_{k+1}\geq 0$, so this inequality simply becomes
\begin{equation*}
    (x_{n-1}-x_n)y_n \leq (y_{n-1}-y_n)x_n
\end{equation*}
by dividing out the $A_n$ term. We can do exactly the same procedure for the other $k$'s, so that $A_i=0$ for $i\leq k$. So the inequality for a $k$th degeneracy has a $g$ of the form
\begin{equation*}
    g(x) = y_{k+1} + \sum_{j=k+2}^n A_jy_j
\end{equation*}
where we have divided out the $A_{k+1}$ term. This concludes the proof of the categorization of the inequalities.

\end{document}